\documentclass [a4paper, 12pt, twoside, onecolumn]{article} 

\usepackage{a4wide, amsmath, amsfonts, amssymb, geometry, fancyhdr}
\usepackage{mathrsfs} 
\usepackage{fancyhdr, dsfont}
\usepackage[latin1]{inputenc}
\usepackage[pdftex]{graphicx}
\usepackage{float}
\usepackage{epstopdf}
\usepackage{slashed}
\usepackage{bm}

\restylefloat{figure}

\newcommand{\nwc}{\newcommand}
\nwc{\ba}  {\begin{array}}
\nwc{\ea}  {\end{array}}
\nwc{\bdm} {\begin{displaymath}}
\nwc{\edm} {\end{displaymath}}
\nwc{\bea} {\begin{equation}\ba{lcl}}
\nwc{\eea} {\ea\end{equation}}
\nwc{\bda} {\bdm\ba{lcl}} 
\nwc{\eda} {\ea\edm}
\nwc{\bc}  {\begin{center}}
\nwc{\ec}  {\end{center}}
\nwc{\ds}  {\displaystyle}
\nwc{\nn} {\nonumber}
\nwc{\nnn} {\nonumber \vspace{.2cm} \\ }
\nwc{\ra}{\rightarrow}
\nwc{\lra}{\longrightarrow}

\newcommand{\ve}{\vec} 
\newcommand{\vecb}{\left(\begin{array}{c}} 
\newcommand{\vece}{\end{array}\right)} 
\newcommand{\ccb}{\left(\begin{array}{cc}} 
\newcommand{\cce}{\end{array}\right)} 
\newcommand{\cccb}{\left(\begin{array}{ccc}} 
\newcommand{\ccce}{\end{array}\right)} 
\newcommand{\ccccb}{\left(\begin{array}{cccc}} 
\newcommand{\cccce}{\end{array}\right)} 
\newcommand{\cccccb}{\left(\begin{array}{ccccc}} 
\newcommand{\ccccce}{\end{array}\right)} 

\newcommand{\al}{\alpha}
\newcommand{\be}{\beta}
\newcommand{\ga}{\gamma}
\newcommand{\de}{\delta}

\newcommand{\vep}{\varepsilon}
\newcommand{\si}{\sigma}
\newcommand{\la}{\lambda}
\newcommand{\ka}{\kappa}
\newcommand{\om}{\omega}

\newcommand{\z}{\theta}
\newcommand{\znu}{\theta_{\nu}}
\newcommand{\tspin}{\Theta^{\vec{a}}_{\vec{b}}}
\newcommand{\ts}{\tilde{s}}
\newcommand{\tpsi}{\tilde{\Psi}}
\newcommand{\spin}{^{\vec{a}}_{\vec{b}}}
\newcommand{\bi}{\bar\imath}
\newcommand{\bj}{\bar\jmath}

\newcommand{\De}{\Delta}

\newcommand{\Om}{\Omega}

\newcommand{\pa}{\partial} 
\newcommand{\dd}{\mathrm{d}} 

\newcommand{\beq}{\begin{equation}} 
\newcommand{\eeq}{\end{equation}} 
\newcommand{\eq}{ \ \ = \ \ } 
\numberwithin{equation}{section} 

\newcommand{\te}{\textrm} 
\newcommand{\co}{\ , \ \ \ \ \ \ } 
\newcommand{\mto}{\rightarrow} 
\newcommand{\ee}{\mathrm{e}} 

\usepackage{color}
\definecolor{dgreen}{rgb}{0,0.70,0.30}
\definecolor{gold}{rgb}{0.85,.66,0}

\newcommand{\dal}{\dot{\alpha}}
\newcommand{\dbe}{\dot{\beta}}
\newcommand{\dga}{\dot{\gamma}}
\newcommand{\dde}{\dot{\delta}}
\newcommand{\dka}{\dot{\kappa}}

\newcommand{\dom}{\dot{\omega}}

\newcommand{\RR}{\mathbb R}
\newcommand{\CC}{\mathbb C}

\newcommand{\ZZ}{\mathbb Z}

\allowdisplaybreaks

\begin{document}

\title{\Huge\textbf{\hskip-1cm Higher Loop Spin Field Correlators} \\ 
\textbf{in $\bm{D=4}$ Superstring Theory} \\[2cm]}
\author{\large O. Schlotterer\\[2cm]}
\date{}
\maketitle
\vskip-2cm
\centerline{\it Max--Planck--Institut f\"ur Physik, 
Werner--Heisenberg--Institut,}
\centerline{\it 80805 M\"unchen, Germany}

\medskip\bigskip
\begin{abstract}
We develop calculational tools to determine higher loop superstring correlators involving massless fermionic and spin fields in four space time dimensions. These correlation functions are basic ingredients for the calculation of loop amplitudes involving both bosons and fermions in $D=4$ heterotic and superstring theories. To obtain the full amplitudes in Lorentz covariant form the loop correlators of fermionic and spin fields have to be expressed in terms of $SO(1,3)$ tensors. This is one of the main achievements in this work.
\end{abstract}

\vskip4cm
\begin{flushright}
{\small  MPP--2009--197}
\end{flushright}


\newpage
\setcounter{tocdepth}{2}
\tableofcontents
\pagebreak

\numberwithin{equation}{section}

\section{Introduction and Summary}
\label{sec:Introduction}




String amplitudes are rich of many symmetries such as target--space
duality as symmetry in the underlying moduli space or symmetries
revealed from the structure of the underlying string worldsheet.
Hence, at the formal level amplitudes play an important role in understanding the structure and symmetries of string theory. Moreover, string amplitudes are of considerable phenomenological interest in describing parton scattering at high energy colliders, for a recent account see \cite{LHC1,LHC2}.

\medskip
In order to compute superstring amplitudes in the Ramond Neveu-Schwarz (RNS) formalism in $D=4$ compactifications one has to deal with the bosonic string coordinate $X^{\mu}$ (and its exponentials), with an $SO(1,3)$ vector of worldsheet fermions $\psi^{\mu}$ interacting with the spacetime part of Ramond spin fields $S_{\al}$, $S_{\dbe}$, the (super--)ghost system and an internal decoupled superconformal field theory (SCFT) describing the compactification details. The $D=4$ spin fields transform as left- and right handed spinors of the four dimensional Lorentz group. Being a free field, the $X^{\mu}$ do not pose any problems in correlation functions. In the presence of spacetime fermions among the external states, on the other hand, the interaction of the worldsheet fermions $\psi^\mu$ with their spin fields turns the computation of higher point correlators into a nontrivial problem -- already at tree level but even more so on higher genus. In \cite{tree} we have derived a general strategy to obtain arbitrary correlation functions involving the $D=4$ fields $\psi^{\mu}, S_{\al}$ and $S_{\dbe}$ at tree level. It is one of the main purpose of this work to generalize these results to higher loop. These RNS correlators are the key ingredients necessary for the computation of general superstring amplitudes on arbitrary genus. The ghost correlation functions on arbitrary genus are treated in reference \cite{ghost}. Finally, the internal fields and their higher loop interactions depend on the compactification details.

\medskip
To make these statements a bit more precise, let us give the schematic form of an $N$ point multiloop amplitude ${\cal M}_g(\Phi_1,..., \Phi_N)$ of open string states $\Phi_{i}$ in a four dimensional setting at genus $g$ (for more details see \cite{VV1,Olaf3,HP}):
\begin{align}
{\cal M}_g(\Phi_1,...,\Phi_N) \eq &\int \frac{\dd^{N_g} \Om}{\det \Om}  \int \frac{ \prod_{i=1}^N \dd z^i}{{\cal V}_{\te{CKG}}^g} \; \sum_{I} {\cal K}_I(\Phi_i) \, C_{X}^I(z_i,\Phi_i,\Om) \notag \\
& \times \ \sum_{(\ve{a},\ve{b})} {\cal Z}^{(\ve{a},\ve{b})}(\Om) \, C^{I;(\ve{a},\ve{b})}_{\psi,S}(z_i,\Phi_i,\Om)  \, C^{(\ve{a},\ve{b})}_{\te{ghost}}(z_i, \Phi_i,\Om) \, C^{(\ve{a},\ve{b})}_{\te{int}}(z_i, \Phi_i,\Om)
\label{amp}
\end{align}
There are various tasks to perform towards the final result of the amplitude:
\begin{itemize}
\item Compute the vacuum amplitude (genus $g$ partition function)
${\cal Z}^{(\ve{a},\ve{b})}$ and the correlation functions of the following four decoupled CFTs: the bosonic $X^\mu$ correlator, the RNS contribution from $\psi^\mu,S_\al, S^{\dbe}$ as well as the correlation functions $C_{\te{ghost}}$ and $C_{\te{int}}$ due to (super-)ghosts and the internal degrees of freedom, respectively. The signatures of the detailled compactification model only enter through ${\cal Z}^{(\ve{a},\ve{b})}$ and the internal CFT correlators $C_{\te{int}}$. The latter may be built in and described by some character valued partition function or elliptic genus \cite{EG1,EG2,EG3,EG4,EG5,EG6}.

\item The index $I$ refers to a set of spacetime kinematics ${\cal K}_I$ originating from contractions of the spacetime fields $X^\mu, \psi^\mu, S_\al, S^{\dbe}$, i.e. from correlators with non-trivial Lorentz structure. We denote the $(z_i,\Om)$ dependence associated with the contraction ${\cal K}_I$ by $C_{X}^I$ and $C_{\psi,S}^I$.
 
\item Perform the sum over spin structures $(\ve{a},\ve{b})$ of the partition function ${\cal Z}^{(\ve{a},\ve{b})}$ together with the three $(\ve{a},\ve{b})$ dependent correlation functions. This is achieved by means of generalized Riemann identities
\cite{fay,mum}. For more details, in particular for higher point applications, see \cite{Olaf3,Sloop1,Sloop2}.

\item Integrate the spin summed correlators over worldsheet positions $z_i$ modulo (genus dependent) conformal Killing group of volume ${\cal V}_{\te{CKG}}^g$.

\item Integrate the intermediate result over the moduli space of genus $g$ Riemann surfaces (of $g$ dependent dimension $N_g$) i.e. over inequivalent $g \times g$ period matrices $\Om$.
\end{itemize}
This paper takes a closer look at the RNS correlation function $C_{\psi,S}$, an important part of the first step in the above list. Because of the interacting nature of the fields $\psi^\mu$ and $S_{\al,\dbe}$ this piece is much more involved to evaluate than the free field correlators of the $X^\mu$ coordinate and the superghosts.

\medskip
The remaining steps towards the full amplitude -- spin structure sum, 
integration over world sheet positions $z_i$ and modular integration -- will be left for future work. In particular, the last two points still require a unified treatment. 

\medskip
Let us now focus on the $D=4$ version of the RNS CFT. The $SO(1,3)$ spin fields factorize into two independent copies of an $SO(2)$ spin model, which is a system of one complex Weyl fermion $\Psi^{\pm}(z)$ and its associated spin fields $s^{\pm}(z)$. The left- and right handed $D=4$ spinors correspond to alike and different Ramond charges respectively:
\beq
S_{\al} \eq s^{\pm} \, \otimes \, \ts^{\pm} \co S_{\dbe} \eq s^{\pm} \, \otimes \, \ts^{\mp}
\label{1,1}
\eeq
Starting point for pushing the results of \cite{tree} to loop level is \cite{AS1} where closed formulae are given for torus correlation functions with spin fields of a single $SO(2)$ system. In \cite{AS3}, the discussion was extended to correlators with both spin fields and fermions of the $SO(2)$ spin model. The analysis is carried over to higher loops -- i.e. to Riemann surfaces of arbitrary genus $g$ -- in \cite{AS2}. All the three references discuss applications to fermion amplitudes in ten dimensional settings by means of a five-fold factorization generalizing (\ref{1,1}). However, the problem of finding $SO(1,9)$ covariant expressions for the correlator as such is not solved in a systematic way. The generalization of the $D=4$ results in this work to higher spacetime dimensions $D=6,8,10$ (in terms of the corresponding $SO(1,D-1)$ tensors) is in progress \cite{OSDH}.

\medskip

Unluckily, the technique of bosonizing spin fields becomes more subtle beyond tree level \cite{AG1,AG2,AG3}. At nonzero genus, the partition function for fermions of definite spin structure arises from projecting the associated bosonic partition function onto sectors of certain soliton- or winding numbers. Only the sum over all the fermionic spin structures can establish equivalence to a bosonic theory with any winding numbers $(\vec{m} , \vec{n}) \in \ZZ^g \times \ZZ^g$ around the $2g$ cycles of the maximal torus allowed. Hence, there is no point in identifying spin fields $s^{\pm}$ within a fixed spin structure sector with a free exponential $e^{\pm iH/2}$ of a worldsheet boson $H$. Another method which becomes less powerful beyond tree level are Lorentz Ward identities. They allow to neatly reduce tree level correlators with $SO(1,3)$ current insertions $\psi^{[\mu} \psi^{\nu]}$ to smaller ones without the current \cite{LE1,LE2}. At genus $g \neq 0$, however, one is faced with inhomogeneities in form of $\al$ cycle integrals over the ``bigger'' correlation including the current \cite{EO}.

\medskip 
There is an alternative approach to the superstring which avoids the interacting CFT and the spin structure sums of the RNS framework -- namely the hybrid formalism \cite{hybrid1}. It is based on some non-trivial field redefinitions which replace the interacting RNS fields $\psi^\mu$, $S_\al$ and $S_{\dot \beta}$ by a new set of free worldsheet fields. First steps towards loop amplitudes in the hybrid formalism have been performed in \cite{hybrid2}.


\medskip
In the following, we will combine the two $SO(2)$ constituents of the $D=4$ CFT to $SO(1,3)$ covariant expressions for various correlation functions -- separately in each spin structure. They pave the way to compute various multileg open- and closed string amplitudes. In particular, the closed formula we will give for $\langle \psi^{\mu_{1}} \, ... \, \psi^{\mu_{2k+2g-1}} S_\al S_{\dbe} \rangle$ enables to compute scattering of $k$ gluons with two spacetime fermions at $g$ loops. Processes with four or more fermions (or RR forms) require knowledge of correlators with the corresponding number of spin fields; a collection of these is systematically derived.

\medskip
This paper is organized as follows: We start by reviewing various aspects of the four dimensional RNS CFT, $SO(2)$ spin models and theta functions in the following section \ref{sec:Review}. These techniques are used to express correlation functions with $SO(1,3)$ spin fields in terms of Lorentz tensors in section \ref{sec:cov}. Closed formulae are found for correlators with arbitrary number of left handed spin fields $S_{\al_i}$ as long as the number of right handed spinors $S_{\dbe_j}$ does not exceed four. The final section \ref{sec:mix} contains mixed correlators with NS fermions present. In particular, the physically most relevant cases with two spin fields and an arbitrary number of $\psi$'s are given in a closed formula.

\medskip
The main part is followed by two appendices. The first one gives some technical details about theta functions, the second one is devoted to the proofs by induction for our expressions claimed for the correlators $\langle S_{\al_1} ... S_{\al_{2M}} \rangle$, $\langle S_{\al_1} ... S_{\al_{2M}} S_{\dga} S_{\dde} \rangle$, $\langle S_{\al_1} ... S_{\al_{2M}} S_{\dga} S_{\dde} S_{\dka} S_{\dom} \rangle$ and $\langle \psi^{\mu_1} ... \psi^{\mu_{2n-1}} S_\al S_{\dbe} \rangle$, $\langle \psi^{\mu_1} ... \psi^{\mu_{2n-2}} S_\al S_{\be} \rangle$.

\section{Review}
\label{sec:Review}

This section takes a closer look at the material we will build upon in the following. We need both the techniques adapted to $D=4$ dimensions \cite{tree} and the one- or higher loop results of \cite{AS1,AS3,AS2} for a single $SO(2)$ spin model. Also, some conventions have to be fixed concerning the generalized theta functions and the precise dictionary beween $SO(1,3)$ spin fields $S_{\al,\dbe}$ and their $SO(2)$ constituents $s^{\pm},\tilde{s}^{\pm}$.

\subsection[The RNS CFT in $D=4$ dimensions]{The RNS CFT in $\bm{D=4}$ dimensions}
\label{sec:D=4}

Let us first of all give the OPEs of the relevant $SO(1,3)$ covariant fields $\psi^{\mu}$, $S_{\al}$ and $S_{\dbe}$ in the convention of \cite{tree}. OPEs are local statements, so they contain information relevant on Riemann surfaces of arbitrary genus.
\begin{subequations}
\begin{align}
\psi^{\mu}(z) \, \psi^{\nu} (w) \ \ &= \ \ \frac{\eta^{\mu \nu}}{z \, - \, w} \ + \ ... \label{rv,1a} \\
S_{\al}(z) \, S_{\dbe}(w) \ \ &= \ \ \frac{1}{\sqrt{2}} \; (z \, - \, w)^{0} \, \si^{\mu}_{\al \dbe} \, \psi_{\mu}(w) \ + \ ... \label{rv,1b} \\
S_{\al}(z) \, S_{\be}(w) \ \ &= \ \ - \, ( z \, - \, w)^{-1/2} \, \vep_{\al \be} \ + \ ... \label{rv,1d} \\
S_{\dal}(z) \, S_{\dbe}(w) \ \ &= \ \ +  \, ( z \, - \, w)^{-1/2} \, \vep_{\dal \dbe}  \ + \ ... \label{rv,1e} \\
\psi^{\mu}(z) \, S_{\al}(w) \ \ &= \ \ \frac{1}{\sqrt{2}} \; (z \, - \, w)^{-1/2} \, \si^{\mu}_{\al \dbe} \, S^{\dbe}(w) \ + \ ... \label{rv,2a} \\
\psi^{\mu}(z) \, S^{\dal}(w) \ \ &= \ \ \frac{1}{\sqrt{2}} \; (z \, - \, w)^{-1/2} \, \bar{\si}^{\mu \dal \be} \, S_{\be}(w) \ + \ ...
\label{rv,2b}
\end{align}
\end{subequations}
In particular, the zero $(z-w)$ power in the OPE (\ref{rv,1b}) is specific to $D=4$ dimensions and allows to factorize the fermions $\psi^{\mu}$ into two spin fields by setting $z=w$:
\beq
\psi^{\mu}(z) \eq - \, \frac{1}{ \sqrt{2} } \; \bar{\si}^{\mu \dbe \al} \, S_{\al}(z) \, S_{\dbe}(z)
\label{2,1}
\eeq
These OPEs can be derived by constructing the fields $\bigl\{ \psi^\mu , S_\al, S_{\dbe} \bigr\}$ from two independent copies of an $SO(2)$ spin system $\bigl\{ \Psi^{\pm}(z), s^{\pm}(z) \bigr\}$ and $\bigl\{ \tpsi^{\pm}(z), \ts^{\pm}(z) \bigr\}$ where $s^{\pm},\tilde{s}^{\pm}$ creates branch cuts for the associated fermion. According to the signs in the superscripts, one associates Ramond charge $\pm 1$ to the $\Psi^{\pm}$ and $\pm \frac{1}{2}$ to $s^{\pm}$. Singularities occur whenever fields of opposite charge approach each other,
\begin{subequations}
\begin{align}
\Psi^{\pm}(z) \, \Psi^{\mp} (w) \ \ &= \ \ \frac{1}{z \, - \, w} \ + \ ... \label{2,1a} \\
s^{\pm}(z) \, s^{\mp}(w) \ \ &= \ \ \frac{1}{(z \, - \, w)^{1/4}}  \ + \ ... \label{2,1b} \\
\Psi^{\pm}(z) \, s^{\mp}(w) \ \ &= \ \ \frac{s^{\pm}(w)}{(z \, - \, w)^{1/2}}  \ + \ ...  \ .\label{2,1c} 
\end{align}
\end{subequations}
whereas fields of alike charge exhibit regular behaviour:
\begin{subequations}
\begin{align}
\Psi^{\pm}(z) \, \Psi^{\pm} (w) \ \ &= \ \ (z \, - \, w) \, \Psi^{\pm}(w) \, \pa \Psi^{\pm}(w) \ + \ ... \label{2,1d} \\
s^{\pm}(z) \, s^{\pm}(w) \ \ &= \ \ (z \, - \, w)^{1/4} \, \Psi^{\pm}(w)  \ + \ ... \label{2,1e} \\
\Psi^{\pm}(z) \, s^{\pm}(w) \ \ &= \ \ (z \, - \, w)^{1/2} \, \hat{s}^{\pm}(w)  \ + \ ... \label{2,1f} 
\end{align}
\end{subequations}
In the last OPE, $\hat{s}^{\pm}$ denotes an excited spin field of conformal weight $9/8$. Analogous statements hold for the ingredients $\tpsi^{\pm}(z), \ts^{\pm}(z)$ of the second spin model.

\medskip
Identifying $\Psi^{\pm} = \ee^{\pm i H}$ and $s^{\pm} = \ee^{\pm iH/2}$ with some boson subject to $H(z) H(w) \sim \ln (z-w)$ certainly reproduces (\ref{2,1a}) to (\ref{2,1f}) but only yields the average over spin structures. As the short distance behaviour remains valid even without bosonizing $\Psi^{\pm}, s^{\pm}$ we will never make use of it.

\medskip
We pick the following convention for $SO(2)$ factorizing the $SO(1,3)$ fields (up to cocycles)
\begin{align} 
S_{\al=(+,+)}(z) \ \ &= \ \ s^{+}(z) \, \otimes \, \ts^{+}(z) \co S_{\al=(-,-)}(z) \eq s^{-}(z) \, \otimes \, \ts^{-}(z) \label{2,1g} \\
 S_{\dbe=(+,-)}(z) \ \ &= \ \ s^{+}(z) \, \otimes \, \ts^{-}(z) \co  S_{\dbe=(-,+)}(z) \eq s^{-}(z) \, \otimes \, \ts^{+}(z) 
\label{2,1h}
\end{align}
The entries of the two dimensional $\vep$ tensor are taken to be
\beq
\vep_{(+,+),(-,-)} \eq - \, \vep_{(-,-),(+,+)} \eq \vep_{(+,-),(-,+)} \eq - \, \vep_{(-,+),(+,-)} \eq + \, 1 \ .
\label{vep}
\eeq
Given the explicit form $\si^\mu = (-\mathds{1} ,\si^i)$ and $\bar{\si}^\mu = (-\mathds{1},-\si^i)$ of the sigma matrices in (\ref{rv,1b}), (\ref{rv,2a}), (\ref{rv,2b}) in terms of the standard Pauli matrices $\si^i$, the four vector $\psi^\mu$ turns out to decompose as
\beq
\vecb \psi^{0}(z) \\ \psi^{1}(z) \\ \psi^{2}(z) \\ \psi^{3}(z) \vece \eq \frac{1}{\sqrt{2}} \; \vecb \Psi^{+}(z) + \Psi^{-}(z) \\  \tpsi^{+}(z) + \tpsi^{-}(z) \\  i\tpsi^{+}(z) - i\tpsi^{-}(z) \\ \Psi^{+}(z) - \Psi^{-}(z) \vece \ .
\label{2,1i}
\eeq

\subsection{Generalized Theta functions}
\label{sec:Gen}

The aim of this work is to compute correlation functions on Riemann surfaces of arbitrary genus $g$. The natural ingredients to implement the required periodicities along the homology cycles $\al_I$, $\be_J$ (with $I,J=1,2,...,g$) are the {\em generalized theta functions} \cite{fay,mum,igu}. They are obtained from the archetype 
\beq
\Theta( \ve{x} \, | \, \Om) \ \ := \ \ \sum_{\ve{n} \in \ZZ^g} \exp \left[ 2\pi i \, \left( \tfrac{1}{2} \; \ve{n} \, \Om \, \ve{n} \ + \ \ve{n} \, \ve{x} \right) \right]
\label{defT1}
\eeq
by shifting the $\ve{x} \in \CC^g$ argument according to some characteristics or spin structure $(\ve{a},\ve{b})$:
\begin{align}
\Theta \biggl[ \begin{array}{c} \ve{a}
    \\ \ve{b} \end{array} \biggr] ( \ve{x} \, | \, \Om) \ \ &:= \ \ \exp \left[ 2\pi i \, \left( \tfrac{1}{8} \; \ve{a} \, \Om \, \ve{a} \ + \ \tfrac{1}{2} \; \ve{a} \, \ve{x}  \ + \ \tfrac{1}{4} \; \ve{a}\, \ve{b} \right) \right] \, \Theta  \left( \ve{x} \, + \, \tfrac{\ve{b}}{2}  \, + \, \tfrac{\Om \, \ve{a}}{2} \, | \, \Om \right) \notag \\
    & \; = \ \ \sum_{\ve{n} \in \ZZ^g} \exp \left[ \pi i \, \left(  \ve{n} \, + \, \tfrac{\ve{a}}{2} \right) \, \Om \, \left(  \ve{n} \, + \, \tfrac{\ve{a}}{2} \right) \ + \ 2\pi i \, \left(  \ve{n} \, + \, \tfrac{\ve{a}}{2} \right) \, \left( \ve{x} \, + \, \tfrac{\ve{b}}{2} \right) \right]
\label{defT2}
\end{align}
The period matrix $\Om$ is defined by the integrals of the $g$ normalized Abelian differentials $\om_I$ along the $\be$ cycles:
\beq
\oint_{\al_J} \om_K \eq \de_{JK} \co \oint_{\be_J} \om_K \eq \Om_{JK}
\label{defT3}
\eeq
It enters the $\Theta$ functions as a second argument and will be suppressed in later sections of this paper.

\medskip
We will parametrize the two dimensional genus $g$ worldsheet by a complex coordinate $z$. The Abel map $z \mapsto \smallint^z_{p} \ve{\om}$ (with some reference point $p$ which drops out in all applications) lifts $z$ to the worldsheet's Jacobian variety $\cong \CC^g / (\ZZ^g + \Om \ZZ^g)$. These integrals are then natural arguments for $\Theta$ functions. This is the way $z$ enters the correlation functions.
The (pseudo-) periodicity properties of the theta functions (\ref{defT2}) under transport of $z$ around a homology cycle are listed in appendix \ref{appB}. 

\medskip
The most important version of the generalized theta functions is the prime form $E$, the unique holomorphic bidifferential of weight $\left( -\frac{1}{2} , - \frac{1}{2} \right)$ with a single zero at $z=w$:
\beq
E(z,w) \ \ := \ \ \frac{ \Theta \biggl[ \begin{array}{c} \ve{a}_0
    \\ \ve{b}_0 \end{array} \biggr] \left( \int^z_w \ve{\om} \, | \, \Om \right) }{ h \biggl[ \begin{array}{c} \ve{a}_0
    \\ \ve{b}_0 \end{array} \biggr](z) \ h \biggl[ \begin{array}{c} \ve{a}_0
    \\ \ve{b}_0 \end{array} \biggr] (w) } 
\label{defE}
\eeq
In this definition, $(\ve{a}_0,\ve{b}_0)$ represent an arbitrary odd spin structure (i.e. $\ve{a}_0 \ve{b}_0$ is odd and therefore $E(z,w) = -E(w,z)$). The half differentials $h$ in the denominator ensure that the prime form is in fact independent on the choice of the odd spin structure $(\ve{a}_0,\ve{b}_0)$. Explicitly, $h$ is given by
\beq
h \biggl[ \begin{array}{c} \ve{a}_0
    \\ \ve{b}_0 \end{array} \biggr](z) \ \ := \ \ \sqrt{ \sum_{j=1}^g \, \om_j(z) \, \pa_j \Theta \biggl[ \begin{array}{c} \ve{a}_0
    \\ \ve{b}_0 \end{array} \biggr] \left( \ve{0} \, | \, \Om \right)  } \ .
\eeq
The essential property of the prime form is its singularity structure
\beq
E(z,w) \eq \frac{1}{z \, - \, w} \ + \ {\cal O}(z-w) \ .
\eeq
Several simplfications occur at genus $g=1$, i.e. on the torus. The period matrix $\Om$ reduces to a single complex number $\tau$, and the associated theta functions are the standard ones:
\beq
\theta_{1} \ \ := \ \ \Theta \biggl[ \begin{array}{c} 1
    \\ 1 \end{array} \biggr] \co \theta_{2} \ \ := \ \ \Theta \biggl[ \begin{array}{c} 1
    \\ 0 \end{array} \biggr] \co \theta_{3} \ \ := \ \ \Theta \biggl[ \begin{array}{c} 0
    \\ 0 \end{array} \biggr] \co \theta_{4} \ \ := \ \ \Theta \biggl[ \begin{array}{c} 0
    \\ 1 \end{array} \biggr] 
    \label{dict}
\eeq
Due to the simple structure $\om = \dd z$ of the holomorphic differential, the prime form coincides with the unique odd theta function:
\beq
E(z,w) \, \Bigl. \Bigr|_{g=1} \eq \frac{\theta_1(z-w \, | \, \tau) }{ \pa_z \theta_1(0 \, | \, \tau)}
\label{tor}
\eeq

\subsection{Loop correlators of a single spin system}
\label{sec:Spinloop}

In a series of papers \cite{AS1,AS3,AS2} the $SO(2)$ spin system was completely solved on Riemann surfaces of arbitrary genus. We reexpress the main result for the correlation functions in terms of the prime form (\ref{defE}) rather than some odd reference spin structure. The equivalence to the formula (10) in \cite{AS2} can be easily verified by counting half differentials.

\medskip
First of all, the $2n$ point correlation function of spin fields with spin structure $(\ve{a},\ve{b})$ for the fermions is given by: 
\beq
\left \langle \prod_{i=1}^n s^{+}(z_i) \, s^{-}(w_i) \right \rangle^{\ve{a}}_{\ve{b}} \eq \frac{\Theta \biggl[ \begin{array}{c} \ve{a}   \\ \ve{b} \end{array} \biggr] \left( \frac{1}{2} \, \sum_{i=1}^n \smallint^{z_i}_{w_i} \ve{\om} \right)}{ \Theta \biggl[ \begin{array}{c} \ve{a}   \\ \ve{b} \end{array} \biggr] (  \ve{0} ) } \; \left( \frac{ \prod_{i<j}^n E(z_i,z_j) \, E(w_i, w_j) }{ \prod_{i,j=1}^n E(z_i,w_j) } \right)^{1/4}
\label{at1}
\eeq
Note that the $\tspin$ functions of even spin structures do not vanish at zero argument.

\medskip
The most general and most important building block for our results is the correlator with both fermions $\Psi^{\pm}$ and spin fields $s^{\pm}$ involved:
\begin{align}
&\left \langle \prod_{i=1}^{N_1} s^{+}(y_i) \, \prod_{j=1}^{N_2} s^{-}(z_j) \, \prod_{k=1}^{N_3} \Psi^{-}(u_k) \, \prod_{l=1}^{N_4} \Psi^{+}(v_l) \right \rangle^{\ve{a}}_{\ve{b}} \eq \left( \Theta \biggl[ \begin{array}{c} \ve{a}   \\ \ve{b} \end{array} \biggr] (\ve{0}) \right)^{-1}  \notag \\
& \ \ \ \times \, \left( \frac{\prod_{r<s}^{N_1} E(y_r,y_s) \, \prod_{r<s}^{N_2} E(z_r,z_s) }{ \prod_{i=1}^{N_1} \prod_{j=1}^{N_2} E(z_j,y_i) } \right)^{1/4} \,
\left( \frac{ \prod_{r<s}^{N_3} E(u_r,u_s) \, \prod_{r<s}^{N_4} E(v_r,v_s)  }{ \prod_{k=1}^{N_3} \prod_{l=1}^{N_4} E(v_l,u_k) }\right) \notag \\
& \ \ \ \times \, \left( \frac{ \prod_{j=1}^{N_2} \prod_{k=1}^{N_3} E(u_k,z_j) \,  \prod_{i=1}^{N_1} \prod_{l=1}^{N_4} E(v_l,y_i) }{  \prod_{i=1}^{N_1} \prod_{k=1}^{N_3} E(u_k,y_i) \, \prod_{j=1}^{N_2} \prod_{l=1}^{N_4} E(v_l,z_j)} \right)^{1/2}    \notag \\
& \ \ \ \times \, \Theta \biggl[ \begin{array}{c} \ve{a}   \\ \ve{b} \end{array} \biggr] \left( \tfrac{1}{2}  \sum_{i=1}^{N_1} \smallint^{y_i}_{p} \ve{\om} \, - \, \tfrac{1}{2}  \sum_{j=1}^{N_2} \smallint^{z_j}_{p} \ve{\om} \, - \, \sum_{k=1}^{N_3} \smallint^{u_k}_{p} \ve{\om} \, + \,  \sum_{l=1}^{N_4} \smallint^{v_l}_{p} \ve{\om}\right)
\label{at2}
\end{align}
The arbitrary reference point $p$ in the integral within the $\tspin$ function drops out due to Ramond charge conservation $\frac{1}{2}(N_1-N_2) +N_4- N_3=0$.

\medskip
In the following sections, we will label any field's argument by $z_i$ rather than $u_i,v_i,y_i$, so it makes sense to introduce the shorthand $E_{ij} := E(z_i,z_j)$. Also, the spin structure dependence will be denoted more economically by $\tspin(\ve{x}) := \Theta \Bigl[ \begin{array}{c} \ve{a}   \\ \ve{b} \end{array} \Bigr](\ve{x} \, | \, \Om)$.


\section{Covariant spin field correlators in four dimensions}
\label{sec:cov}

In this section, we collect correlation functions of $SO(1,3)$ spin fields. As we have explained above, they are basic building block for any loop correlators including $\psi^{\mu}$- and spin fields. Repeated use of (\ref{2,1}) leads to the prescription
 \begin{align}
   &\langle \psi^{\mu_1}(z_1) \, ... \, \psi^{\mu_n}(z_n)\,S_{\al_1}(x_1)\, ... \, S_{\al_{r}}(x_{r})\,S_{\dbe_1}(y_1)\, ... \, S_{\dbe_{s}}(y_{s}) \rangle \spin \eq
   \prod_{i=1}^n \left(- \, \frac{\bar{\si}^{\mu_{i}\,\dka_i \ka_i}}{\sqrt{2}}\right)\notag\\
   &\ \ \times\ \langle S_{\ka_1}(z_1) \, ... \, S_{\ka_{n}}(z_{n})\,S_{\al_1}(x_1)\, ... \, S_{\al_{r}}(x_{r})\,S_{\dka_1}(z_1)\, ... \, S_{\dka_{n}}(z_{n})\,
     S_{\dbe_1}(y_1)\, ... \, S_{\dbe_{s}}(y_{s}) \rangle \spin \ .
\label{3,0}
\end{align}
to eliminate NS fermions.

\medskip
One big obstacle on the way to higher point spin field correlation functions on the torus or higher genus Riemann surfaces is their failure to factorize into left- and right handed parts: the argument of \cite{tree} for the identity
\beq
\langle S_{\al_1}(z_1)\, ... \, S_{\al_{r}}(z_{r})\,S_{\dbe_1}(w_1)\, ... \, S_{\dbe_{s}}(w_{s}) \rangle \eq \langle S_{\al_1}(z_1)\, ... \, S_{\al_r}(z_r) \rangle \,\cdot\,\langle S_{\dbe_1}(w_1)\, ... \, S_{\dbe_{s}}(w_{s}) \rangle
\label{3,1}
\eeq
valid at tree level relies on bosonization techniques which are not directly applicable at nonzero genus. The coupling between the left- and right handed sector will turn out to sit exclusively in the spin structure dependent $\Theta$ functions. But still, the prime forms $E(z_{i},z_{j})$ carrying the singularities mimic the tree level factorization.

\medskip
Because of this problem, we did not succeed in finding a nice expressions for correlators with six left handed {\em and} six right handed spin fields at the same time. In the following some lower order results for spin field correlators are given, arranged by number of alike chiralities. Then, inspired by the result for the two-, four- and six point functions, a general formula for $2M$ spin fields of uniform chirality is written down and proven by induction in appendix \ref{sec:appC}. Moreover, correlations of any number of spin fields of alike chirality with up to four spin fields of the opposite type are given in closed from.

\subsection{Two alike chiralities}
\label{sec:2}

As long as at most two spin fields of each chirality are present, there is only one possible Lorentz tensor structure, namely the $\vep_{\al \be}, \vep_{\dal \dbe}$ symbols. They are nonzero if the indices take distinct values $\al \neq \be$ and $\dal \neq \dbe$ i.e. if the R charge is conserved. The $z$ dependence is determined by (\ref{at1}) in subsection \ref{sec:Spinloop}. Hence, the left handed two point function in the nonzero configuration reads
\beq
\langle S_{\al=(+,+)}(z_{1}) \, S_{\be=(-,-)}(z_{2}) \rangle \spin \eq \langle s^+(z_1)  \, s^-(z_2) \rangle \spin \  \langle \ts ^+(z_1) \, \ts^-(z_2) \rangle \spin \notag \ \ = \ \ \left( \frac{ \tspin \left( \tfrac{1}{2} \smallint^{z_1}_{z_2} \ve{\om} \right) }{ ( E_{12})^{1/4} \, \tspin(\ve{0}) } \right)^2
\label{3,2}
\eeq
with covariant generalization
\beq
\langle S_{\al}(z_{1}) \, S_{\be}(z_{2}) \rangle \spin \eq - \, \vep_{\al \be} \; \frac{ \left[ \tspin \left( \tfrac{1}{2} \smallint^{z_1}_{z_2} \ve{\om} \right) \right]^2 }{ (E_{12})^{1/2} \, \tspin (\ve{0}) \, \tspin (\ve{0}) } \ .
\label{(2,0)}
\eeq
The prefactor is enforced by the OPE (\ref{rv,1d}). The right handed analogue is obtained similarly by means of $\langle s_1 ^+ \, s_2^- \rangle \,  \langle \ts_1 ^- \, \ts_2^+ \rangle$ and the OPE (\ref{rv,1e}):
\beq
\langle S_{\dal}(z_{1}) \, S_{\dbe}(z_{2}) \rangle \spin \eq + \,  \vep_{\dal \dbe} \; \frac{ \left[ \tspin \left( \tfrac{1}{2} \smallint^{z_1}_{z_2} \ve{\om} \right) \right]^2 }{ (E_{12})^{1/2} \, \tspin (\ve{0}) \, \tspin (\ve{0}) }
\label{(0,2)}
\eeq
Let us give these two correlators explicitly for the torus case $g=1$ with the simple expression (\ref{tor}) for the prime form and with spin structure $\nu=2,3,4$ according to (\ref{dict}):
\begin{subequations}
\begin{align}
\langle S_{\al}(z_{1}) \, S_{\be}(z_{2}) \rangle_{\nu} \ \ &= \ \ - \,  \vep_{\al \be} \; \frac{\bigl( \z_1'(0) \bigr)^{1/2}}{ \bigl(\theta_1(z_{12}) \bigr)^{1/2}} \, \cdot \, \frac{\znu^2 \left( \tfrac{1}{2} \, z_{12} \right) }{ \znu^2(0) } \\
 \langle S_{\dal}(z_{1}) \, S_{\dbe}(z_{2}) \rangle_{\nu} \ \ &= \ \ + \,  \vep_{\dal \dbe} \; \frac{\bigl( \z_1'(0) \bigr)^{1/2}}{ \bigl(\theta_1(z_{12}) \bigr)^{1/2}} \, \cdot \, \frac{\znu^2 \left( \tfrac{1}{2} \, z_{12} \right) }{ \znu^2(0) }
\label{2tor}
\end{align}
\end{subequations}
The mixed four point function represents a first example of our non-factorization statement from the beginning of this section: One can first of all check that
\beq
\langle S_{(+,+)}(z_{1}) \, S_{(-,-)}(z_{2}) \, S_{(+,-)}(z_{3}) \, S_{(-,+)}(z_{4}) \rangle \spin \ \ \sim \ \  \frac{ \tspin \left( \tfrac{1}{2} \smallint^{z_1}_{z_2} \ve{\om} \, + \, \tfrac{1}{2} \smallint^{z_3}_{z_4}\ve{\om} \right) \, \tspin \left( \tfrac{1}{2} \smallint^{z_1}_{z_2}\ve{\om} \, - \, \tfrac{1}{2} \smallint^{z_3}_{z_4}\ve{\om} \right)}{ (E_{12} \, E_{34})^{1/2} \, \, \tspin (\ve{0}) \, \tspin (\ve{0})}
\label{3,3}
\eeq
and then conclude by $SO(1,3)$ covariance that
\beq
\langle S_{\al}(z_{1}) \, S_{\be}(z_{2}) \, S_{\dga}(z_{3}) \, S_{\dde}(z_{4}) \rangle \spin \eq - \, \vep_{\al \be}  \,  \vep_{\dga \dde} \; \frac{ \tspin \left( \tfrac{1}{2} \smallint^{z_1}_{z_2} \ve{\om} \, + \, \tfrac{1}{2} \smallint^{z_3}_{z_4} \ve{\om}\right) \, \tspin \left( \tfrac{1}{2} \smallint^{z_1}_{z_2}\ve{\om} \, - \, \tfrac{1}{2} \smallint^{z_3}_{z_4}\ve{\om} \right)}{ (E_{12} \, E_{34})^{1/2} \, \tspin (\ve{0}) \, \tspin (\ve{0})} \ .
\label{(2,2)}
\eeq
As claimed above, the arguments $z_{1,2}$ of the left handed fields couple to the right handed ones $z_{3,4}$ through the $\tspin$ functions as claimed above. The prime forms $E_{ij}$ simply replace the $z_{ij}$ from the tree level result \cite{tree}.

\subsection{Four alike chiralities}
\label{sec:4}

In correlators with four spin fields $S_{\al} S_{\be} S_{\ga}S_{\de}$ of the same chirality, one can find two independent Clebsch Gordan coefficients $\vep_{\al \be}  \vep_{\ga \de}$ and $\vep_{\al \de}  \vep_{\ga \be}$. A third possibility can be reduced to the former ones by means of the Fierz identity $\vep_{\al \ga}  \vep_{\be \de} = \vep_{\al \be}  \vep_{\ga \de} - \vep_{\al \de}  \vep_{\ga \be}$, see \cite{tree} for the group theoretical background.

\medskip
We anticipate $E_{ij}$ functions analogous to the $z_{ij}$ at tree level, hence our ansatz is
\begin{align}
\langle S_{\al}(z_{1}) \, S_{\be}(z_{2}) \, S_{\ga}(z_{3}) \, S_{\de}(z_{4}) \rangle \spin \eq \frac{  \vep_{\al \be}  \, \vep_{\ga \de} \, E_{14} \, E_{23} \, F \spin (z_{ij}) \ + \ \vep_{\al \de} \, \vep_{\ga \be} \, E_{12} \, E_{34} \, G \spin (z_{ij}) }{ \left( E_{12} \, E_{13} \, E_{14} \, E_{23} \, E_{24} \, E_{34} \right)^{1/2} \, \tspin (\ve{0}) \, \tspin (\ve{0})}  \ .
\label{3,4}
\end{align}
The spin structure dependent coefficients $F \spin , G \spin$ can be obtained by testing the following two index configurations:
\begin{subequations}
\begin{align}
\al = \de = &(+,+) \co \be = \ga = (-,-) \notag \\ 
&\Rightarrow \ \ \ F \spin (z_{ij}) \eq \left[ \tspin \left( \tfrac{1}{2} \smallint^{z_1}_{z_2} \ve{\om} \, - \, \tfrac{1}{2} \smallint^{z_3}_{z_4} \ve{\om} \right) \right]^2 \label{3,5a} \\
\al = \be = &(+,+) \co \ga = \de = (-,-) \notag \\ 
&\Rightarrow \ \ \ G \spin (z_{ij}) \eq \left[ \tspin \left( \tfrac{1}{2} \smallint^{z_1}_{z_4} \ve{\om} \, - \, \tfrac{1}{2} \smallint^{z_3}_{z_2} \ve{\om} \right) \right]^2 \label{3,5b}
\end{align}
\end{subequations}
Putting everything together, one arrives at
\begin{align}
&\langle S_{\al}(z_{1}) \, S_{\be}(z_{2}) \, S_{\ga}(z_{3}) \, S_{\de}(z_{4}) \rangle \spin \eq \frac{1}{(E_{12} \, E_{13} \, E_{14} \, E_{23} \, E_{24} \, E_{34} )^{1/2} \,  \tspin (\ve{0}) \, \tspin (\ve{0})} \notag \\
&\ \ \ \times \, \left\{ \vep_{\al \be} \, \vep_{\ga \de} \, E_{14} \, E_{23} \, \left[ \tspin \left( \tfrac{1}{2} \smallint^{z_1}_{z_2} \ve{\om} \, - \, \tfrac{1}{2} \smallint^{z_3}_{z_4} \ve{\om} \right) \right]^2 \ + \ \vep_{\al \de} \, \vep_{\ga \be} \, E_{12} \, E_{34} \, \left[ \tspin \left( \tfrac{1}{2} \smallint^{z_1}_{z_4} \ve{\om} \, - \, \tfrac{1}{2} \smallint^{z_3}_{z_2} \ve{\om} \right) \right]^2 \right\} 
\label{(4,0)}
\end{align}
which again reduces to the tree level result under $E_{ij} \mapsto z_{ij}$ and $\tspin \mapsto 1$.

\medskip
One might wonder what happens in the third non-trivial index choice $\al = \ga = (+,+)$ and $\be = \de = (-,-) $ where both $\vep_{\al \be}  \vep_{\ga \de}$ and $\vep_{\al \de}  \vep_{\ga \be}$ are nonzero. Straightforward evaluation of $\langle S_{(+,+)}(z_1) S_{(-,-)}(z_2) S_{(+,+)}(z_3) S_{(-,-)}(z_4) \rangle$ by means of (\ref{at1}) leads to an expression where (\ref{(4,0)}) is not recognizable at first glance. The non-trivial consistency is based on a Fay trisecant identity, in particular on (\ref{ft2}) at $\ve{\De} = \ve{0}$ of appendix \ref{appA}. This appendix explains and collects this addition theorem for generalized $\Theta$ functions and generalizations thereof.

\medskip
Next we include right handed spin fields into the left handed four point function. Using the index configurations (\ref{3,5a}) and (\ref{3,5b}) in the left handed sector as well as $\dal =(+,-)$ and $\dbe=(-,+)$ on the right handed side, we find
\begin{align}
&\langle S_{\al}(z_{1}) \, S_{\be}(z_{2}) \, S_{\ga}(z_{3}) \, S_{\de}(z_{4}) \, S_{\dal}(z_5) \, S_{\dbe}(z_6) \rangle \spin \eq \frac{\vep_{\dal \dbe}}{(E_{12} \, E_{13} \, E_{14} \, E_{23} \, E_{24} \, E_{34} \, E_{56})^{1/2} \,  \tspin (\ve{0}) \, \tspin (\ve{0})} \notag \\
&\ \ \ \times \, \left\{ \vep_{\al \be} \, \vep_{\ga \de} \, E_{14} \, E_{23} \,  \tspin \left( \tfrac{1}{2} \smallint^{z_1}_{z_2} \ve{\om} \, - \, \tfrac{1}{2} \smallint^{z_3}_{z_4} \ve{\om} \, + \, \tfrac{1}{2} \smallint^{z_5}_{z_6} \ve{\om} \right) \,  \tspin \left( \tfrac{1}{2} \smallint^{z_1}_{z_2} \ve{\om} \, - \, \tfrac{1}{2} \smallint^{z_3}_{z_4} \ve{\om} \, - \, \tfrac{1}{2} \smallint^{z_5}_{z_6} \ve{\om} \right)  \right. \notag \\
& \ \ \ \ \ \ \! \left. + \  \vep_{\al \de} \, \vep_{\ga \be} \, E_{12} \, E_{34} \, \tspin \left( \tfrac{1}{2} \smallint^{z_1}_{z_4} \ve{\om} \, - \, \tfrac{1}{2} \smallint^{z_3}_{z_2} \ve{\om} \, + \, \tfrac{1}{2} \smallint^{z_5}_{z_6} \ve{\om} \right) \, \tspin \left( \tfrac{1}{2} \smallint^{z_1}_{z_4} \ve{\om} \, - \, \tfrac{1}{2} \smallint^{z_3}_{z_2} \ve{\om} \, - \, \tfrac{1}{2} \smallint^{z_5}_{z_6} \ve{\om} \right)  \right\} \ .
\label{(4,2)}
\end{align}
With four spin fields of both chiralities, there are four choices of the indices to plug into (\ref{at1}):
\begin{align}
&\langle S_{\al}(z_{1}) \, S_{\be}(z_{2}) \, S_{\ga}(z_{3}) \, S_{\de}(z_{4}) \, S_{\dal}(z_5) \, S_{\dbe}(z_6) \, S_{\dal}(z_5) \, S_{\dbe}(z_8) \rangle \spin \notag \\
&= \ \ \frac{1}{(E_{12} \, E_{13} \, E_{14} \, E_{23} \, E_{24} \, E_{34})^{1/2} \, (E_{56} \, E_{57} \, E_{58} \, E_{67} \, E_{68} \, E_{78})^{1/2} \,  \tspin (\ve{0}) \, \tspin (\ve{0})} \notag \\
&\ \ \ \ \ \times \, \biggl\{ \vep_{\al \be} \, \vep_{\ga \de} \, \vep_{\dal \dbe} \, \vep_{\dga \dde} \, E_{14} \, E_{23} \, E_{58} \, E_{67} \, \tspin \left( \tfrac{1}{2} \smallint^{z_1}_{z_2} \ve{\om} \, - \, \tfrac{1}{2} \smallint^{z_3}_{z_4} \ve{\om} \, + \, \tfrac{1}{2} \smallint^{z_5}_{z_6} \ve{\om} \, - \, \tfrac{1}{2} \smallint^{z_7}_{z_8} \ve{\om}  \right) \biggr. \notag \\
& \ \ \ \ \ \ \ \ \ \ \ \ \ \ \ \ \ \ \ \ \ \  \biggl.   \tspin \left( \tfrac{1}{2} \smallint^{z_1}_{z_2} \ve{\om} \, - \, \tfrac{1}{2} \smallint^{z_3}_{z_4} \ve{\om} \, - \, \tfrac{1}{2} \smallint^{z_5}_{z_6} \ve{\om} \, + \, \tfrac{1}{2} \smallint^{z_7}_{z_8} \ve{\om}  \right)  \biggr. \notag \\
&\ \ \ \ \ \ \ \ \ \ \  \biggl. + \ \vep_{\al \be} \, \vep_{\ga \de} \, \vep_{\dal \dde} \, \vep_{\dga \dbe} \, E_{14} \, E_{23} \, E_{56} \, E_{78} \, \tspin \left( \tfrac{1}{2} \smallint^{z_1}_{z_2} \ve{\om} \, - \, \tfrac{1}{2} \smallint^{z_3}_{z_4} \ve{\om} \, + \, \tfrac{1}{2} \smallint^{z_5}_{z_8} \ve{\om} \, - \, \tfrac{1}{2} \smallint^{z_7}_{z_6} \ve{\om}  \right) \biggr. \notag \\
& \ \ \ \ \ \ \ \ \ \ \ \ \ \ \ \ \ \ \ \ \ \  \biggl.  \tspin \left( \tfrac{1}{2} \smallint^{z_1}_{z_2} \ve{\om} \, - \, \tfrac{1}{2} \smallint^{z_3}_{z_4} \ve{\om} \, - \, \tfrac{1}{2} \smallint^{z_5}_{z_8} \ve{\om} \, + \, \tfrac{1}{2} \smallint^{z_7}_{z_6} \ve{\om}  \right)  \biggr. \notag \\
&\ \ \ \ \ \ \ \ \ \ \  \biggl. + \ \vep_{\al \de} \, \vep_{\ga \be} \, \vep_{\dal \dbe} \, \vep_{\dga \dde} \, E_{12} \, E_{34} \, E_{58} \, E_{67} \, \tspin \left( \tfrac{1}{2} \smallint^{z_1}_{z_4} \ve{\om} \, - \, \tfrac{1}{2} \smallint^{z_3}_{z_2} \ve{\om} \, + \, \tfrac{1}{2} \smallint^{z_5}_{z_6} \ve{\om} \, - \, \tfrac{1}{2} \smallint^{z_7}_{z_8} \ve{\om}  \right) \biggr. \notag \\
& \ \ \ \ \ \ \ \ \ \ \ \ \ \ \ \ \ \ \ \ \ \  \biggl.   \tspin \left( \tfrac{1}{2} \smallint^{z_1}_{z_2} \ve{\om} \, - \, \tfrac{1}{2} \smallint^{z_3}_{z_4} \ve{\om} \, - \, \tfrac{1}{2} \smallint^{z_5}_{z_6} \ve{\om} \, + \, \tfrac{1}{2} \smallint^{z_7}_{z_8} \ve{\om}  \right)  \biggr. \notag \\
&\ \ \ \ \ \ \ \ \ \ \  \biggl. + \ \vep_{\al \de} \, \vep_{\ga \be} \, \vep_{\dal \dde} \, \vep_{\dga \dbe} \, E_{12} \, E_{34} \, E_{56} \, E_{78} \, \tspin \left( \tfrac{1}{2} \smallint^{z_1}_{z_4} \ve{\om} \, - \, \tfrac{1}{2} \smallint^{z_3}_{z_2} \ve{\om} \, + \, \tfrac{1}{2} \smallint^{z_5}_{z_8} \ve{\om} \, - \, \tfrac{1}{2} \smallint^{z_7}_{z_6} \ve{\om}  \right) \biggr. \notag \\
& \ \ \ \ \ \ \ \ \ \ \ \ \ \ \ \ \ \ \ \ \ \  \biggl.  \tspin \left( \tfrac{1}{2} \smallint^{z_1}_{z_2} \ve{\om} \, - \, \tfrac{1}{2} \smallint^{z_3}_{z_4} \ve{\om} \, - \, \tfrac{1}{2} \smallint^{z_5}_{z_8} \ve{\om} \, + \, \tfrac{1}{2} \smallint^{z_7}_{z_6} \ve{\om}  \right)  \biggr\} \ .
\label{(4,4)}
\end{align}
As before, one or two applications of (\ref{ft2}) with nonzero $\ve{\De}$ (e.g. $\ve{\De} = \pm \tfrac{1}{2} \smallint^{z_5}_{z_6} \ve{\om}$ for (\ref{(4,2)})) guarantees consistency with those index configurations such as $\al = \ga$ or $\dal = \dga$ where two or more terms are nonzero.

\subsection{Six alike chiralities}
\label{sec:6}

The next step is computing the six point function $\langle S_{\al}  S_{\be}  S_{\ga}  S_{\de}  S_{\ka}  S_{\om} \rangle \spin$ of uniform chirality. From our experience with the tree level result \cite{tree}, it seems worthwhile to use a non-minimal basis of $\vep_{\cdot \cdot} \vep_{\cdot \cdot} \vep_{\cdot \cdot}$ tensor as listed in the following table to ensure manifest symmetry under exchange of spin fields. As before we first of all test some configurations of the six spinor indices: 
\begin{center}
\begin{tabular}{|r|c|c|c|c|c|c|}\hline charges $\left( \begin{smallmatrix} \al &\be &\ga \\ \de &\ka &\om \end{smallmatrix} \right)$
& $\left( \begin{smallmatrix} + &- &- \\ + &- &+ \end{smallmatrix} \right)$ 
& $\left( \begin{smallmatrix} + &- &- \\ - &+ &+ \end{smallmatrix} \right)$ 
& $\left( \begin{smallmatrix} + &+ &- \\ - &- &+ \end{smallmatrix} \right)$
 & $\left( \begin{smallmatrix} + &+ &- \\ - &+ &- \end{smallmatrix} \right)$
 & $\left( \begin{smallmatrix} + & + &- \\ + &- &- \end{smallmatrix} \right)$
 & $\left( \begin{smallmatrix} + &- &+ \\ -&+ &- \end{smallmatrix} \right)$ \\ \hline \hline 
$\vep_{\al \be} \, \vep_{\ga \de} \, \vep_{\ka \om}$ & + 1 & 0 & 0 & 0 & 0 & + 1  \\ \hline
$\vep_{\al \be} \, \vep_{\ga \om} \, \vep_{\ka \de}$ & + 1 & - 1 & 0 & 0 & 0 & + 1  \\ \hline
$\vep_{\al \de} \, \vep_{\ga \om} \, \vep_{\ka \be}$ & 0 & - 1 & + 1  & 0 & 0 & + 1 \\ \hline
$\vep_{\al \de} \, \vep_{\ga \be} \, \vep_{\ka \om}$ & 0 & 0 & + 1 & - 1 & 0 & + 1 \\\hline
$\vep_{\al \om} \, \vep_{\ga \be} \, \vep_{\ka \de}$ & 0 & 0 & 0 & - 1 & + 1 & + 1 \\ \hline 
$\vep_{\al \om} \, \vep_{\ga \de} \, \vep_{\ka \be}$ &0 &0 &0 &0 & + 1 & + 1 \\ \hline
\end{tabular}
\begin{flushleft}{\bf Table 1}. {\it Each non-trivial charge assignment (with three $(++)$ and $(-,-)$ indices each) yields a certain set of (at least two) nonzero $\vep$ combinations. Since both $SO(2)$ spin fields of $S_{\al=(\pm,\pm)} = s^{\pm} \ts^{\pm}$ carry the same charge, the vales of $\al,\be,...$ are denoted by a single sign $\pm$ for each index in the headline.}\\[3mm]
\end{flushleft}
\end{center}
Each column contains at least two nonzero entries, so it is not possible to directly probe the $z$ dependence associated with a single $\vep$ product using (\ref{at1}) as it was the case for the four point function (\ref{3,5a}) and (\ref{3,5b}). This is quite natural in view of the identity
\begin{align}
  0 \eq &\vep_{\al \be} \,  \vep_{\ga \de} \, \vep_{\ka \om} \ - \ \vep_{\al \be} \,  \vep_{\ga \om} \, \vep_{\ka \de} \ - \ \vep_{\al \de}  \, \vep_{\ga \be} \,  \vep_{\ka \om} \notag \\
   + \ &\vep_{\al \de} \, \vep_{\ga \om} \, \vep_{\ka \be} \ + \ \vep_{\al \om} \,  \vep_{\ga \be} \,  \vep_{\ka \de} \ - \ \vep_{\al \om} \, \vep_{\ga \de} \, \vep_{\ka \be} \ .
\label{sigspin1}
\end{align}
Apart from this rather technical point, the true punishment for choosing the non-minimal basis is the appearance of an additional nontrivial $\tspin$ function in denominator:
\begin{align}
&\langle S_{\al}(z_{1}) \, S_{\be}(z_{2}) \, S_{\ga}(z_{3}) \, S_{\de}(z_{4}) \, S_{\ka}(z_5) \, S_{\om}(z_6) \rangle \spin \notag \\
&= \ \ \frac{- \, (E_{12} \, E_{14} \, E_{16} \, E_{23} \, E_{25} \, E_{34} \, E_{36}\, E_{45} \, E_{56})^{1/2} }{ (E_{13} \, E_{15} \, E_{35} \, E_{24} \, E_{26} \, E_{46} )^{1/2} \, \tspin \left( \tfrac{1}{2} \smallint^{z_1}_{z_2} \ve{\om} \, + \, \tfrac{1}{2} \smallint^{z_3}_{z_4} \ve{\om} \, + \, \tfrac{1}{2} \smallint^{z_5}_{z_6} \ve{\om} \right) \, \tspin (\ve{0}) \, \tspin (\ve{0})} \notag \\
& \ \ \ \times \ \left\{ \frac{\vep_{\al \be} \, \vep_{\ga \de} \, \vep_{\ka \om}}{E_{12} \, E_{34} \, E_{56}} \; \tspin \left( -\tfrac{1}{2} \smallint^{z_1}_{z_2} \ve{\om} \, + \, \tfrac{1}{2} \smallint^{z_3}_{z_4} \ve{\om} \, + \, \tfrac{1}{2} \smallint^{z_5}_{z_6} \ve{\om} \right) \right. \notag \\
& \ \ \ \ \ \ \ \ \ \ \ \ \ \ \left. \tspin \left( \tfrac{1}{2} \smallint^{z_1}_{z_2} \ve{\om} \, - \, \tfrac{1}{2} \smallint^{z_3}_{z_4} \ve{\om} \, + \, \tfrac{1}{2} \smallint^{z_5}_{z_6} \ve{\om} \right) \, \tspin \left( \tfrac{1}{2} \smallint^{z_1}_{z_2} \ve{\om} \, + \, \tfrac{1}{2} \smallint^{z_3}_{z_4} \ve{\om} \, - \, \tfrac{1}{2} \smallint^{z_5}_{z_6} \ve{\om} \right) 
 \right. \notag \\
& \ \ \ \ \ \ \ \left. - \ \frac{\vep_{\al \be} \, \vep_{\ga \om} \, \vep_{\ka \de}}{E_{12} \, E_{36} \, E_{54}} \; \tspin \left( -\tfrac{1}{2} \smallint^{z_1}_{z_2} \ve{\om} \, + \, \tfrac{1}{2} \smallint^{z_3}_{z_6} \ve{\om} \, + \, \tfrac{1}{2} \smallint^{z_5}_{z_4} \ve{\om} \right) \right. \notag \\
& \ \ \ \ \ \ \ \ \ \ \ \ \ \ \left. \tspin \left( \tfrac{1}{2} \smallint^{z_1}_{z_2} \ve{\om} \, - \, \tfrac{1}{2} \smallint^{z_3}_{z_6} \ve{\om} \, + \, \tfrac{1}{2} \smallint^{z_5}_{z_4} \ve{\om} \right) \, \tspin \left( \tfrac{1}{2} \smallint^{z_1}_{z_2} \ve{\om} \, + \, \tfrac{1}{2} \smallint^{z_3}_{z_6} \ve{\om} \, - \, \tfrac{1}{2} \smallint^{z_5}_{z_4} \ve{\om} \right) 
 \right. \notag \\
& \ \ \ \ \ \ \ \left. + \ \frac{\vep_{\al \de} \, \vep_{\ga \om} \, \vep_{\ka \be}}{E_{14} \, E_{36} \, E_{52}} \; \tspin \left( -\tfrac{1}{2} \smallint^{z_1}_{z_4} \ve{\om} \, + \, \tfrac{1}{2} \smallint^{z_3}_{z_6} \ve{\om} \, + \, \tfrac{1}{2} \smallint^{z_5}_{z_2} \ve{\om} \right) \right. \notag \\
& \ \ \ \ \ \ \ \ \ \ \ \ \ \ \left. \tspin \left( \tfrac{1}{2} \smallint^{z_1}_{z_4} \ve{\om} \, - \, \tfrac{1}{2} \smallint^{z_3}_{z_6} \ve{\om} \, + \, \tfrac{1}{2} \smallint^{z_5}_{z_2} \ve{\om} \right) \, \tspin \left( \tfrac{1}{2} \smallint^{z_1}_{z_4} \ve{\om} \, + \, \tfrac{1}{2} \smallint^{z_3}_{z_6} \ve{\om} \, - \, \tfrac{1}{2} \smallint^{z_5}_{z_2} \ve{\om} \right) 
 \right. \notag \\
& \ \ \ \ \ \ \ \left. - \ \frac{\vep_{\al \de} \, \vep_{\ga \be} \, \vep_{\ka \om}}{E_{14} \, E_{32} \, E_{56}} \; \tspin \left( -\tfrac{1}{2} \smallint^{z_1}_{z_4} \ve{\om} \, + \, \tfrac{1}{2} \smallint^{z_3}_{z_2} \ve{\om} \, + \, \tfrac{1}{2} \smallint^{z_5}_{z_6} \ve{\om} \right) \right. \notag \\
& \ \ \ \ \ \ \ \ \ \ \ \ \ \ \left. \tspin \left( \tfrac{1}{2} \smallint^{z_1}_{z_4} \ve{\om} \, - \, \tfrac{1}{2} \smallint^{z_3}_{z_2} \ve{\om} \, + \, \tfrac{1}{2} \smallint^{z_5}_{z_6} \ve{\om} \right) \, \tspin \left( \tfrac{1}{2} \smallint^{z_1}_{z_4} \ve{\om} \, + \, \tfrac{1}{2} \smallint^{z_3}_{z_2} \ve{\om} \, - \, \tfrac{1}{2} \smallint^{z_5}_{z_6} \ve{\om} \right) 
 \right. \notag \\
& \ \ \ \ \ \ \ \left. + \ \frac{\vep_{\al \om} \, \vep_{\ga \be} \, \vep_{\ka \de}}{E_{16} \, E_{32} \, E_{54}} \; \tspin \left( -\tfrac{1}{2} \smallint^{z_1}_{z_6} \ve{\om} \, + \, \tfrac{1}{2} \smallint^{z_3}_{z_2} \ve{\om} \, + \, \tfrac{1}{2} \smallint^{z_5}_{z_4} \ve{\om} \right) \right. \notag \\
& \ \ \ \ \ \ \ \ \ \ \ \ \ \ \left. \tspin \left( \tfrac{1}{2} \smallint^{z_1}_{z_6} \ve{\om} \, - \, \tfrac{1}{2} \smallint^{z_3}_{z_2} \ve{\om} \, + \, \tfrac{1}{2} \smallint^{z_5}_{z_4} \ve{\om} \right) \, \tspin \left( \tfrac{1}{2} \smallint^{z_1}_{z_6} \ve{\om} \, + \, \tfrac{1}{2} \smallint^{z_3}_{z_2} \ve{\om} \, - \, \tfrac{1}{2} \smallint^{z_5}_{z_4} \ve{\om} \right) 
 \right. \notag \\
& \ \ \ \ \ \ \ \left. - \ \frac{\vep_{\al \om} \, \vep_{\ga \de} \, \vep_{\ka \be}}{E_{16} \, E_{34} \, E_{52}} \; \tspin \left( -\tfrac{1}{2} \smallint^{z_1}_{z_6} \ve{\om} \, + \, \tfrac{1}{2} \smallint^{z_3}_{z_4} \ve{\om} \, + \, \tfrac{1}{2} \smallint^{z_5}_{z_2} \ve{\om} \right) \right. \notag \\
& \ \ \ \ \ \ \ \ \ \ \ \ \ \ \left. \tspin \left( \tfrac{1}{2} \smallint^{z_1}_{z_6} \ve{\om} \, - \, \tfrac{1}{2} \smallint^{z_3}_{z_4} \ve{\om} \, + \, \tfrac{1}{2} \smallint^{z_5}_{z_2} \ve{\om} \right) \, \tspin \left( \tfrac{1}{2} \smallint^{z_1}_{z_6} \ve{\om} \, + \, \tfrac{1}{2} \smallint^{z_3}_{z_4} \ve{\om} \, - \, \tfrac{1}{2} \smallint^{z_5}_{z_2} \ve{\om} \right) 
 \right\} \label{(6,0)}
\end{align}
In the first five index configurations of the table, (\ref{(6,0)}) can be easily shown to agree with the result of (\ref{at1}) using the simplest Fay trisecant identity (\ref{ft2}). But checking consistency in the last case $\al = \ga = \ka = (+,+)$ and $\be = \de = \om = (-,-)$ where all the six terms in (\ref{(6,0)}) contribute requires a higher order Fay trisecant identity (\ref{ft3}) (with $\ve{\De}=\ve{0}$).

\medskip
Next we add right handed spin fields similar to section \ref{sec:4}:
\begin{align}
&\langle S_{\al}(z_{1}) \, S_{\be}(z_{2}) \, S_{\ga}(z_{3}) \, S_{\de}(z_{4})  \, S_{\ka}(z_5) \, S_{\om}(z_6) \, S_{\dal}(z_7) \, S_{\dbe}(z_8) \rangle \spin \notag \\
&= \ \ \frac{- \, (E_{12} \, E_{14} \, E_{16} \, E_{23} \, E_{25} \, E_{34} \, E_{36}\, E_{45} \, E_{56})^{1/2} \, \vep_{\dal \dbe} }{ (E_{13} \, E_{15} \, E_{35} \, E_{24} \, E_{26} \, E_{46} \, E_{78} )^{1/2} \, \tspin \left( \tfrac{1}{2} \smallint^{z_1}_{z_2} \ve{\om} \, + \, \tfrac{1}{2} \smallint^{z_3}_{z_4} \ve{\om} \, + \, \tfrac{1}{2} \smallint^{z_5}_{z_6} \ve{\om} \, \pm \, \tfrac{1}{2} \smallint^{z_7}_{z_8} \ve{\om} \right) \, \tspin (\ve{0}) \, \tspin (\ve{0})} \notag \\
& \ \ \ \times \ \left\{ \frac{\vep_{\al \be} \, \vep_{\ga \de} \, \vep_{\ka \om}}{E_{12} \, E_{34} \, E_{56}} \; \tspin \left( -\tfrac{1}{2} \smallint^{z_1}_{z_2} \ve{\om} \, + \, \tfrac{1}{2} \smallint^{z_3}_{z_4} \ve{\om} \, + \, \tfrac{1}{2} \smallint^{z_5}_{z_6} \ve{\om} \, \pm \, \tfrac{1}{2} \smallint^{z_7}_{z_8} \ve{\om}\right) \right. \notag \\
& \ \ \ \ \ \ \ \ \ \ \ \ \ \ \left. \tspin \left( \tfrac{1}{2} \smallint^{z_1}_{z_2} \ve{\om} \, - \, \tfrac{1}{2} \smallint^{z_3}_{z_4} \ve{\om} \, + \, \tfrac{1}{2} \smallint^{z_5}_{z_6} \ve{\om} \, \pm \, \tfrac{1}{2} \smallint^{z_7}_{z_8} \ve{\om}\right) \, \tspin \left( \tfrac{1}{2} \smallint^{z_1}_{z_2} \ve{\om} \, + \, \tfrac{1}{2} \smallint^{z_3}_{z_4} \ve{\om} \, - \, \tfrac{1}{2} \smallint^{z_5}_{z_6} \ve{\om}\, \pm \, \tfrac{1}{2} \smallint^{z_7}_{z_8} \ve{\om} \right) 
 \right. \notag \\
& \ \ \ \ \ \ \ \left. - \ \frac{\vep_{\al \be} \, \vep_{\ga \om} \, \vep_{\ka \de}}{E_{12} \, E_{36} \, E_{54}} \; \tspin \left( -\tfrac{1}{2} \smallint^{z_1}_{z_2} \ve{\om} \, + \, \tfrac{1}{2} \smallint^{z_3}_{z_6} \ve{\om} \, + \, \tfrac{1}{2} \smallint^{z_5}_{z_4} \ve{\om}\, \pm \, \tfrac{1}{2} \smallint^{z_7}_{z_8} \ve{\om} \right) \right. \notag \\
& \ \ \ \ \ \ \ \ \ \ \ \ \ \ \left. \tspin \left( \tfrac{1}{2} \smallint^{z_1}_{z_2} \ve{\om} \, - \, \tfrac{1}{2} \smallint^{z_3}_{z_6} \ve{\om} \, + \, \tfrac{1}{2} \smallint^{z_5}_{z_4} \ve{\om} \, \pm \, \tfrac{1}{2} \smallint^{z_7}_{z_8} \ve{\om}\right) \, \tspin \left( \tfrac{1}{2} \smallint^{z_1}_{z_2} \ve{\om} \, + \, \tfrac{1}{2} \smallint^{z_3}_{z_6} \ve{\om} \, - \, \tfrac{1}{2} \smallint^{z_5}_{z_4} \ve{\om} \, \pm \, \tfrac{1}{2} \smallint^{z_7}_{z_8} \ve{\om} \right) 
 \right. \notag \\
& \ \ \ \ \ \ \ \left. + \ \frac{\vep_{\al \de} \, \vep_{\ga \om} \, \vep_{\ka \be}}{E_{14} \, E_{36} \, E_{52}} \; \tspin \left( -\tfrac{1}{2} \smallint^{z_1}_{z_4} \ve{\om} \, + \, \tfrac{1}{2} \smallint^{z_3}_{z_6} \ve{\om} \, + \, \tfrac{1}{2} \smallint^{z_5}_{z_2} \ve{\om} \, \pm \, \tfrac{1}{2} \smallint^{z_7}_{z_8} \ve{\om}\right) \right. \notag \\
& \ \ \ \ \ \ \ \ \ \ \ \ \ \ \left. \tspin \left( \tfrac{1}{2} \smallint^{z_1}_{z_4} \ve{\om} \, - \, \tfrac{1}{2} \smallint^{z_3}_{z_6} \ve{\om} \, + \, \tfrac{1}{2} \smallint^{z_5}_{z_2} \ve{\om} \, \pm \, \tfrac{1}{2} \smallint^{z_7}_{z_8} \ve{\om} \right) \, \tspin \left( \tfrac{1}{2} \smallint^{z_1}_{z_4} \ve{\om} \, + \, \tfrac{1}{2} \smallint^{z_3}_{z_6} \ve{\om} \, - \, \tfrac{1}{2} \smallint^{z_5}_{z_2} \ve{\om} \, \pm \, \tfrac{1}{2} \smallint^{z_7}_{z_8} \ve{\om}\right) 
 \right. \notag \\
& \ \ \ \ \ \ \ \left. - \ \frac{\vep_{\al \de} \, \vep_{\ga \be} \, \vep_{\ka \om}}{E_{14} \, E_{32} \, E_{56}} \; \tspin \left( -\tfrac{1}{2} \smallint^{z_1}_{z_4} \ve{\om} \, + \, \tfrac{1}{2} \smallint^{z_3}_{z_2} \ve{\om} \, + \, \tfrac{1}{2} \smallint^{z_5}_{z_6} \ve{\om} \, \pm \, \tfrac{1}{2} \smallint^{z_7}_{z_8} \ve{\om}\right) \right. \notag \\
& \ \ \ \ \ \ \ \ \ \ \ \ \ \ \left. \tspin \left( \tfrac{1}{2} \smallint^{z_1}_{z_4} \ve{\om} \, - \, \tfrac{1}{2} \smallint^{z_3}_{z_2} \ve{\om} \, + \, \tfrac{1}{2} \smallint^{z_5}_{z_6} \ve{\om} \, \pm \, \tfrac{1}{2} \smallint^{z_7}_{z_8} \ve{\om}\right) \, \tspin \left( \tfrac{1}{2} \smallint^{z_1}_{z_4} \ve{\om} \, + \, \tfrac{1}{2} \smallint^{z_3}_{z_2} \ve{\om} \, - \, \tfrac{1}{2} \smallint^{z_5}_{z_6} \ve{\om}\, \pm \, \tfrac{1}{2} \smallint^{z_7}_{z_8} \ve{\om} \right) 
 \right. \notag \\
& \ \ \ \ \ \ \ \left. + \ \frac{\vep_{\al \om} \, \vep_{\ga \be} \, \vep_{\ka \de}}{E_{16} \, E_{32} \, E_{54}} \; \tspin \left( -\tfrac{1}{2} \smallint^{z_1}_{z_6} \ve{\om} \, + \, \tfrac{1}{2} \smallint^{z_3}_{z_2} \ve{\om} \, + \, \tfrac{1}{2} \smallint^{z_5}_{z_4} \ve{\om} \, \pm \, \tfrac{1}{2} \smallint^{z_7}_{z_8} \ve{\om} \right) \right. \notag \\
& \ \ \ \ \ \ \ \ \ \ \ \ \ \ \left. \tspin \left( \tfrac{1}{2} \smallint^{z_1}_{z_6} \ve{\om} \, - \, \tfrac{1}{2} \smallint^{z_3}_{z_2} \ve{\om} \, + \, \tfrac{1}{2} \smallint^{z_5}_{z_4} \ve{\om} \, \pm \, \tfrac{1}{2} \smallint^{z_7}_{z_8} \ve{\om} \right) \, \tspin \left( \tfrac{1}{2} \smallint^{z_1}_{z_6} \ve{\om} \, + \, \tfrac{1}{2} \smallint^{z_3}_{z_2} \ve{\om} \, - \, \tfrac{1}{2} \smallint^{z_5}_{z_4} \ve{\om} \, \pm \, \tfrac{1}{2} \smallint^{z_7}_{z_8} \ve{\om} \right) 
 \right. \notag \\
& \ \ \ \ \ \ \ \left. - \ \frac{\vep_{\al \om} \, \vep_{\ga \de} \, \vep_{\ka \be}}{E_{16} \, E_{34} \, E_{52}} \; \tspin \left( -\tfrac{1}{2} \smallint^{z_1}_{z_6} \ve{\om} \, + \, \tfrac{1}{2} \smallint^{z_3}_{z_4} \ve{\om} \, + \, \tfrac{1}{2} \smallint^{z_5}_{z_2} \ve{\om} \, \pm \, \tfrac{1}{2} \smallint^{z_7}_{z_8} \ve{\om} \right) \right. \notag \\
& \ \ \ \ \ \ \ \ \ \ \ \ \ \ \left. \tspin \left( \tfrac{1}{2} \smallint^{z_1}_{z_6} \ve{\om} \, - \, \tfrac{1}{2} \smallint^{z_3}_{z_4} \ve{\om} \, + \, \tfrac{1}{2} \smallint^{z_5}_{z_2} \ve{\om} \, \pm \, \tfrac{1}{2} \smallint^{z_7}_{z_8} \ve{\om}\right) \, \tspin \left( \tfrac{1}{2} \smallint^{z_1}_{z_6} \ve{\om} \, + \, \tfrac{1}{2} \smallint^{z_3}_{z_4} \ve{\om} \, - \, \tfrac{1}{2} \smallint^{z_5}_{z_2} \ve{\om} \, \pm \, \tfrac{1}{2} \smallint^{z_7}_{z_8} \ve{\om}\right) 
 \right\}
\label{(6,2)}
\end{align}
The fact that $\tfrac{1}{2} \smallint^{z_7}_{z_8} \ve{\om}$ enters the $\tspin$ functions with an ambigous sign $\pm$ certainly requires further explanation. Using the simplest version (\ref{ft2}) of the Fay trisecant identity, one can demonstrate that both sign choices are in fact equivalent (as long as they are chosen consistently throughout the whole correlator). Flipping the sign amounts to adding a term proportional to the vanishing sum (\ref{sigspin1}) to this eight point function.

\medskip
Checking the agreement of (\ref{(6,2)}) at $\al = \ga = \ka = (+,+)$ and $\be = \de = \om = (-,-)$ with (\ref{at1}) is again a matter of (\ref{ft3}), this time with $\ve{\De} = \pm  \tfrac{1}{2} \smallint^{z_7}_{z_8} \ve{\om}$.

\medskip
The phenomenon of sign ambiguity occurs once again for the ten point function with six left handed and four right handed spin fields:
\begin{align}
&\langle S_{\al}(z_{1}) \, S_{\be}(z_{2}) \, S_{\ga}(z_{3}) \, S_{\de}(z_{4})  \, S_{\ka}(z_5) \, S_{\om}(z_6) \, S_{\dal}(z_7) \, S_{\dbe}(z_8) \, S_{\dga}(z_9) \, S_{\dde}(z_0) \rangle \spin \notag \\
&= \ \ \frac{- \, (E_{12} \, E_{14} \, E_{16} \, E_{23} \, E_{25} \, E_{34} \, E_{36}\, E_{45} \, E_{56} \, E_{78} \, E_{70} \, E_{89} \, E_{90})^{1/2}  }{ (E_{13} \, E_{15} \, E_{35} \, E_{24} \, E_{26} \, E_{46} \, E_{79} \, E_{80} )^{1/2} \, \tspin (\ve{0}) \, \tspin (\ve{0})} \notag \\
& \times \ \left\{ \frac{ \vep_{\dal \dbe} \, \vep_{\dga \dde} }{ E_{78} \, E_{90} \, \tspin \left( \tfrac{1}{2} \smallint^{z_1}_{z_2} \ve{\om} \, + \, \tfrac{1}{2} \smallint^{z_3}_{z_4} \ve{\om} \, + \, \tfrac{1}{2} \smallint^{z_5}_{z_6} \ve{\om} \, \pm \, \tfrac{1}{2} \smallint^{z_7}_{z_8} \ve{\om} \, \mp \, \tfrac{1}{2} \smallint^{z_9}_{z_0} \ve{\om}\right) } \right. \notag \\
& \ \ \ \ \ \left[ \frac{\vep_{\al \be} \, \vep_{\ga \de} \, \vep_{\ka \om}}{E_{12} \, E_{34} \, E_{56}} \; \tspin \left( \, \frac{1}{2} \, \left[ - \, \smallint^{z_1}_{z_2} \ve{\om} \, + \,  \smallint^{z_3}_{z_4} \ve{\om} \, + \,  \smallint^{z_5}_{z_6} \ve{\om} \, \pm \, \smallint^{z_7}_{z_8} \ve{\om} \, \mp  \, \smallint^{z_9}_{z_0} \ve{\om}  \right] \, \right) \right. \notag \\
& \ \ \ \ \ \left. \tspin \left( \, \frac{1}{2} \, \left[  \smallint^{z_1}_{z_2} \ve{\om} \, - \,  \smallint^{z_3}_{z_4} \ve{\om} \, + \,  \smallint^{z_5}_{z_6} \ve{\om} \, \pm \, \smallint^{z_7}_{z_8} \ve{\om} \, \mp  \, \smallint^{z_9}_{z_0} \ve{\om}  \right] \, \right)\, \tspin \left( \, \frac{1}{2} \, \left[ \smallint^{z_1}_{z_2} \ve{\om} \, + \,  \smallint^{z_3}_{z_4} \ve{\om} \, - \,  \smallint^{z_5}_{z_6} \ve{\om} \, \pm \, \smallint^{z_7}_{z_8} \ve{\om} \, \mp  \, \smallint^{z_9}_{z_0} \ve{\om}  \right] \, \right)
 \right. \notag \\
 & \ \ \  \left. - \ \frac{\vep_{\al \be} \, \vep_{\ga \om} \, \vep_{\ka \de}}{E_{12} \, E_{36} \, E_{54}} \; \tspin \left( \, \frac{1}{2} \, \left[ - \, \smallint^{z_1}_{z_2} \ve{\om} \, + \,  \smallint^{z_3}_{z_6} \ve{\om} \, + \,  \smallint^{z_5}_{z_4} \ve{\om} \, \pm \, \smallint^{z_7}_{z_8} \ve{\om} \, \mp  \, \smallint^{z_9}_{z_0} \ve{\om}  \right] \, \right) \right. \notag \\
& \ \ \ \ \ \left. \tspin \left( \, \frac{1}{2} \, \left[  \smallint^{z_1}_{z_2} \ve{\om} \, - \,  \smallint^{z_3}_{z_6} \ve{\om} \, + \,  \smallint^{z_5}_{z_4} \ve{\om} \, \pm \, \smallint^{z_7}_{z_8} \ve{\om} \, \mp  \, \smallint^{z_9}_{z_0} \ve{\om}  \right] \, \right)\, \tspin \left( \, \frac{1}{2} \, \left[ \smallint^{z_1}_{z_2} \ve{\om} \, + \,  \smallint^{z_3}_{z_6} \ve{\om} \, - \,  \smallint^{z_5}_{z_4} \ve{\om} \, \pm \, \smallint^{z_7}_{z_8} \ve{\om} \, \mp  \, \smallint^{z_9}_{z_0} \ve{\om}  \right] \, \right)
 \right. \notag \\
 & \ \ \  \left. + \ \frac{\vep_{\al \de} \, \vep_{\ga \om} \, \vep_{\ka \be}}{E_{14} \, E_{36} \, E_{52}} \; \tspin \left( \, \frac{1}{2} \, \left[ - \, \smallint^{z_1}_{z_4} \ve{\om} \, + \,  \smallint^{z_3}_{z_6} \ve{\om} \, + \,  \smallint^{z_5}_{z_2} \ve{\om} \, \pm \, \smallint^{z_7}_{z_8} \ve{\om} \, \mp  \, \smallint^{z_9}_{z_0} \ve{\om}  \right] \, \right) \right. \notag \\
& \ \ \ \ \ \left. \tspin \left( \, \frac{1}{2} \, \left[  \smallint^{z_1}_{z_4} \ve{\om} \, - \,  \smallint^{z_3}_{z_6} \ve{\om} \, + \,  \smallint^{z_5}_{z_2} \ve{\om} \, \pm \, \smallint^{z_7}_{z_8} \ve{\om} \, \mp  \, \smallint^{z_9}_{z_0} \ve{\om}  \right] \, \right)\, \tspin \left( \, \frac{1}{2} \, \left[ \smallint^{z_1}_{z_4} \ve{\om} \, + \,  \smallint^{z_3}_{z_6} \ve{\om} \, - \,  \smallint^{z_5}_{z_2} \ve{\om} \, \pm \, \smallint^{z_7}_{z_8} \ve{\om} \, \mp  \, \smallint^{z_9}_{z_0} \ve{\om}  \right] \, \right)
 \right. \notag \\
 & \ \ \  \left. - \ \frac{\vep_{\al \de} \, \vep_{\ga \be} \, \vep_{\ka \om}}{E_{14} \, E_{32} \, E_{56}} \; \tspin \left( \, \frac{1}{2} \, \left[ - \, \smallint^{z_1}_{z_4} \ve{\om} \, + \,  \smallint^{z_3}_{z_2} \ve{\om} \, + \,  \smallint^{z_5}_{z_6} \ve{\om} \, \pm \, \smallint^{z_7}_{z_8} \ve{\om} \, \mp  \, \smallint^{z_9}_{z_0} \ve{\om}  \right] \, \right) \right. \notag \\
& \ \ \ \ \ \left. \tspin \left( \, \frac{1}{2} \, \left[  \smallint^{z_1}_{z_4} \ve{\om} \, - \,  \smallint^{z_3}_{z_2} \ve{\om} \, + \,  \smallint^{z_5}_{z_6} \ve{\om} \, \pm \, \smallint^{z_7}_{z_8} \ve{\om} \, \mp  \, \smallint^{z_9}_{z_0} \ve{\om}  \right] \, \right)\, \tspin \left( \, \frac{1}{2} \, \left[ \smallint^{z_1}_{z_4} \ve{\om} \, + \,  \smallint^{z_3}_{z_2} \ve{\om} \, - \,  \smallint^{z_5}_{z_6} \ve{\om} \, \pm \, \smallint^{z_7}_{z_8} \ve{\om} \, \mp  \, \smallint^{z_9}_{z_0} \ve{\om}  \right] \, \right)
 \right. \notag \\
 & \ \ \  \left. + \ \frac{\vep_{\al \om} \, \vep_{\ga \be} \, \vep_{\ka \de}}{E_{16} \, E_{32} \, E_{54}} \; \tspin \left( \, \frac{1}{2} \, \left[ - \, \smallint^{z_1}_{z_6} \ve{\om} \, + \,  \smallint^{z_3}_{z_2} \ve{\om} \, + \,  \smallint^{z_5}_{z_4} \ve{\om} \, \pm \, \smallint^{z_7}_{z_8} \ve{\om} \, \mp  \, \smallint^{z_9}_{z_0} \ve{\om}  \right] \, \right) \right. \notag \\
& \ \ \ \ \ \left. \tspin \left( \, \frac{1}{2} \, \left[  \smallint^{z_1}_{z_6} \ve{\om} \, - \,  \smallint^{z_3}_{z_2} \ve{\om} \, + \,  \smallint^{z_5}_{z_4} \ve{\om} \, \pm \, \smallint^{z_7}_{z_8} \ve{\om} \, \mp  \, \smallint^{z_9}_{z_0} \ve{\om}  \right] \, \right)\, \tspin \left( \, \frac{1}{2} \, \left[ \smallint^{z_1}_{z_6} \ve{\om} \, + \,  \smallint^{z_3}_{z_2} \ve{\om} \, - \,  \smallint^{z_5}_{z_4} \ve{\om} \, \pm \, \smallint^{z_7}_{z_8} \ve{\om} \, \mp  \, \smallint^{z_9}_{z_0} \ve{\om}  \right] \, \right)
 \right. \notag \\
 & \ \ \  \left. - \ \frac{\vep_{\al \om} \, \vep_{\ga \de} \, \vep_{\ka \be}}{E_{16} \, E_{34} \, E_{52}} \; \tspin \left( \, \frac{1}{2} \, \left[ - \, \smallint^{z_1}_{z_6} \ve{\om} \, + \,  \smallint^{z_3}_{z_4} \ve{\om} \, + \,  \smallint^{z_5}_{z_2} \ve{\om} \, \pm \, \smallint^{z_7}_{z_8} \ve{\om} \, \mp  \, \smallint^{z_9}_{z_0} \ve{\om}  \right] \, \right) \right. \notag \\
& \ \ \ \ \ \left. \tspin \left( \, \frac{1}{2} \, \left[  \smallint^{z_1}_{z_6} \ve{\om} \, - \,  \smallint^{z_3}_{z_4} \ve{\om} \, + \,  \smallint^{z_5}_{z_2} \ve{\om} \, \pm \, \smallint^{z_7}_{z_8} \ve{\om} \, \mp  \, \smallint^{z_9}_{z_0} \ve{\om}  \right] \, \right)\, \tspin \left( \, \frac{1}{2} \, \left[ \smallint^{z_1}_{z_6} \ve{\om} \, + \,  \smallint^{z_3}_{z_4} \ve{\om} \, - \,  \smallint^{z_5}_{z_2} \ve{\om} \, \pm \, \smallint^{z_7}_{z_8} \ve{\om} \, \mp  \, \smallint^{z_9}_{z_0} \ve{\om}  \right] \, \right)
 \right] \notag \\
& \ \left. + \ \frac{ \vep_{\dal \dde} \, \vep_{\dga \dbe} }{ E_{70} \, E_{89} \, \tspin \left( \tfrac{1}{2} \smallint^{z_1}_{z_2} \ve{\om} \, + \, \tfrac{1}{2} \smallint^{z_3}_{z_4} \ve{\om} \, + \, \tfrac{1}{2} \smallint^{z_5}_{z_6} \ve{\om} \, \pm \, \tfrac{1}{2} \smallint^{z_7}_{z_0} \ve{\om} \, \mp \, \tfrac{1}{2} \smallint^{z_9}_{z_8} \ve{\om}\right) } \right. \notag \\
& \ \ \ \ \ \left[ \frac{\vep_{\al \be} \, \vep_{\ga \de} \, \vep_{\ka \om}}{E_{12} \, E_{34} \, E_{56}} \; \tspin \left( \, \frac{1}{2} \, \left[ - \, \smallint^{z_1}_{z_2} \ve{\om} \, + \,  \smallint^{z_3}_{z_4} \ve{\om} \, + \,  \smallint^{z_5}_{z_6} \ve{\om} \, \pm \, \smallint^{z_7}_{z_0} \ve{\om} \, \mp  \, \smallint^{z_9}_{z_8} \ve{\om}  \right] \, \right) \right. \notag \\
& \ \ \ \ \ \left. \tspin \left( \, \frac{1}{2} \, \left[  \smallint^{z_1}_{z_2} \ve{\om} \, - \,  \smallint^{z_3}_{z_4} \ve{\om} \, + \,  \smallint^{z_5}_{z_6} \ve{\om} \, \pm \, \smallint^{z_7}_{z_0} \ve{\om} \, \mp  \, \smallint^{z_9}_{z_8} \ve{\om}  \right] \, \right)\, \tspin \left( \, \frac{1}{2} \, \left[ \smallint^{z_1}_{z_2} \ve{\om} \, + \,  \smallint^{z_3}_{z_4} \ve{\om} \, - \,  \smallint^{z_5}_{z_6} \ve{\om} \, \pm \, \smallint^{z_7}_{z_0} \ve{\om} \, \mp  \, \smallint^{z_9}_{z_8} \ve{\om}  \right] \, \right)
 \right. \notag \\
 & \ \ \  \left. - \ \frac{\vep_{\al \be} \, \vep_{\ga \om} \, \vep_{\ka \de}}{E_{12} \, E_{36} \, E_{54}} \; \tspin \left( \, \frac{1}{2} \, \left[ - \, \smallint^{z_1}_{z_2} \ve{\om} \, + \,  \smallint^{z_3}_{z_6} \ve{\om} \, + \,  \smallint^{z_5}_{z_4} \ve{\om} \, \pm \, \smallint^{z_7}_{z_0} \ve{\om} \, \mp  \, \smallint^{z_9}_{z_8} \ve{\om}  \right] \, \right) \right. \notag \\
& \ \ \ \ \ \left. \tspin \left( \, \frac{1}{2} \, \left[  \smallint^{z_1}_{z_2} \ve{\om} \, - \,  \smallint^{z_3}_{z_6} \ve{\om} \, + \,  \smallint^{z_5}_{z_4} \ve{\om} \, \pm \, \smallint^{z_7}_{z_0} \ve{\om} \, \mp  \, \smallint^{z_9}_{z_8} \ve{\om}  \right] \, \right)\, \tspin \left( \, \frac{1}{2} \, \left[ \smallint^{z_1}_{z_2} \ve{\om} \, + \,  \smallint^{z_3}_{z_6} \ve{\om} \, - \,  \smallint^{z_5}_{z_4} \ve{\om} \, \pm \, \smallint^{z_7}_{z_0} \ve{\om} \, \mp  \, \smallint^{z_9}_{z_8} \ve{\om}  \right] \, \right)
 \right. \notag \\
 & \ \ \  \left. + \ \frac{\vep_{\al \de} \, \vep_{\ga \om} \, \vep_{\ka \be}}{E_{14} \, E_{36} \, E_{52}} \; \tspin \left( \, \frac{1}{2} \, \left[ - \, \smallint^{z_1}_{z_4} \ve{\om} \, + \,  \smallint^{z_3}_{z_6} \ve{\om} \, + \,  \smallint^{z_5}_{z_2} \ve{\om} \, \pm \, \smallint^{z_7}_{z_0} \ve{\om} \, \mp  \, \smallint^{z_9}_{z_8} \ve{\om}  \right] \, \right) \right. \notag \\
& \ \ \ \ \ \left. \tspin \left( \, \frac{1}{2} \, \left[  \smallint^{z_1}_{z_4} \ve{\om} \, - \,  \smallint^{z_3}_{z_6} \ve{\om} \, + \,  \smallint^{z_5}_{z_2} \ve{\om} \, \pm \, \smallint^{z_7}_{z_0} \ve{\om} \, \mp  \, \smallint^{z_9}_{z_8} \ve{\om}  \right] \, \right)\, \tspin \left( \, \frac{1}{2} \, \left[ \smallint^{z_1}_{z_4} \ve{\om} \, + \,  \smallint^{z_3}_{z_6} \ve{\om} \, - \,  \smallint^{z_5}_{z_2} \ve{\om} \, \pm \, \smallint^{z_7}_{z_0} \ve{\om} \, \mp  \, \smallint^{z_9}_{z_8} \ve{\om}  \right] \, \right)
 \right. \notag \\
 & \ \ \  \left. - \ \frac{\vep_{\al \de} \, \vep_{\ga \be} \, \vep_{\ka \om}}{E_{14} \, E_{32} \, E_{56}} \; \tspin \left( \, \frac{1}{2} \, \left[ - \, \smallint^{z_1}_{z_4} \ve{\om} \, + \,  \smallint^{z_3}_{z_2} \ve{\om} \, + \,  \smallint^{z_5}_{z_6} \ve{\om} \, \pm \, \smallint^{z_7}_{z_0} \ve{\om} \, \mp  \, \smallint^{z_9}_{z_8} \ve{\om}  \right] \, \right) \right. \notag \\
& \ \ \ \ \ \left. \tspin \left( \, \frac{1}{2} \, \left[  \smallint^{z_1}_{z_4} \ve{\om} \, - \,  \smallint^{z_3}_{z_2} \ve{\om} \, + \,  \smallint^{z_5}_{z_6} \ve{\om} \, \pm \, \smallint^{z_7}_{z_0} \ve{\om} \, \mp  \, \smallint^{z_9}_{z_8} \ve{\om}  \right] \, \right)\, \tspin \left( \, \frac{1}{2} \, \left[ \smallint^{z_1}_{z_4} \ve{\om} \, + \,  \smallint^{z_3}_{z_2} \ve{\om} \, - \,  \smallint^{z_5}_{z_6} \ve{\om} \, \pm \, \smallint^{z_7}_{z_0} \ve{\om} \, \mp  \, \smallint^{z_9}_{z_8} \ve{\om}  \right] \, \right)
 \right. \notag \\
 & \ \ \  \left. + \ \frac{\vep_{\al \om} \, \vep_{\ga \be} \, \vep_{\ka \de}}{E_{16} \, E_{32} \, E_{54}} \; \tspin \left( \, \frac{1}{2} \, \left[ - \, \smallint^{z_1}_{z_6} \ve{\om} \, + \,  \smallint^{z_3}_{z_2} \ve{\om} \, + \,  \smallint^{z_5}_{z_4} \ve{\om} \, \pm \, \smallint^{z_7}_{z_0} \ve{\om} \, \mp  \, \smallint^{z_9}_{z_8} \ve{\om}  \right] \, \right) \right. \notag \\
& \ \ \ \ \ \left. \tspin \left( \, \frac{1}{2} \, \left[  \smallint^{z_1}_{z_6} \ve{\om} \, - \,  \smallint^{z_3}_{z_2} \ve{\om} \, + \,  \smallint^{z_5}_{z_4} \ve{\om} \, \pm \, \smallint^{z_7}_{z_0} \ve{\om} \, \mp  \, \smallint^{z_9}_{z_8} \ve{\om}  \right] \, \right)\, \tspin \left( \, \frac{1}{2} \, \left[ \smallint^{z_1}_{z_6} \ve{\om} \, + \,  \smallint^{z_3}_{z_2} \ve{\om} \, - \,  \smallint^{z_5}_{z_4} \ve{\om} \, \pm \, \smallint^{z_7}_{z_0} \ve{\om} \, \mp  \, \smallint^{z_9}_{z_8} \ve{\om}  \right] \, \right)
 \right. \notag \\
 & \ \ \  \left. - \ \frac{\vep_{\al \om} \, \vep_{\ga \de} \, \vep_{\ka \be}}{E_{16} \, E_{34} \, E_{52}} \; \tspin \left( \, \frac{1}{2} \, \left[ - \, \smallint^{z_1}_{z_6} \ve{\om} \, + \,  \smallint^{z_3}_{z_4} \ve{\om} \, + \,  \smallint^{z_5}_{z_2} \ve{\om} \, \pm \, \smallint^{z_7}_{z_0} \ve{\om} \, \mp  \, \smallint^{z_9}_{z_8} \ve{\om}  \right] \, \right) \right. \notag \\
& \ \ \ \ \ \left. \left. \tspin \left( \, \frac{1}{2} \, \left[  \smallint^{z_1}_{z_6} \ve{\om} \, - \,  \smallint^{z_3}_{z_4} \ve{\om} \, + \,  \smallint^{z_5}_{z_2} \ve{\om} \, \pm \, \smallint^{z_7}_{z_0} \ve{\om} \, \mp  \, \smallint^{z_9}_{z_8} \ve{\om}  \right] \, \right)\, \tspin \left( \, \frac{1}{2} \, \left[ \smallint^{z_1}_{z_6} \ve{\om} \, + \,  \smallint^{z_3}_{z_4} \ve{\om} \, - \,  \smallint^{z_5}_{z_2} \ve{\om} \, \pm \, \smallint^{z_7}_{z_0} \ve{\om} \, \mp  \, \smallint^{z_9}_{z_8} \ve{\om}  \right] \, \right)
 \right]  \right\}
\label{(6,4)}
\end{align}
At this stage it would be nice to complete the list with the twelve point function:
\begin{align}
\langle S_{\al}(z_{1}) \, S_{\be}(z_{2}) \, S_{\ga}(z_{3}) \, S_{\de}(z_{4})  \, S_{\ka}(z_5) \, S_{\om}(z_6) \, S_{\dal}(z_A) \, S_{\dbe}(z_B) \, S_{\dga}(z_C) \, S_{\dde}(z_D)   \, S_{\dka}(z_E) \, S_{\dom}(z_F) \rangle \spin
\label{(6,6)}
\end{align}
The natural candidate for the $z$ dependence associated with the tensor $\vep_{\al \be} \vep_{\ga \de} \vep_{\ka \om}  \vep_{\dal \dbe}  \vep_{\dga \dde}  \vep_{\dka \dom}$ is the tree-level-like bunch of prime forms $(E_{12} E_{34} E_{56} E_{AB} E_{CD} E_{EF})^{-1}$ together with a combination of various factors
\beq
 \tspin \left( \, \frac{1}{2} \, \left[ \pm \, \smallint^{z_1}_{z_2} \ve{\om} \, \pm \,  \smallint^{z_3}_{z_4} \ve{\om} \, \pm \,  \smallint^{z_5}_{z_6} \ve{\om} \, \pm \, \smallint^{z_A}_{z_B} \ve{\om} \, \pm \,  \smallint^{z_C}_{z_D} \ve{\om} \, \pm \,  \smallint^{z_E}_{z_F} \ve{\om} \right] \, \right) \ .
\label{6,6a}
\eeq
But no combination of these $\Theta$ functions seems to be compatible with symmetry, periodicity properties and limits $z_i \mto z_j$ at the same time.

\subsection[$2M$ alike chiralities]{$\bm{2M}$ alike chiralities}
\label{sec:8}

Given the explicit results (\ref{(2,0)}), (\ref{(4,0)}) and (\ref{(6,0)}) for correlators with left handed spin fields only, we are led to the following guess for $2M$ insertions:
\bigskip

\framebox{\begin{minipage}{6.3in}
\begin{align}
&\langle S_{\al_1}(z_1) \, S_{\al_2}(z_2)\, ... \, S_{\al_{2M-1}}(z_{2M-1})\,S_{\al_{2M}}(z_{2M}) \rangle\spin \eq \frac{ (-1)^M}{\bigl[\tspin(\ve{0}) \bigr]^2} \; \left[ \tspin \left( \frac{1}{2} \, \sum_{i=1}^M \smallint^{z_{2i-1}}_{z_{2i}} \ve{\om}  \right) \right]^{2-M} \notag \\
 &  \ \ \times \ \bigg(\prod_{i \leq j}^M E_{2i-1,2j}\,\prod_{\bi<\bj}^{M}  E_{2\bi,2\bj-1}\bigg)^{1/2} \, \left(\prod_{k<l}^M E_{2k-1,2l-1}\, E_{2k,2l}\right)^{-1/2} \notag \\
  & \ \ \times \ \sum_{\rho \in S_M}\te{sgn}(\rho)\,
  \prod_{m=1}^M\frac{\vep_{\al_{2m-1}\al_{\rho(2m)}}}{E_{2m-1,\rho(2m)}} \; \tspin \left( \frac{1}{2} \, \sum_{i=1}^M \smallint^{z_{2i-1}}_{z_{2i}} \ve{\om} \, - \smallint^{z_{2m-1}}_{z_{\rho(2m)}} \ve{\om} \right)  \label{(2M,0)}
\end{align}
\end{minipage}}

\bigskip
\noindent
The eight point function according to this formula reads
\begin{align}
&\langle S_{\al}(z_{1}) \, S_{\be}(z_{2}) \, S_{\ga}(z_{3}) \, S_{\de}(z_{4})  \, S_{\ka}(z_5) \, S_{\om}(z_6) \, S_{\pi}(z_7) \, S_{\chi}(z_8) \rangle \spin \eq \frac{1}{( E_{13} \, E_{15} \, E_{17} \, E_{35} \, E_{37} \, E_{57} )^{1/2} } \notag \\
&\times \  \frac{(E_{12} \, E_{14} \, E_{16} \, E_{18} \, E_{23} \, E_{25} \, E_{27} \, E_{34} \, E_{36} \, E_{38} \, E_{45} \, E_{47} \, E_{56} \, E_{58} \, E_{67} \, E_{78} )^{1/2}}{( E_{24} \, E_{26} \, E_{28} \, E_{46} \, E_{48} \, E_{68}  )^{1/2} \, \left[ \tspin(\ve{0}) \, \tspin \left(  \frac{1}{2} \, \left[ \smallint^{z_1}_{z_2} \ve{\om} \, + \,  \smallint^{z_3}_{z_4} \ve{\om} \, + \,  \smallint^{z_5}_{z_6} \ve{\om} \, + \, \smallint^{z_7}_{z_8} \ve{\om} \right]  \right) \right]^2 } \notag \\
&\times \ \biggl\{ \frac{ \vep_{\al \be} \, \vep_{\ga \de} \, \vep_{\ka \om} \, \vep_{\pi \chi} }{ E_{12} \, E_{34} \, E_{56} \, E_{78} } \; \tspin \left(  \frac{1}{2}  \left[ - \smallint^{z_1}_{z_2} \ve{\om} + \smallint^{z_3}_{z_4} \ve{\om} +  \smallint^{z_5}_{z_6} \ve{\om} + \smallint^{z_7}_{z_8} \ve{\om} \right]  \right) \, \tspin \left(  \frac{1}{2} \, \left[ \smallint^{z_1}_{z_2} \ve{\om} -  \smallint^{z_3}_{z_4} \ve{\om} +   \smallint^{z_5}_{z_6} \ve{\om} + \smallint^{z_7}_{z_8} \ve{\om} \right]  \right) \biggr. \notag \\
& \ \ \ \ \ \ \ \ \tspin \left(  \frac{1}{2} \, \left[ \smallint^{z_1}_{z_2} \ve{\om} +  \smallint^{z_3}_{z_4} \ve{\om} -  \smallint^{z_5}_{z_6} \ve{\om} + \smallint^{z_7}_{z_8} \ve{\om} \right]  \right)  \, \tspin \left(  \frac{1}{2} \, \left[ \smallint^{z_1}_{z_2} \ve{\om} + \smallint^{z_3}_{z_4} \ve{\om} +   \smallint^{z_5}_{z_6} \ve{\om} - \smallint^{z_7}_{z_8} \ve{\om} \right]  \right) \notag \\
& \ \ \ - \ \frac{ \vep_{\al \be} \, \vep_{\ga \de} \, \vep_{\ka \chi} \, \vep_{\pi \om} }{ E_{12} \, E_{34} \, E_{58} \, E_{76} } \; \tspin \left(  \frac{1}{2} \, \left[ - \smallint^{z_1}_{z_2} \ve{\om} + \smallint^{z_3}_{z_4} \ve{\om} +  \smallint^{z_5}_{z_8} \ve{\om} + \smallint^{z_7}_{z_6} \ve{\om} \right]  \right) \, \tspin \left(  \frac{1}{2} \, \left[ \smallint^{z_1}_{z_2} \ve{\om} -  \smallint^{z_3}_{z_4} \ve{\om} +   \smallint^{z_5}_{z_8} \ve{\om} + \smallint^{z_7}_{z_6} \ve{\om} \right]  \right) \biggr. \notag \\
& \ \ \ \ \ \ \ \ \tspin \left(  \frac{1}{2} \, \left[ \smallint^{z_1}_{z_2} \ve{\om} +  \smallint^{z_3}_{z_4} \ve{\om} -  \smallint^{z_5}_{z_8} \ve{\om} + \smallint^{z_7}_{z_6} \ve{\om} \right]  \right)  \, \tspin \left(  \frac{1}{2} \, \left[ \smallint^{z_1}_{z_2} \ve{\om} + \smallint^{z_3}_{z_4} \ve{\om} +   \smallint^{z_5}_{z_8} \ve{\om} - \smallint^{z_7}_{z_6} \ve{\om} \right]  \right) \notag \\
& \ \ \ \ \biggl. \pm \  \ 22 \ \te{further permutations of} \, \bigl[ (z_2,\be) , (z_4,\de), (z_6,\om) , (z_8,\chi) \bigr] \biggr\} \ .
\end{align}
Depending on the choice of indices, either four, six or all of the 24 terms can be nonzero. Let us give one example of each case and quote the Fay trisecant identity necessary to check consistency with the result computed via (\ref{at1}):
\begin{itemize}
\item $\al = \be = \ga = \de = (+,+)$ and $\ka = \om = \pi = \chi = (-,-)$:
\begin{align*}
\Rightarrow \ \ &\vep_{\al \om} \vep_{\ga \chi} \vep_{\ka \be} \vep_{\pi \de} , \ \vep_{\al \om} \vep_{\ga \chi} \vep_{\ka \de} \vep_{\pi \be}, \ \vep_{\al \chi} \vep_{\ga \om} \vep_{\ka \be} \vep_{\pi \de} , \ \vep_{\al \chi} \vep_{\ga \om} \vep_{\ka \de} \vep_{\pi \be} \ \neq \ 0 \notag \\
\Rightarrow \ \ &\te{need (\ref{ft1a}) at $N=2$, i.e. (\ref{ft2})}
\end{align*}
\item $\al = \be = \ga = \ka = (+,+)$ and $\de = \om = \pi = \chi = (-,-)$:
\begin{align*}
\Rightarrow \ \ &\left. \begin{array}{l} \vep_{\al \de} \vep_{\ga \om} \vep_{\ka \chi} \vep_{\pi \be} 
, \ \vep_{\al \de} \vep_{\ga \chi} \vep_{\ka \om} \vep_{\pi \be}
, \ \vep_{\al \om} \vep_{\ga \de} \vep_{\ka \chi} \vep_{\pi \be} \\
\vep_{\al \om} \vep_{\ga \chi} \vep_{\ka \de} \vep_{\pi \be}
, \ \vep_{\al \chi} \vep_{\ga \de} \vep_{\ka \om} \vep_{\pi \be}
, \ \vep_{\al \chi} \vep_{\ga \om} \vep_{\ka \de} \vep_{\pi \be} \end{array} \right\} \ \neq \ 0 \notag \\
\Rightarrow \ \ &\te{need (\ref{ft1a}) at $N=3$, i.e. (\ref{ft3})}
\end{align*}
\item $\al = \ga = \ka = \pi = (+,+)$ and $\be = \de = \om = \chi = (-,-)$:
\begin{align*}
\Rightarrow \ \ &\te{any} \ \vep_{\al \cdot} \vep_{\ga \cdot} \vep_{\ka \cdot} \vep_{\pi \cdot} \ \neq \ 0 \notag \\
\Rightarrow \ \ &\te{need (\ref{ft1a}) at $N=4$, i.e. (\ref{ft4})}
\end{align*}
\end{itemize}
When adding two right handed spin fields, we can easily mimic the mechanism of (\ref{(2,2)}), (\ref{(4,2)}) and in particular of (\ref{(6,2)}) to include the arguments $z_{C,D}$ of the right handed fields into the $\tspin$ functions:
\bigskip

\framebox{\begin{minipage}{6.3in}
\begin{align}
&\langle S_{\al_1}(z_1) \, S_{\al_2}(z_2)\, ... \, S_{\al_{2M-1}}(z_{2M-1})\,S_{\al_{2M}}(z_{2M}) \, S_{\dga}(z_C) \, S_{\dde}(z_D) \rangle\spin   \notag \\
&= \ \ \frac{ (-1)^M}{\bigl[\tspin(\ve{0}) \bigr]^2} \; \left[ \tspin \left( \frac{1}{2} \, \sum_{i=1}^M \smallint^{z_{2i-1}}_{z_{2i}} \ve{\om} \, \pm \, \frac{1}{2} \smallint^{z_{C}}_{z_{D}} \ve{\om} \right) \right]^{2-M} \; \frac{\vep_{\dga \dde}}{E_{CD}^{1/2}} \notag \\
 & \ \ \times \ \bigg(\prod_{i \leq j}^M E_{2i-1,2j}\,\prod_{\bi<\bj}^{M}  E_{2\bi,2\bj-1}\bigg)^{1/2} \, \left(\prod_{k<l}^M E_{2k-1,2l-1}\, E_{2k,2l}\right)^{-1/2} \notag \\
  & \ \ \times \ \sum_{\rho \in S_M}\te{sgn}(\rho)\,
  \prod_{m=1}^M\frac{\vep_{\al_{2m-1}\al_{\rho(2m)}}}{E_{2m-1,\rho(2m)}} \; \tspin \left( \frac{1}{2} \, \sum_{i=1}^M \smallint^{z_{2i-1}}_{z_{2i}} \ve{\om} \, - \smallint^{z_{2m-1}}_{z_{\rho(2m)}} \ve{\om} \, \pm \, \frac{1}{2} \smallint^{z_{C}}_{z_{D}} \ve{\om} \right)  \label{(2M,2)}
\end{align}
\end{minipage}}

\bigskip
\noindent
The sign ambiguity vanishes at $M \leq 2$. The cases with $M \geq 3$ offer an increasing number of possibilities to add zeros in the form $0 = \sum_{\rho \in S_m} \te{sgn}(\rho) \prod_{k=1}^m \vep_{\al_{2k-1} \al_{\rho(2k)}}$ for any $3 \leq m \leq M$ which then allow to identify both sign choices.

\medskip
Similarly, as a generalization of (\ref{(4,4)}) and (\ref{(6,4)}) with four right handed spin fields, we claim that
\bigskip

\framebox{\begin{minipage}{6.3in}
\begin{align}
&\langle S_{\al_1}(z_1) \, S_{\al_2}(z_2)\, ... \, S_{\al_{2M}}(z_{2M}) \, S_{\dga}(z_C) \, S_{\dde}(z_D) \, S_{\dka}(z_E) \, S_{\dom}(z_F) \rangle\spin  \eq \frac{ (-1)^M}{\bigl[\tspin(\ve{0}) \bigr]^2} \notag \\
& \ \ \times \;  \left( \frac{E_{CD} \, E_{CF} \, E_{DE} \, E_{EF}}{E_{CE} \, E_{DF}}\right)^{1/2} \, \left(\prod_{i \leq j}^M E_{2i-1,2j}\,\prod_{\bi<\bj}^{M}  E_{2\bi,2\bj-1}\right)^{1/2} \, \left(\prod_{k<l}^M E_{2k-1,2l-1}\, E_{2k,2l}\right)^{-1/2} \notag \\
  & \ \ \times \ \Biggl\{ \frac{\vep_{\dga \dde} \, \vep_{\dka \dom}}{E_{CD} \, E_{EF}} \; \left[ \tspin \left( \frac{1}{2} \, \sum_{i=1}^M \smallint^{z_{2i-1}}_{z_{2i}} \ve{\om} \, \pm \, \frac{1}{2} \smallint^{z_{C}}_{z_{D}} \ve{\om} \, \mp \, \frac{1}{2} \smallint^{z_{E}}_{z_{F}} \ve{\om} \right) \right]^{2-M} \, \sum_{\rho \in S_M}\te{sgn}(\rho)\Biggr. \notag \\
  & \ \ \ \ \ \ \ \ \ \ \ \ \ \ \ \ \ \Biggl. \prod_{m=1}^M\frac{\vep_{\al_{2m-1}\al_{\rho(2m)}}}{E_{2m-1,\rho(2m)}} \; \tspin \left( \frac{1}{2} \, \sum_{i=1}^M \smallint^{z_{2i-1}}_{z_{2i}} \ve{\om} \, - \smallint^{z_{2m-1}}_{z_{\rho(2m)}} \ve{\om} \, \pm \, \frac{1}{2} \smallint^{z_{C}}_{z_{D}} \ve{\om} \, \mp \, \frac{1}{2} \smallint^{z_{E}}_{z_{F}} \ve{\om} \right) \Biggr. \notag \\
  & \ \ \ \ \ - \; \frac{\vep_{\dga \dom} \, \vep_{\dka \dde}}{E_{CF} \, E_{ED}} \; \left[ \tspin \left( \frac{1}{2} \, \sum_{i=1}^M \smallint^{z_{2i-1}}_{z_{2i}} \ve{\om} \, \pm \, \frac{1}{2} \smallint^{z_{C}}_{z_{F}} \ve{\om} \, \mp \, \frac{1}{2} \smallint^{z_{E}}_{z_{D}} \ve{\om} \right) \right]^{2-M} \, \sum_{\rho \in S_M}\te{sgn}(\rho)\Biggr. \notag \\
  & \ \ \ \ \ \ \ \ \ \ \ \ \ \ \ \ \ \Biggl. \prod_{m=1}^M\frac{\vep_{\al_{2m-1}\al_{\rho(2m)}}}{E_{2m-1,\rho(2m)}} \; \tspin \left( \frac{1}{2} \, \sum_{i=1}^M \smallint^{z_{2i-1}}_{z_{2i}} \ve{\om} \, - \smallint^{z_{2m-1}}_{z_{\rho(2m)}} \ve{\om} \, \pm \, \frac{1}{2} \smallint^{z_{C}}_{z_{F}} \ve{\om} \, \mp \, \frac{1}{2} \smallint^{z_{E}}_{z_{D}} \ve{\om} \right) \Biggr\} \ . \label{(2M,4)}
\end{align}
\end{minipage}}

\bigskip
\noindent
These expressions will be proven in appendix \ref{sec:appC}.

\medskip
With these correlation functions at hand, it is in principle possible to derive a cornucopia of correlators with NS fields included via $\psi^\mu$ factorization (\ref{3,0}):
\begin{itemize}
\item starting from $\langle S_{\al_1} \, ... \, S_{\al_{2M}} \, S_{\dga} \, S_{\dde} \rangle \spin$
\[ \Rightarrow \ \ \langle \psi^\mu \, S_{\al_1} \, ... \, S_{\al_{2M-1}} \, S_{\dga} \,  \rangle \spin \ , \ \ \langle \psi^\mu \, \psi^\nu \, S_{\al_1} \, ... \, S_{\al_{2M-2}} \rangle \spin 
\]
\item starting from $\langle S_{\al_1} \, ... \, S_{\al_{2M}} \, S_{\dga} \, S_{\dde} \, S_{\dka} \, S_{\dom} \rangle \spin$
\[ \Rightarrow \ \ \left\{ \ \begin{array}{l} \langle \psi^\mu \, S_{\al_1} \, ... \, S_{\al_{2M-1}} \, S_{\dga} \, S_{\dde} \, S_{\dka}  \rangle \spin \ , \ \ \langle \psi^\mu \, \psi^\nu \, S_{\al_1} \, ... \, S_{\al_{2M-2}} \, S_{\dga} \, S_{\dde}  \rangle \spin \\
\langle \psi^\mu \, \psi^\nu \, \psi^\la \, S_{\al_1} \, ... \, S_{\al_{2M-3}} \, S_{\dga}   \rangle \spin \ , \ \ \langle \psi^\mu \, \psi^\nu \, \psi^\la \, \psi^\rho \, S_{\al_1} \, ... \, S_{\al_{2M-4}}  \rangle \spin
\end{array} \right.
\]
\end{itemize}


\section{Mixed correlators with fermions and spin fields}
\label{sec:mix}

In this section, we will derive correlation functions with all types of fields $\{ \psi^\mu, S_\al , S_{\dbe} \}$ of the $D=4$ RNS CFT involved. These are the correlators of physical interest since scattering amplitudes with many external states and on higher genus involve plenty of NS fermions. The number of spin fields, on the other hand, precisely equals the number of spacetime fermions, independent on the ghost pictures chosen and on the loop order.

\medskip
An important class of amplitudes are those with two spacetime fermions. In this section, the required correlators $\langle \psi^{\mu_1} ... \psi^{\mu_n} S_{\al} S_{\be , \dbe} \rangle \spin$ are given in closed form for arbitrary genus $g$ and worldsheet fermion number $n$, similar to the preceding work \cite{tree} on tree level. At one loop, for instance, computing a massless gauge multiplet amplitude ${\cal M}_{g=1}(A^{k},\chi,\bar{\chi})$ with $k$ gluons $A$ and two gauginos $\chi,\bar{\chi}$ requires knowledge of the correlator $\langle \psi^{\mu_1}...\psi^{\mu_{2k+1}} S_\al S_{\dbe} \rangle \spin$ and lower order relatives. The increased CFT effort compared to the tree level computations in \cite{LHC1,LHC2} is due to the vanishing superghost background charge at one loop which enforces higher ghost pictures for the partons' vertex operators.

\medskip
It would be interesting to extend the results of \cite{MHV} on supersymmetric Ward identities at tree level to nonzero genus. The correlation functions given in this section are essential for exploring and checking relations as e.g. between ${\cal M}_{g}(A^{k},\chi^{\ell},\bar{\chi}^{\ell})$ and ${\cal M}_{g}(A^{k+2},\chi^{\ell-1},\bar{\chi}^{\ell-1})$ which are well-established at tree level.

\medskip
In presence of four spin fields, correlation functions could not be written down for arbitrary number of $\psi$ insertions. We have restricted our attention to those correlators necessary for computing a four fermion amplitude at one loop.

\medskip
There are in general three avenues to obtain the $n+r+s$ point function
\beq
\langle \psi^{\mu_1}(z_1)\dots\psi^{\mu_n}(z_n)\,S_{\al_1}(x_1)\dots S_{\al_{r}}(x_{r})\,S_{\dbe_1}(y_1)\dots S_{\dbe_{s}}(y_{s}) \rangle \spin
\label{gen}
\eeq
\begin{itemize}
\item [(i)] The factorization prescription $\psi^{\mu} = -\frac{ \bar{\si}^{\mu \dka \ka}}{\sqrt{2}} S_\ka S_{\dka}$. This is a straightforward method which works extremely well in lower order examples. However, the complicated nature of the higer point spin field correlators starting with (\ref{(6,0)}) spoils the efficiency of this method as soon as the factorized correlator reaches $n+r \geq 6$ or $n+s \geq 6$ in the notation of (\ref{gen}). In particular, we have not found a compact expression for (\ref{(6,6)}) with $n=0$ and $r=s=6$, i.e. the cases where both $n+r \geq 6$ and $n+s \geq 6$ at the same time cannot be solved in the factorization approach.
\item [(ii)] Matching the singularity structures with the OPEs (\ref{rv,1a}) to (\ref{rv,2b}) and adjusting the $\tspin$ arguments by periodicity considerations as well as limiting behaviour. Here we have a recursive algorithm which determines $N$ point functions by matching poles in $z_{ij}$ with $N-1$ and $N-2$ point functions related by OPEs. In principle, there are no limitations on the number of fields (our failure to find (\ref{(6,6)}) should be considered as a rather technical obstruction here), but the computational effort increases enormously with $n,r,s$.
\item [(iii)] Reduce the correlator (\ref{gen}) to its $SO(2)$ spin system constituents and compute it in certain configurations of the indices $\mu_i, \al_i, \dbe_i$ using (\ref{at2}). This is the method we will use in this section beyond the five point function $\langle \psi^\mu \psi^\nu \psi^\la S_\al S_\be \rangle \spin$. One has to start with a convenient basis of $SO(1,3)$ index structures namely antisymmetrized $\si$ matrix products. Then one can find values of the indices where only one of the tensors is non-zero and read off the associated $z_{ij}$ function by applying (\ref{at2}). Converting antisymmetric $\si$ products into ordered $\si$ chains (in terms of which our final results will be given) is a matter of the Dirac algebra $\si^{\mu} \bar{\si}^\nu + \si^\nu \bar{\si}^\mu = -2\eta^{\mu \nu}$.
\end{itemize}

\subsection{Correlation functions with two spin fields}
\label{sec:2sp}

The simple correlator $\langle \psi^\mu S_\al S_{\dbe} \rangle \spin$ -- the $n=r=s=1$ case of (\ref{gen}) -- can be neatly reduced to the four point function (\ref{(2,2)}) by factorizing the fermion:
\beq
\langle \psi^{\mu}(z_{1}) \, S_{\al}(z_{2}) \, S_{\dbe}(z_{3}) \rangle \spin \ \ = \ \ \frac{ \si^{\mu}_{\al \dbe}}{\sqrt{2} \, (E_{12} \, E_{13})^{1/2}} \; \frac{ \tspin \left( \frac{1}{2} \int_{z_1}^{z_2} \ve{\om} \, + \, \frac{1}{2} \int_{z_1}^{z_3} \ve{\om} \right) \, \tspin \left( \frac{1}{2} \int_{z_3}^{z_2}   \ve{\om} \right) }{ \tspin ( \ve{0} ) \, \tspin (\ve{0}) }
\label{3pt2}
\eeq
The four point function $\langle \psi^\mu \psi^\nu S_\al S_\be \rangle \spin$ can be expressed in two ways:
\begin{subequations}
\begin{align}
\langle &\psi^{\mu}(z_{1}) \, \psi^{\nu} (z_{2}) \, S_{\al}(z_{3}) \, S_{\be}(z_{4}) \rangle \spin \eq \frac{- \, 1}{(E_{13} \, E_{14} \, E_{23} \, E_{24} \, E_{34})^{1/2} \, \tspin ( \ve{0} ) \, \tspin (\ve{0}) } \notag \\
& \ \times \, \left\{ \frac{\eta^{\mu \nu} }{E_{12}} \; \vep_{\al \be} \, E_{13} \, E_{24} \, \tspin \left(  \smallint_{z_2}^{z_1} \ve{\om} \, + \, \tfrac{1}{2} \smallint_{z_4}^{z_3} \ve{\om} \right) \, \tspin \left( \tfrac{1}{2} \smallint_{z_4}^{z_3}   \ve{\om} \right) \right. \notag \\
& \ \ \ \ \ \ \ \ \left. + \ (\si^\mu \, \bar{\si}^\nu \, \vep)_{\al \be} \; \frac{E_{34}}{2} \; \tspin \left( \tfrac{1}{2} \smallint_{z_1}^{z_3} \ve{\om} \, + \, \tfrac{1}{2} \smallint_{z_1}^{z_4} \ve{\om} \right) \, \tspin \left( \tfrac{1}{2} \smallint_{z_2}^{z_3} \ve{\om} \, + \, \tfrac{1}{2} \smallint_{z_2}^{z_4} \ve{\om} \right) 
 \right\}
\label{4pt2} \\
&= \ \ \frac{- \, 1}{2 \, (E_{13} \, E_{14} \, E_{23} \, E_{24} \, E_{34})^{1/2} \, \tspin ( \ve{0} ) \, \tspin (\ve{0})} \notag \\
& \ \times \, \left\{ \frac{\eta^{\mu \nu} }{E_{12}} \; \vep_{\al \be} \, \tspin \left( \tfrac{1}{2} \smallint_{z_4}^{z_3}   \ve{\om} \right) \, \left[ E_{13} \, E_{24} \, \tspin \left(  \smallint_{z_2}^{z_1} \ve{\om} \, + \, \tfrac{1}{2} \smallint_{z_4}^{z_3} \ve{\om} \right) \ + \ E_{14} \, E_{23} \, \tspin \left(  -\smallint_{z_2}^{z_1} \ve{\om} \, + \, \tfrac{1}{2} \smallint_{z_4}^{z_3} \ve{\om} \right)  \right] \right. \notag \\
& \ \ \ \ \ \ \ \ \left. + \ (\si^{\mu \nu} \, \vep)_{\al \be} \, E_{34} \, \tspin \left( \tfrac{1}{2} \smallint_{z_1}^{z_3} \ve{\om} \, + \, \tfrac{1}{2} \smallint_{z_1}^{z_4} \ve{\om} \right) \, \tspin \left( \tfrac{1}{2} \smallint_{z_2}^{z_3} \ve{\om} \, + \, \tfrac{1}{2} \smallint_{z_2}^{z_4} \ve{\om} \right) 
 \right\} \label{4pt2b}
\end{align}
\end{subequations}
Reducing this correlator down to the 4+2 point function (\ref{(4,2)}) first of all gives the form (\ref{4pt2}), and the second equality follows from the $\si$ matrix identity $\si^{\mu} \bar{\si}^{\nu} = \si^{\mu \nu} - \eta^{\mu \nu}$. The expression (\ref{4pt2b}) is manifestly antisymmetric under exchange of $\psi^{\mu}(z_{1}) \leftrightarrow \psi^{\nu} (z_{2})$. A similar mechanism applies to higher point correlators given in the following, see the corresponding section in \cite{tree} for the details. To keep the results short, we stick to a presentation in terms of ordered $\si$ matrix chains in the following.

\medskip
Computing the two fermion -- one boson vertex on the torus requires knowledge of the five point function $\langle \psi^\mu \psi^\nu \psi^\la S_\al S_\be \rangle \spin$ with $n=3$ and $r=s=1$. This is the last time where the factorization algorithm can be applied with a reasonable effort. Starting point is the spin field eight point function (\ref{(4,4)}).
\begin{align}
\langle &\psi^{\mu}(z_{1}) \, \psi^{\nu} (z_{2}) \, \psi^{\la}(z_{3}) \, S_{\al}(z_{4}) \, S_{\dbe}(z_{5}) \rangle \spin \ \ = \ \ \frac{1}{\sqrt{2} \, (E_{14} \, E_{15} \, E_{24} \, E_{25} \, E_{34} \, E_{35})^{1/2} \,  \tspin ( \ve{0} ) \, \tspin ( \ve{0} )}  \notag \\
&\times \, \left\{  \frac{\eta^{\mu \nu}}{E_{12}} \; \si^{\la}_{\al \dbe} \; E_{14} \, E_{25} \,  \tspin \left( \smallint_{z_2}^{z_1} \ve{\om} \, + \, \tfrac{1}{2} \smallint_{z_5}^{z_4} \ve{\om} \right)  \, \tspin \left( \tfrac{1}{2} \smallint_{z_3}^{z_4} \ve{\om} \, + \, \tfrac{1}{2} \smallint_{z_3}^{z_5} \ve{\om} \right) \right. \notag \\
& \ \ \left. - \ \frac{\eta^{\mu \la}}{E_{13}} \; \si^{\nu}_{\al \dbe} \; E_{14} \, E_{35} \,  \tspin \left( \smallint_{z_3}^{z_1} \ve{\om} \, + \, \tfrac{1}{2} \smallint_{z_5}^{z_4} \ve{\om} \right)  \, \tspin \left( \tfrac{1}{2} \smallint_{z_2}^{z_4} \ve{\om} \, + \, \tfrac{1}{2} \smallint_{z_2}^{z_5} \ve{\om} \right) \right. \notag \\
& \ \ \left. + \ \frac{\eta^{\nu \la}}{E_{23}} \; \si^{\mu}_{\al \dbe} \; E_{24} \, E_{35} \,  \tspin \left( \smallint_{z_3}^{z_2} \ve{\om} \, + \, \tfrac{1}{2} \smallint_{z_5}^{z_4} \ve{\om} \right)  \, \tspin \left( \tfrac{1}{2} \smallint_{z_1}^{z_4} \ve{\om} \, + \, \tfrac{1}{2} \smallint_{z_1}^{z_5} \ve{\om} \right) \right. \notag \\ 
& \ \ \left. + \ (\si^{\mu} \, \bar{\si}^{\nu} \, \si^{\la})_{\al \dbe} \; \frac{E_{45}}{2 \, \tspin \left( \tfrac{1}{2} \smallint_{z_5}^{z_4} \ve{\om} \right) } \; \tspin \left( \tfrac{1}{2} \smallint_{z_1}^{z_4} \ve{\om} \, + \, \tfrac{1}{2} \smallint_{z_1}^{z_5} \ve{\om} \right) \right. \notag \\
& \ \ \ \ \ \ \ \ \ \ \ \ \ \ \ \ \ \ \ \ \ \ \left.   \tspin \left( \tfrac{1}{2} \smallint_{z_2}^{z_4} \ve{\om} \, + \, \tfrac{1}{2} \smallint_{z_2}^{z_5} \ve{\om} \right) \, \tspin \left( \tfrac{1}{2} \smallint_{z_3}^{z_4} \ve{\om} \, + \, \tfrac{1}{2} \smallint_{z_3}^{z_5} \ve{\om} \right) \right\}
\label{5pt2}
\end{align}
On the step towards $n=4$, $r=2$ and $s=0$ in (\ref{gen}), it makes sense to switch gears. From this point on, we will rely on (\ref{at2}) to compute the $z$ dependent prefactors. A suitable $SO(1,3)$ covariant basis for $\langle \psi^\mu \psi^\nu \psi^\la \psi^\rho S_\al S_\be \rangle \spin$ is displayed in the following table:
\begin{center}
\begin{tabular}{|r|c|c|c|c|c|}\hline $\left( \begin{smallmatrix} \mu &\nu &\la &\rho \\  &\al &\be \end{smallmatrix} \right)$
& $\left( \begin{smallmatrix} 0 &0 &1 &1 \\ &+ &- \end{smallmatrix} \right)$ 
& $\left( \begin{smallmatrix} 0 &1 &0 &1 \\ &+ &- \end{smallmatrix} \right)$ 
& $\left( \begin{smallmatrix} 0 &1 &1 &0 \\ &+ &- \end{smallmatrix} \right)$
 & $\left( \begin{smallmatrix} 0 &0 &1 &2 \\ &+ &- \end{smallmatrix} \right)$
 & $\left( \begin{smallmatrix} 0 &1 &2 &3 \\ &+ &- \end{smallmatrix} \right)$
  \\ \hline \hline 
$\eta^{\mu \nu} \, \eta^{\la \rho} \, \vep_{\al \be}$ &1 & 0 & 0 & 0 & 0   \\ \hline
$\eta^{\mu \la} \, \eta^{\nu \rho} \, \vep_{\al \be}$ &0 & 1 & 0 & 0 & 0  \\ \hline
$\eta^{\mu \rho} \, \eta^{\nu \la} \, \vep_{\al \be}$ &0 & 0 & 1  & 0 & 0 \\ \hline
$\eta^{\mu \nu} \, (\si^{\la \rho} \, \vep)_{\al \be}$ & 0 & 0 &0 & 1 & 0  \\ \hline
$\eta^{\mu \la} \, (\si^{\nu \rho} \, \vep)_{\al \be}$ & 0 & 0 & 0 &0 &0 \\ \hline 
$\eta^{\mu \rho} \, (\si^{\nu \la} \, \vep)_{\al \be}$ & 0 & 0 & 0 &0 &0 \\ \hline 
$\eta^{\nu \la} \, (\si^{\mu \rho} \, \vep)_{\al \be}$ & 0 & 0 & 0 &0 &0 \\ \hline 
$\eta^{\nu \rho} \, (\si^{\mu \la} \, \vep)_{\al \be}$ & 0 & 0 & 0 &0 &0 \\ \hline 
$\eta^{ \la \rho } \, (\si^{\mu \nu} \, \vep)_{\al \be}$ & 0 & 0 & 0 &0 &0 \\ \hline 
$\vep^{\mu \nu \la \rho} \, \vep_{\al \be}$ & 0 & 0 & 0 &0 &1 \\ \hline 
\end{tabular}
\begin{flushleft}{\bf Table 2}. {\it The correlator $\langle \psi^\mu \psi^\nu \psi^\la \psi^\rho S_\al S_\be \rangle$ can be obtained in the given basis of tensor structures by plugging the index configurations shown in the head line (and permutations thereof) into (\ref{at2}) using the $SO(2) \leftrightarrow SO(1,3)$ dictionary (\ref{2,1g}), (\ref{2,1h}), (\ref{2,1i}). The numbers '1' should not be taken literally, they represent phases $\pm 1$ or $\pm i$ depending on the active $\si^\mu$ entries.}\\[3mm]
\end{flushleft}
\end{center}
If we now convert the antisymmetric $\si$ products into ordered ones using $\si^{\mu \nu} = \si^{\mu} \bar{\si}^\nu + \eta^{\mu \nu}$ and
\begin{align}
-i \, \vep^{\mu \nu \la \rho} \, \vep_{\al \be} \ \ = \ \ (\si^\mu \, \bar{\si}^\nu \, \si^\la \, \bar{\si}^\rho \, \vep )_{\al \be} \ &+ \ \eta^{\mu \nu} \, (\si^\la \, \bar{\si}^\rho \, \vep )_{\al \be} \ - \ \eta^{\mu \la} \, (\si^\nu \, \bar{\si}^\rho \, \vep )_{\al \be} \ + \ \eta^{\mu \rho} \, (\si^\nu \, \bar{\si}^\la \, \vep )_{\al \be}  \notag \\
&+ \ \eta^{ \nu \la } \, (\si^\mu \, \bar{\si}^\rho \, \vep )_{\al \be} \ - \ \eta^{ \nu \rho } \, (\si^\mu \, \bar{\si}^\la \, \vep )_{\al \be} \ + \ \eta^{\la \rho} \, (\si^\mu \, \bar{\si}^\nu \, \vep )_{\al \be} \notag \\
& + \ \bigl( \eta^{\mu \nu} \, \eta^{\la \rho} \ - \ \eta^{\mu \la} \, \eta^{\nu \rho} \ + \ \eta^{\mu \rho} \, \eta^{\nu \la} \bigl) \, \vep_{\al \be} \ , \label{si4}
\end{align}
then the $N=2$ Fay trisecant identity (\ref{ft2}) yields the shortest form of the correlator:
\begin{align}
&\langle \psi^{\mu} (z_{1}) \, \psi^{\nu}(z_{2}) \, \psi^{\la}(z_{3}) \, \psi^{\rho}(z_{4}) \, S_{\al}(z_{5}) \, S_{\be}(z_{6}) \rangle \spin \notag \\
&= \ \ \frac{- \, 1}{ (E_{15} \, E_{16} \, E_{25} \, E_{26} \, E_{35} \, E_{36} \, E_{45} \, E_{46} \, E_{56})^{1/2} \,  \tspin ( \ve{0} ) \, \tspin ( \ve{0} )} \notag \\
&\ \ \ \times \, \left\{ \frac{\eta^{\mu \nu} \, \eta^{\la \rho}}{E_{12} \, E_{34}} \; \vep_{\al \be} \, E_{15} \, E_{26} \, E_{35} \, E_{46} \, \tspin \left( \smallint_{z_2}^{z_1} \ve{\om} \, + \, \tfrac{1}{2} \smallint_{z_6}^{z_5} \ve{\om} \right) \, \tspin \left( \smallint_{z_4}^{z_3} \ve{\om} \, + \, \tfrac{1}{2} \smallint_{z_6}^{z_5} \ve{\om} \right) \right. \notag  \\
&\ \ \ \ \ \left. - \ \frac{\eta^{\mu \la} \, \eta^{\nu \rho}}{E_{13} \, E_{24}} \; \vep_{\al \be} \, E_{15} \, E_{36} \, E_{25} \, E_{46} \, \tspin \left( \smallint_{z_3}^{z_1} \ve{\om} \, + \, \tfrac{1}{2} \smallint_{z_6}^{z_5} \ve{\om} \right) \, \tspin \left( \smallint_{z_4}^{z_2} \ve{\om} \, + \, \tfrac{1}{2} \smallint_{z_6}^{z_5} \ve{\om} \right) \right. \notag  \\
& \ \ \ \ \ \left. + \ \frac{\eta^{\mu \rho} \, \eta^{\nu \la}}{E_{14} \, E_{23}} \; \vep_{\al \be} \, E_{15} \, E_{46} \, E_{25} \, E_{36} \, \tspin \left( \smallint_{z_4}^{z_1} \ve{\om} \, + \, \tfrac{1}{2} \smallint_{z_6}^{z_5} \ve{\om} \right) \, \tspin \left( \smallint_{z_3}^{z_2} \ve{\om} \, + \, \tfrac{1}{2} \smallint_{z_6}^{z_5} \ve{\om} \right) \right. \notag  \\
& \ \ \ \ \ \left. + \ \frac{\eta^{\mu \nu} }{E_{12} } \; ( \si^{\la } \, \bar{\si}^{\rho} \, \vep)_{\al \be} \, E_{15} \, E_{26}  \, \tspin \left( \smallint_{z_2}^{z_1} \ve{\om} \, + \, \tfrac{1}{2} \smallint_{z_6}^{z_5} \ve{\om} \right)  \right. \notag \\
& \ \ \ \ \ \ \ \ \ \ \ \ \ \  \frac{E_{56}}{2\, \tspin \left( \tfrac{1}{2} \smallint_{z_6}^{z_5} \ve{\om} \right) } \; \tspin \left( \tfrac{1}{2} \smallint_{z_3}^{z_5} \ve{\om} \, + \, \tfrac{1}{2} \smallint_{z_3}^{z_6} \ve{\om} \right) \, \tspin \left( \tfrac{1}{2} \smallint_{z_4}^{z_5} \ve{\om} \, + \, \tfrac{1}{2} \smallint_{z_4}^{z_6} \ve{\om} \right)   \notag \\
& \ \ \ \ \ \left. - \ \frac{\eta^{\mu \la} }{E_{13} } \; ( \si^{\nu } \, \bar{\si}^{\rho} \, \vep)_{\al \be} \, E_{15} \, E_{36}  \, \tspin \left( \smallint_{z_3}^{z_1} \ve{\om} \, + \, \tfrac{1}{2} \smallint_{z_6}^{z_5} \ve{\om} \right)  \right. \notag \\
& \ \ \ \ \ \ \ \ \ \ \ \ \ \  \frac{E_{56}}{2 \, \tspin \left( \tfrac{1}{2} \smallint_{z_6}^{z_5} \ve{\om} \right)} \;  \tspin \left( \tfrac{1}{2} \smallint_{z_2}^{z_5} \ve{\om} \, + \, \tfrac{1}{2} \smallint_{z_2}^{z_6} \ve{\om} \right) \, \tspin \left( \tfrac{1}{2} \smallint_{z_4}^{z_5} \ve{\om} \, + \, \tfrac{1}{2} \smallint_{z_4}^{z_6} \ve{\om} \right)  \notag \\
& \ \ \ \ \ \left. + \ \frac{\eta^{\mu \rho} }{E_{14} } \; ( \si^{\nu } \, \bar{\si}^{\la} \, \vep)_{\al \be} \, E_{15} \, E_{46}  \, \tspin \left( \smallint_{z_4}^{z_1} \ve{\om} \, + \, \tfrac{1}{2} \smallint_{z_6}^{z_5} \ve{\om} \right)  \right. \notag \\
& \ \ \ \ \ \ \ \ \ \ \ \ \ \  \frac{E_{56}}{2 \, \tspin \left( \tfrac{1}{2} \smallint_{z_6}^{z_5} \ve{\om} \right)} \; \tspin \left( \tfrac{1}{2} \smallint_{z_2}^{z_5} \ve{\om} \, + \, \tfrac{1}{2} \smallint_{z_2}^{z_6} \ve{\om} \right) \, \tspin \left( \tfrac{1}{2} \smallint_{z_3}^{z_5} \ve{\om} \, + \, \tfrac{1}{2} \smallint_{z_3}^{z_6} \ve{\om} \right)   \notag \\
& \ \ \ \ \ \left. + \ \frac{\eta^{\nu \la} }{E_{23} } \; ( \si^{\mu } \, \bar{\si}^{\rho} \, \vep)_{\al \be} \, E_{25} \, E_{36}  \, \tspin \left( \smallint_{z_3}^{z_2} \ve{\om} \, + \, \tfrac{1}{2} \smallint_{z_6}^{z_5} \ve{\om} \right)  \right. \notag \\
& \ \ \ \ \ \ \ \ \ \ \ \ \ \  \frac{E_{56}}{2 \,  \tspin \left( \tfrac{1}{2} \smallint_{z_6}^{z_5} \ve{\om} \right) } \;  \tspin \left( \tfrac{1}{2} \smallint_{z_1}^{z_5} \ve{\om} \, + \, \tfrac{1}{2} \smallint_{z_1}^{z_6} \ve{\om} \right) \, \tspin \left( \tfrac{1}{2} \smallint_{z_4}^{z_5} \ve{\om} \, + \, \tfrac{1}{2} \smallint_{z_4}^{z_6} \ve{\om} \right)   \notag \\
& \ \ \ \ \ \left. - \ \frac{\eta^{\nu \rho} }{E_{24} } \; ( \si^{\mu } \, \bar{\si}^{\la} \, \vep)_{\al \be} \, E_{25} \, E_{46}  \, \tspin \left( \smallint_{z_4}^{z_2} \ve{\om} \, + \, \tfrac{1}{2} \smallint_{z_6}^{z_5} \ve{\om} \right)  \right. \notag \\
& \ \ \ \ \ \ \ \ \ \ \ \ \ \  \frac{E_{56}}{2 \, \tspin \left( \tfrac{1}{2} \smallint_{z_6}^{z_5} \ve{\om} \right) } \;  \tspin \left( \tfrac{1}{2} \smallint_{z_1}^{z_5} \ve{\om} \, + \, \tfrac{1}{2} \smallint_{z_1}^{z_6} \ve{\om} \right) \, \tspin \left( \tfrac{1}{2} \smallint_{z_3}^{z_5} \ve{\om} \, + \, \tfrac{1}{2} \smallint_{z_3}^{z_6} \ve{\om} \right)   \notag \\
&\ \ \ \ \ \left. + \ \frac{\eta^{\la \rho} }{E_{34} } \; ( \si^{\mu } \, \bar{\si}^{\nu} \, \vep)_{\al \be} \, E_{35} \, E_{46}  \, \tspin \left( \smallint_{z_4}^{z_3} \ve{\om} \, + \, \tfrac{1}{2} \smallint_{z_6}^{z_5} \ve{\om} \right)  \right. \notag \\
& \ \ \ \ \ \ \ \ \ \ \ \ \ \  \frac{E_{56}}{2 \,  \tspin \left( \tfrac{1}{2} \smallint_{z_6}^{z_5} \ve{\om} \right)} \; \tspin \left( \tfrac{1}{2} \smallint_{z_1}^{z_5} \ve{\om} \, + \, \tfrac{1}{2} \smallint_{z_1}^{z_6} \ve{\om} \right) \, \tspin \left( \tfrac{1}{2} \smallint_{z_2}^{z_5} \ve{\om} \, + \, \tfrac{1}{2} \smallint_{z_2}^{z_6} \ve{\om} \right)  \notag \\
& \ \ \ \ \ \left. + \  ( \si^\mu \, \bar{\si}^{\nu} \, \si^{\la } \, \bar{\si}^{\rho} \, \vep)_{\al \be} \; \left( \frac{E_{56}}{2 \, \tspin \left( \tfrac{1}{2} \smallint_{z_6}^{z_5} \ve{\om} \right)} \right)^2 \; \tspin \left( \tfrac{1}{2} \smallint_{z_1}^{z_5} \ve{\om} \, + \, \tfrac{1}{2} \smallint_{z_1}^{z_6} \ve{\om} \right) \right. \notag \\
& \ \ \ \ \ \ \ \ \ \ \ \ \ \ \left. \, \tspin \left( \tfrac{1}{2} \smallint_{z_2}^{z_5} \ve{\om} \, + \, \tfrac{1}{2} \smallint_{z_2}^{z_6} \ve{\om} \right) \, \tspin \left( \tfrac{1}{2} \smallint_{z_3}^{z_5} \ve{\om} \, + \, \tfrac{1}{2} \smallint_{z_3}^{z_6} \ve{\om} \right) \, \tspin \left( \tfrac{1}{2} \smallint_{z_4}^{z_5} \ve{\om} \, + \, \tfrac{1}{2} \smallint_{z_4}^{z_6} \ve{\om} \right)  \right\} \label{6pt2}
\end{align}
A similar procedure can be applied to $\langle \psi^\mu \psi^\nu \psi^\la \psi^\rho \psi^\tau S_\al S_{\dbe} \rangle \spin$: Take the 25 dimensional basis of Lorentz tensors to consist of $\eta^{\mu \nu} \eta^{\la \rho} \si^\tau$, $\eta^{\mu \nu} \si^{[\la} \bar{\si}^\rho \si^{\tau]}$ and permutations, then the coefficient of $\eta^{\mu \nu} \eta^{\la \rho} \si^\tau$ follows from (\ref{at2}) via $(\mu,\nu,\la,\rho,\tau) = (0,0,1,1,3)$ and $\al = \dbe = 1$. Similarly, the $\eta^{\mu \nu} \si^{[\la} \bar{\si}^\rho \si^{\tau]}$ coefficient arises from $(\mu,\nu,\la,\rho,\tau) = (0,0,1,2,3)$ and $\al = \dbe = 1$. In both cases, it is helpful to use the identity
\begin{align}
&E_{15}^2 \, E_{26} \, E_{27} \, \tspin \left( \smallint^{p}_{z_1} \ve{\om} \, + \, \smallint^{z_2}_{p} \ve{\om} \, + \, \smallint^{p}_{z_5} \ve{\om} \, + \, \tfrac{1}{2} \smallint^{z_6}_{p} \ve{\om} \, + \, \tfrac{1}{2} \smallint^{z_7}_{p} \ve{\om} \right) \notag \\
+ \ &E_{25}^2 \, E_{16} \, E_{17} \, \tspin \left( \smallint^{z_1}_p \ve{\om} \, + \, \smallint_{z_2}^{p} \ve{\om} \, + \, \smallint^{p}_{z_5} \ve{\om} \, + \, \tfrac{1}{2} \smallint^{z_6}_{p} \ve{\om} \, + \, \tfrac{1}{2} \smallint^{z_7}_{p} \ve{\om} \right) \notag \\
- \ &E_{12}^2 \, E_{56} \, E_{57} \, \tspin \left( \smallint^{p}_{z_1} \ve{\om} \, + \, \smallint_{z_2}^{p} \ve{\om} \, + \, \smallint_{p}^{z_5} \ve{\om} \, + \, \tfrac{1}{2} \smallint^{z_6}_{p} \ve{\om} \, + \, \tfrac{1}{2} \smallint^{z_7}_{p} \ve{\om} \right) \notag \\
= \ \ &E_{15} \, E_{25} \, \frac{\tspin \left(\tfrac{1}{2} \smallint^{z_6}_{z_5} \ve{\om} \, + \, \tfrac{1}{2} \smallint^{z_7}_{z_5} \ve{\om} \right)}{\tspin \left( \tfrac{1}{2} \smallint^{z_6}_{z_7} \ve{\om} \right)} \notag \\
& \ \ \ \ \times \ \left\{ E_{16} \, E_{27} \, \tspin \left(  \smallint^{z_1}_{z_2} \ve{\om} \, + \, \tfrac{1}{2} \smallint^{z_6}_{z_7} \ve{\om} \right) \ + \ E_{17} \, E_{26} \, \tspin \left( -\smallint^{z_1}_{z_2} \ve{\om} \, + \, \tfrac{1}{2} \smallint^{z_6}_{z_7} \ve{\om} \right) \right\}
\end{align}
which is already necessary to demonstrate consistency of (\ref{6pt2}) with all choices of the indices. Putting everything together, we find
\begin{align}
&\langle \psi^{\mu}(z_{1}) \, \psi^{\nu} (z_{2}) \, \psi^{\la}(z_{3}) \, \psi^{\rho}(z_{4}) \, \psi^{\tau}(z_{5}) \, S_{\al}(z_{6}) \, S_{\dbe}(z_{7}) \rangle \spin \notag \\
&= \ \ \frac{1}{4 \, (E_{16} \, E_{17} \, E_{26} \, E_{27} \, E_{36} \, E_{37} \, 
E_{46} \, E_{47} \, E_{56} \, E_{57} )^{1/2} \, \tspin \left( \tfrac{1}{2} \smallint_{z_7}^{z_6} \ve{\om} \right) \,  \tspin ( \ve{0} ) \, \tspin ( \ve{0} )} \notag \\
&\times \, \biggl\{ + \ \si^{\mu}_{\al \dbe} \, \tspin \left( \tfrac{1}{2} \smallint_{z_1}^{z_6} \ve{\om} \, + \, \tfrac{1}{2} \smallint_{z_1}^{z_7} \ve{\om} \right) \; \frac{\eta^{\nu \la} \, \eta^{\rho \tau}}{E_{23} \, E_{45}} \biggr. \notag \\
& \ \ \ \ \ \ \ \ \ \ \ \left[ E_{26} \, E_{37} \,  \tspin \left( \smallint_{z_3}^{z_2} \ve{\om} \, + \, \tfrac{1}{2} \smallint_{z_7}^{z_6} \ve{\om} \right) \ + \ E_{27} \, E_{36} \,  \tspin \left(- \, \smallint_{z_3}^{z_2} \ve{\om} \, + \, \tfrac{1}{2} \smallint_{z_7}^{z_6} \ve{\om} \right) \right] \notag \\
& \ \ \ \ \ \ \ \ \ \ \ \left[ E_{46} \, E_{57} \, \tspin \left( \smallint_{z_5}^{z_4} \ve{\om} \, + \, \tfrac{1}{2} \smallint_{z_7}^{z_6} \ve{\om} \right) \ + \ E_{47} \, E_{56} \, \tspin \left( - \, \smallint_{z_5}^{z_4} \ve{\om} \, + \, \tfrac{1}{2} \smallint_{z_7}^{z_6} \ve{\om} \right) \right]  \notag \\
&\ \ \ \ \ \left. \pm \ 14 \ \te{permutations} \right. \notag \\
&\ \ \ \ \ \left. + \  (\si^{[\la} \, \bar{\si}^{\rho} \, \si^{\tau]})_{\al \dbe}   \, \tspin \left( \tfrac{1}{2} \smallint_{z_3}^{z_6} \ve{\om} \, + \, \tfrac{1}{2} \smallint_{z_3}^{z_7} \ve{\om} \right) \, \tspin \left( \tfrac{1}{2} \smallint_{z_4}^{z_6} \ve{\om} \, + \, \tfrac{1}{2} \smallint_{z_4}^{z_7} \ve{\om} \right) \, \tspin \left( \tfrac{1}{2} \smallint_{z_5}^{z_6} \ve{\om} \, + \, \tfrac{1}{2} \smallint_{z_5}^{z_7} \ve{\om} \right) \right. \notag \\
& \ \ \ \ \ \ \ \ \ \ \ \frac{E_{67}}{ \tspin \left( \tfrac{1}{2} \smallint_{z_7}^{z_6} \ve{\om} \right) } \; \frac{\eta^{\mu \nu}}{E_{12}} \; \left[ E_{16} \, E_{27} \, \tspin \left( \smallint_{z_2}^{z_1} \ve{\om} \, + \, \tfrac{1}{2} \smallint_{z_7}^{z_6} \ve{\om} \right) \ + \ E_{17} \, E_{26} \, \tspin \left( - \, \smallint_{z_2}^{z_1} \ve{\om} \, + \, \tfrac{1}{2} \smallint_{z_7}^{z_6} \ve{\om} \right) \right] \notag \\
&\ \ \ \ \ \biggl. \pm \ 9 \ \te{permutations} \biggr\} \ .
\label{7pt2a}
\end{align}
Conversions of type $\si^{[\la} \bar{\si}^\rho \si^{\tau]} = \si^{\la} \bar{\si}^\rho \si^{\tau} + \eta^{\la \rho} \si^\tau - \eta^{\la \tau} \si^{\rho} + \eta^{\rho \tau} \si^\la$ allow to recast the lengthy result (\ref{7pt2a}) in a shorter form in terms of ordered $\si$ matrix products
\begin{align}
&\langle \psi^{\mu}(z_{1}) \, \psi^{\nu} (z_{2}) \, \psi^{\la}(z_{3}) \, \psi^{\rho}(z_{4}) \, \psi^{\tau}(z_{5}) \, S_{\al}(z_{6}) \, S_{\dbe}(z_{7}) \rangle \spin \notag \\
&= \ \ \frac{1}{ (E_{16} \, E_{17} \, E_{26} \, E_{27} \, E_{36} \, E_{37} \, 
E_{46} \, E_{47} \, E_{56} \, E_{57} )^{1/2} \, \tspin \left( \tfrac{1}{2} \smallint_{z_7}^{z_6} \ve{\om} \right) \,  \tspin ( \ve{0} ) \, \tspin ( \ve{0} )} \notag \\
&\times \, \left\{ + \ \si^{\mu}_{\al \dbe} \, \tspin \left( \tfrac{1}{2} \smallint_{z_1}^{z_6} \ve{\om} \, + \, \tfrac{1}{2} \smallint_{z_1}^{z_7} \ve{\om} \right) \right. \notag \\
& \ \ \ \ \ \ \ \ \ \ \ \, \left[ \frac{\eta^{\nu \la} \, \eta^{\rho \tau}}{E_{23} \, E_{45}} \; E_{26} \, E_{37} \, E_{46} \, E_{57} \, \tspin \left( \smallint_{z_3}^{z_2} \ve{\om} \, + \, \tfrac{1}{2} \smallint_{z_7}^{z_6} \ve{\om} \right) \, \tspin \left( \smallint_{z_5}^{z_4} \ve{\om} \, + \, \tfrac{1}{2} \smallint_{z_7}^{z_6} \ve{\om} \right)  \right. \notag \\
& \ \ \ \ \ \ \ \ \ \  \left. - \; \frac{  \eta^{\nu \rho} \, \eta^{\la \tau} }{ E_{24} \, E_{35} } \; E_{26} \, E_{47} \, E_{36} \, E_{57} \, \tspin \left( \smallint_{z_4}^{z_2} \ve{\om} \, + \, \tfrac{1}{2} \smallint_{z_7}^{z_6} \ve{\om} \right) \, \tspin \left( \smallint_{z_5}^{z_3} \ve{\om} \, + \, \tfrac{1}{2} \smallint_{z_7}^{z_6} \ve{\om} \right) \right.   \notag \\
& \ \ \ \ \ \ \ \ \ \  \left. + \; \frac{  \eta^{\nu \tau} \, \eta^{\la \rho} }{ E_{25} \, E_{34} } \; E_{26} \, E_{57} \, E_{36} \, E_{47} \, \tspin \left( \smallint_{z_5}^{z_2} \ve{\om} \, + \, \tfrac{1}{2} \smallint_{z_7}^{z_6} \ve{\om} \right) \, \tspin \left( \smallint_{z_4}^{z_3} \ve{\om} \, + \, \tfrac{1}{2} \smallint_{z_7}^{z_6} \ve{\om} \right) \right]  \notag \\
&\ \ \ \ \ \left. - \ \si^{\nu}_{\al \dbe} \, \tspin \left( \tfrac{1}{2} \smallint_{z_2}^{z_6} \ve{\om} \, + \, \tfrac{1}{2} \smallint_{z_2}^{z_7} \ve{\om} \right) \right. \notag \\
& \ \ \ \ \ \ \ \ \ \ \ \, \left[ \frac{\eta^{\mu \la} \, \eta^{\rho \tau}}{E_{13} \, E_{45}} \; E_{16} \, E_{37} \, E_{46} \, E_{57} \, \tspin \left( \smallint_{z_3}^{z_1} \ve{\om} \, + \, \tfrac{1}{2} \smallint_{z_7}^{z_6} \ve{\om} \right) \, \tspin \left( \smallint_{z_5}^{z_4} \ve{\om} \, + \, \tfrac{1}{2} \smallint_{z_7}^{z_6} \ve{\om} \right)  \right. \notag \\
& \ \ \ \ \ \ \ \ \ \  \left. - \; \frac{  \eta^{\mu \rho} \, \eta^{\la \tau} }{ E_{14} \, E_{35} } \; E_{16} \, E_{47} \, E_{36} \, E_{57} \, \tspin \left( \smallint_{z_4}^{z_1} \ve{\om} \, + \, \tfrac{1}{2} \smallint_{z_7}^{z_6} \ve{\om} \right) \, \tspin \left( \smallint_{z_5}^{z_3} \ve{\om} \, + \, \tfrac{1}{2} \smallint_{z_7}^{z_6} \ve{\om} \right) \right.   \notag \\
& \ \ \ \ \ \ \ \ \ \  \left. + \; \frac{  \eta^{\mu \tau} \, \eta^{\la \rho} }{ E_{15} \, E_{34} } \; E_{16} \, E_{57} \, E_{36} \, E_{47} \, \tspin \left( \smallint_{z_5}^{z_1} \ve{\om} \, + \, \tfrac{1}{2} \smallint_{z_7}^{z_6} \ve{\om} \right) \, \tspin \left( \smallint_{z_4}^{z_3} \ve{\om} \, + \, \tfrac{1}{2} \smallint_{z_7}^{z_6} \ve{\om} \right) \right]  \notag \\
&\ \ \ \ \ \left. + \ \si^{\la}_{\al \dbe} \, \tspin \left( \tfrac{1}{2} \smallint_{z_3}^{z_6} \ve{\om} \, + \, \tfrac{1}{2} \smallint_{z_3}^{z_7} \ve{\om} \right) \right. \notag \\
& \ \ \ \ \ \ \ \ \ \ \ \, \left[ \frac{\eta^{\mu \nu} \, \eta^{\rho \tau}}{E_{12} \, E_{45}} \; E_{16} \, E_{27} \, E_{46} \, E_{57} \, \tspin \left( \smallint_{z_2}^{z_1} \ve{\om} \, + \, \tfrac{1}{2} \smallint_{z_7}^{z_6} \ve{\om} \right) \, \tspin \left( \smallint_{z_5}^{z_4} \ve{\om} \, + \, \tfrac{1}{2} \smallint_{z_7}^{z_6} \ve{\om} \right)  \right. \notag \\
& \ \ \ \ \ \ \ \ \ \  \left. - \; \frac{  \eta^{\mu \rho} \, \eta^{\nu \tau} }{ E_{14} \, E_{25} } \; E_{16} \, E_{47} \, E_{26} \, E_{57} \, \tspin \left( \smallint_{z_4}^{z_1} \ve{\om} \, + \, \tfrac{1}{2} \smallint_{z_7}^{z_6} \ve{\om} \right) \, \tspin \left( \smallint_{z_5}^{z_2} \ve{\om} \, + \, \tfrac{1}{2} \smallint_{z_7}^{z_6} \ve{\om} \right) \right.   \notag \\
& \ \ \ \ \ \ \ \ \ \  \left. + \; \frac{  \eta^{\mu \tau} \, \eta^{\nu \rho} }{ E_{15} \, E_{24} } \; E_{16} \, E_{57} \, E_{26} \, E_{47} \, \tspin \left( \smallint_{z_5}^{z_1} \ve{\om} \, + \, \tfrac{1}{2} \smallint_{z_7}^{z_6} \ve{\om} \right) \, \tspin \left( \smallint_{z_4}^{z_2} \ve{\om} \, + \, \tfrac{1}{2} \smallint_{z_7}^{z_6} \ve{\om} \right) \right]  \notag \\
&\ \ \ \ \ \left. - \ \si^{\rho}_{\al \dbe} \, \tspin \left( \tfrac{1}{2} \smallint_{z_4}^{z_6} \ve{\om} \, + \, \tfrac{1}{2} \smallint_{z_4}^{z_7} \ve{\om} \right) \right. \notag \\
& \ \ \ \ \ \ \ \ \ \ \ \, \left[ \frac{\eta^{\mu \nu} \, \eta^{\la \tau}}{E_{12} \, E_{35}} \; E_{16} \, E_{27} \, E_{36} \, E_{57} \, \tspin \left( \smallint_{z_2}^{z_1} \ve{\om} \, + \, \tfrac{1}{2} \smallint_{z_7}^{z_6} \ve{\om} \right) \, \tspin \left( \smallint_{z_5}^{z_3} \ve{\om} \, + \, \tfrac{1}{2} \smallint_{z_7}^{z_6} \ve{\om} \right)  \right. \notag \\
& \ \ \ \ \ \ \ \ \ \  \left. - \; \frac{  \eta^{\mu \la} \, \eta^{\nu \tau} }{ E_{13} \, E_{25} } \; E_{16} \, E_{37} \, E_{26} \, E_{57} \, \tspin \left( \smallint_{z_3}^{z_1} \ve{\om} \, + \, \tfrac{1}{2} \smallint_{z_7}^{z_6} \ve{\om} \right) \, \tspin \left( \smallint_{z_5}^{z_2} \ve{\om} \, + \, \tfrac{1}{2} \smallint_{z_7}^{z_6} \ve{\om} \right) \right.   \notag \\
& \ \ \ \ \ \ \ \ \ \  \left. + \; \frac{  \eta^{\mu \tau} \, \eta^{\nu \la} }{ E_{15} \, E_{23} } \; E_{16} \, E_{57} \, E_{26} \, E_{37} \, \tspin \left( \smallint_{z_5}^{z_1} \ve{\om} \, + \, \tfrac{1}{2} \smallint_{z_7}^{z_6} \ve{\om} \right) \, \tspin \left( \smallint_{z_3}^{z_2} \ve{\om} \, + \, \tfrac{1}{2} \smallint_{z_7}^{z_6} \ve{\om} \right) \right]  \notag \\
&\ \ \ \ \ \left. + \ \si^{\tau}_{\al \dbe} \, \tspin \left( \tfrac{1}{2} \smallint_{z_5}^{z_6} \ve{\om} \, + \, \tfrac{1}{2} \smallint_{z_5}^{z_7} \ve{\om} \right) \right. \notag \\
& \ \ \ \ \ \ \ \ \ \ \ \, \left[ \frac{\eta^{\mu \nu} \, \eta^{\la \rho}}{E_{12} \, E_{34}} \; E_{16} \, E_{27} \, E_{36} \, E_{47} \, \tspin \left( \smallint_{z_2}^{z_1} \ve{\om} \, + \, \tfrac{1}{2} \smallint_{z_7}^{z_6} \ve{\om} \right) \, \tspin \left( \smallint_{z_4}^{z_3} \ve{\om} \, + \, \tfrac{1}{2} \smallint_{z_7}^{z_6} \ve{\om} \right)  \right. \notag \\
& \ \ \ \ \ \ \ \ \ \  \left. - \; \frac{  \eta^{\mu \la} \, \eta^{\nu \rho} }{ E_{13} \, E_{24} } \; E_{16} \, E_{37} \, E_{26} \, E_{47} \, \tspin \left( \smallint_{z_3}^{z_1} \ve{\om} \, + \, \tfrac{1}{2} \smallint_{z_7}^{z_6} \ve{\om} \right) \, \tspin \left( \smallint_{z_4}^{z_2} \ve{\om} \, + \, \tfrac{1}{2} \smallint_{z_7}^{z_6} \ve{\om} \right) \right.   \notag \\
& \ \ \ \ \ \ \ \ \ \  \left. + \; \frac{  \eta^{\mu \rho} \, \eta^{\nu \la} }{ E_{14} \, E_{23} } \; E_{16} \, E_{47} \, E_{26} \, E_{37} \, \tspin \left( \smallint_{z_4}^{z_1} \ve{\om} \, + \, \tfrac{1}{2} \smallint_{z_7}^{z_6} \ve{\om} \right) \, \tspin \left( \smallint_{z_3}^{z_2} \ve{\om} \, + \, \tfrac{1}{2} \smallint_{z_7}^{z_6} \ve{\om} \right) \right]  \notag \\
&\ \ \ \ \ \left. + \ \frac{\eta^{\mu \nu}}{E_{12}} \; (\si^{\la} \, \bar{\si}^{\rho} \, \si^{\tau})_{\al \dbe} \ E_{16} \, E_{27} \, \tspin \left( \smallint_{z_2}^{z_1} \ve{\om} \, + \, \tfrac{1}{2} \smallint_{z_7}^{z_6} \ve{\om} \right) \, \frac{E_{67}}{2 \, \tspin \left( \tfrac{1}{2} \smallint_{z_7}^{z_6} \ve{\om} \right) } \right. \notag \\
& \ \ \ \ \ \ \ \ \ \ \ \ \  \tspin \left( \tfrac{1}{2} \smallint_{z_3}^{z_6} \ve{\om} \, + \, \tfrac{1}{2} \smallint_{z_3}^{z_7} \ve{\om} \right) \, \tspin \left( \tfrac{1}{2} \smallint_{z_4}^{z_6} \ve{\om} \, + \, \tfrac{1}{2} \smallint_{z_4}^{z_7} \ve{\om} \right) \, \tspin \left( \tfrac{1}{2} \smallint_{z_5}^{z_6} \ve{\om} \, + \, \tfrac{1}{2} \smallint_{z_5}^{z_7} \ve{\om} \right) \notag \\
&\ \ \ \ \ \left. - \ \frac{\eta^{\mu \la}}{E_{13}} \; (\si^{\nu} \, \bar{\si}^{\rho} \, \si^{\tau})_{\al \dbe} \ E_{16} \, E_{37} \, \tspin \left( \smallint_{z_3}^{z_1} \ve{\om} \, + \, \tfrac{1}{2} \smallint_{z_7}^{z_6} \ve{\om} \right) \, \frac{E_{67}}{2 \, \tspin \left( \tfrac{1}{2} \smallint_{z_7}^{z_6} \ve{\om} \right) } \right. \notag \\
& \ \ \ \ \ \ \ \ \ \ \ \ \  \tspin \left( \tfrac{1}{2} \smallint_{z_2}^{z_6} \ve{\om} \, + \, \tfrac{1}{2} \smallint_{z_2}^{z_7} \ve{\om} \right) \, \tspin \left( \tfrac{1}{2} \smallint_{z_4}^{z_6} \ve{\om} \, + \, \tfrac{1}{2} \smallint_{z_4}^{z_7} \ve{\om} \right) \, \tspin \left( \tfrac{1}{2} \smallint_{z_5}^{z_6} \ve{\om} \, + \, \tfrac{1}{2} \smallint_{z_5}^{z_7} \ve{\om} \right) \notag \\
&\ \ \ \ \ \left. + \ \frac{\eta^{\mu \rho}}{E_{14}} \; (\si^{\nu} \, \bar{\si}^{\la} \, \si^{\tau})_{\al \dbe} \ E_{16} \, E_{47} \, \tspin \left( \smallint_{z_4}^{z_1} \ve{\om} \, + \, \tfrac{1}{2} \smallint_{z_7}^{z_6} \ve{\om} \right) \, \frac{E_{67}}{2 \, \tspin \left( \tfrac{1}{2} \smallint_{z_7}^{z_6} \ve{\om} \right) } \right. \notag \\
& \ \ \ \ \ \ \ \ \ \ \ \ \  \tspin \left( \tfrac{1}{2} \smallint_{z_2}^{z_6} \ve{\om} \, + \, \tfrac{1}{2} \smallint_{z_2}^{z_7} \ve{\om} \right) \, \tspin \left( \tfrac{1}{2} \smallint_{z_3}^{z_6} \ve{\om} \, + \, \tfrac{1}{2} \smallint_{z_3}^{z_7} \ve{\om} \right) \, \tspin \left( \tfrac{1}{2} \smallint_{z_5}^{z_6} \ve{\om} \, + \, \tfrac{1}{2} \smallint_{z_5}^{z_7} \ve{\om} \right) \notag \\
&\ \ \ \ \ \left. - \ \frac{\eta^{\mu \tau}}{E_{15}} \; (\si^{\nu} \, \bar{\si}^{\la} \, \si^{\rho})_{\al \dbe} \ E_{16} \, E_{57} \, \tspin \left( \smallint_{z_5}^{z_1} \ve{\om} \, + \, \tfrac{1}{2} \smallint_{z_7}^{z_6} \ve{\om} \right) \, \frac{E_{67}}{2 \, \tspin \left( \tfrac{1}{2} \smallint_{z_7}^{z_6} \ve{\om} \right) } \right. \notag \\
& \ \ \ \ \ \ \ \ \ \ \ \ \  \tspin \left( \tfrac{1}{2} \smallint_{z_2}^{z_6} \ve{\om} \, + \, \tfrac{1}{2} \smallint_{z_2}^{z_7} \ve{\om} \right) \, \tspin \left( \tfrac{1}{2} \smallint_{z_3}^{z_6} \ve{\om} \, + \, \tfrac{1}{2} \smallint_{z_3}^{z_7} \ve{\om} \right) \, \tspin \left( \tfrac{1}{2} \smallint_{z_4}^{z_6} \ve{\om} \, + \, \tfrac{1}{2} \smallint_{z_4}^{z_7} \ve{\om} \right) \notag \\
&\ \ \ \ \ \left. + \ \frac{\eta^{\nu \la}}{E_{23}} \; (\si^{\mu} \, \bar{\si}^{\rho} \, \si^{\tau})_{\al \dbe} \ E_{26} \, E_{37} \, \tspin \left( \smallint_{z_3}^{z_2} \ve{\om} \, + \, \tfrac{1}{2} \smallint_{z_7}^{z_6} \ve{\om} \right) \, \frac{E_{67}}{2 \, \tspin \left( \tfrac{1}{2} \smallint_{z_7}^{z_6} \ve{\om} \right) } \right. \notag \\
& \ \ \ \ \ \ \ \ \ \ \ \ \  \tspin \left( \tfrac{1}{2} \smallint_{z_1}^{z_6} \ve{\om} \, + \, \tfrac{1}{2} \smallint_{z_1}^{z_7} \ve{\om} \right) \, \tspin \left( \tfrac{1}{2} \smallint_{z_4}^{z_6} \ve{\om} \, + \, \tfrac{1}{2} \smallint_{z_4}^{z_7} \ve{\om} \right) \, \tspin \left( \tfrac{1}{2} \smallint_{z_5}^{z_6} \ve{\om} \, + \, \tfrac{1}{2} \smallint_{z_5}^{z_7} \ve{\om} \right) \notag \\
&\ \ \ \ \ \left. - \ \frac{\eta^{\nu \rho}}{E_{24}} \; (\si^{\mu} \, \bar{\si}^{\la} \, \si^{\tau})_{\al \dbe} \ E_{26} \, E_{47} \, \tspin \left( \smallint_{z_4}^{z_2} \ve{\om} \, + \, \tfrac{1}{2} \smallint_{z_7}^{z_6} \ve{\om} \right) \, \frac{E_{67}}{2 \, \tspin \left( \tfrac{1}{2} \smallint_{z_7}^{z_6} \ve{\om} \right) } \right. \notag \\
& \ \ \ \ \ \ \ \ \ \ \ \ \  \tspin \left( \tfrac{1}{2} \smallint_{z_1}^{z_6} \ve{\om} \, + \, \tfrac{1}{2} \smallint_{z_1}^{z_7} \ve{\om} \right) \, \tspin \left( \tfrac{1}{2} \smallint_{z_3}^{z_6} \ve{\om} \, + \, \tfrac{1}{2} \smallint_{z_3}^{z_7} \ve{\om} \right) \, \tspin \left( \tfrac{1}{2} \smallint_{z_5}^{z_6} \ve{\om} \, + \, \tfrac{1}{2} \smallint_{z_5}^{z_7} \ve{\om} \right) \notag \\
&\ \ \ \ \ \left. + \ \frac{\eta^{\nu \tau}}{E_{25}} \; (\si^{\mu} \, \bar{\si}^{\la} \, \si^{\rho})_{\al \dbe} \ E_{26} \, E_{57} \, \tspin \left( \smallint_{z_5}^{z_2} \ve{\om} \, + \, \tfrac{1}{2} \smallint_{z_7}^{z_6} \ve{\om} \right) \, \frac{E_{67}}{2 \, \tspin \left( \tfrac{1}{2} \smallint_{z_7}^{z_6} \ve{\om} \right) } \right. \notag \\
& \ \ \ \ \ \ \ \ \ \ \ \ \  \tspin \left( \tfrac{1}{2} \smallint_{z_1}^{z_6} \ve{\om} \, + \, \tfrac{1}{2} \smallint_{z_1}^{z_7} \ve{\om} \right) \, \tspin \left( \tfrac{1}{2} \smallint_{z_3}^{z_6} \ve{\om} \, + \, \tfrac{1}{2} \smallint_{z_3}^{z_7} \ve{\om} \right) \, \tspin \left( \tfrac{1}{2} \smallint_{z_4}^{z_6} \ve{\om} \, + \, \tfrac{1}{2} \smallint_{z_4}^{z_7} \ve{\om} \right) \notag \\
&\ \ \ \ \ \left. + \ \frac{\eta^{\la \rho}}{E_{34}} \; (\si^{\mu} \, \bar{\si}^{\nu} \, \si^{\tau})_{\al \dbe} \ E_{36} \, E_{47} \, \tspin \left( \smallint_{z_4}^{z_3} \ve{\om} \, + \, \tfrac{1}{2} \smallint_{z_7}^{z_6} \ve{\om} \right) \, \frac{E_{67}}{2 \, \tspin \left( \tfrac{1}{2} \smallint_{z_7}^{z_6} \ve{\om} \right) } \right. \notag \\
& \ \ \ \ \ \ \ \ \ \ \ \ \  \tspin \left( \tfrac{1}{2} \smallint_{z_1}^{z_6} \ve{\om} \, + \, \tfrac{1}{2} \smallint_{z_1}^{z_7} \ve{\om} \right) \, \tspin \left( \tfrac{1}{2} \smallint_{z_2}^{z_6} \ve{\om} \, + \, \tfrac{1}{2} \smallint_{z_2}^{z_7} \ve{\om} \right) \, \tspin \left( \tfrac{1}{2} \smallint_{z_5}^{z_6} \ve{\om} \, + \, \tfrac{1}{2} \smallint_{z_5}^{z_7} \ve{\om} \right) \notag \\
&\ \ \ \ \ \left. - \ \frac{\eta^{\la \tau}}{E_{35}} \; (\si^{\mu} \, \bar{\si}^{\nu} \, \si^{\rho})_{\al \dbe} \ E_{36} \, E_{57} \, \tspin \left( \smallint_{z_5}^{z_3} \ve{\om} \, + \, \tfrac{1}{2} \smallint_{z_7}^{z_6} \ve{\om} \right) \, \frac{E_{67}}{2 \, \tspin \left( \tfrac{1}{2} \smallint_{z_7}^{z_6} \ve{\om} \right) } \right. \notag \\
& \ \ \ \ \ \ \ \ \ \ \ \ \  \tspin \left( \tfrac{1}{2} \smallint_{z_1}^{z_6} \ve{\om} \, + \, \tfrac{1}{2} \smallint_{z_1}^{z_7} \ve{\om} \right) \, \tspin \left( \tfrac{1}{2} \smallint_{z_2}^{z_6} \ve{\om} \, + \, \tfrac{1}{2} \smallint_{z_2}^{z_7} \ve{\om} \right) \, \tspin \left( \tfrac{1}{2} \smallint_{z_4}^{z_6} \ve{\om} \, + \, \tfrac{1}{2} \smallint_{z_4}^{z_7} \ve{\om} \right) \notag \\
&\ \ \ \ \ \left. + \ \frac{\eta^{\rho \tau}}{E_{45}} \; (\si^{\mu} \, \bar{\si}^{\nu} \, \si^{\la})_{\al \dbe} \ E_{46} \, E_{57} \, \tspin \left( \smallint_{z_5}^{z_4} \ve{\om} \, + \, \tfrac{1}{2} \smallint_{z_7}^{z_6} \ve{\om} \right) \, \frac{E_{67}}{2 \, \tspin \left( \tfrac{1}{2} \smallint_{z_7}^{z_6} \ve{\om} \right) } \right. \notag \\
& \ \ \ \ \ \ \ \ \ \ \ \ \  \tspin \left( \tfrac{1}{2} \smallint_{z_1}^{z_6} \ve{\om} \, + \, \tfrac{1}{2} \smallint_{z_1}^{z_7} \ve{\om} \right) \, \tspin \left( \tfrac{1}{2} \smallint_{z_2}^{z_6} \ve{\om} \, + \, \tfrac{1}{2} \smallint_{z_2}^{z_7} \ve{\om} \right) \, \tspin \left( \tfrac{1}{2} \smallint_{z_3}^{z_6} \ve{\om} \, + \, \tfrac{1}{2} \smallint_{z_3}^{z_7} \ve{\om} \right) \notag \\
&\ \ \ \ \ \left. + \ (\si^{\mu} \, \bar{\si}^{\nu} \, \si^{\la} \, \bar{\si}^{\rho} \, \si^{\tau})_{\al \dbe} \; \left( \frac{E_{67}}{2 \, \tspin \left( \tfrac{1}{2} \smallint_{z_7}^{z_6} \ve{\om} \right) } \right)^2 \, \tspin \left( \tfrac{1}{2} \smallint_{z_1}^{z_6} \ve{\om} \, + \, \tfrac{1}{2} \smallint_{z_1}^{z_7} \ve{\om} \right) \, \tspin \left( \tfrac{1}{2} \smallint_{z_2}^{z_6} \ve{\om} \, + \, \tfrac{1}{2} \smallint_{z_2}^{z_7} \ve{\om} \right)  \right. \notag \\
& \ \ \ \ \ \ \ \ \ \ \ \ \  \left. \tspin \left( \tfrac{1}{2} \smallint_{z_3}^{z_6} \ve{\om} \, + \, \tfrac{1}{2} \smallint_{z_3}^{z_7} \ve{\om} \right) \, \tspin \left( \tfrac{1}{2} \smallint_{z_4}^{z_6} \ve{\om} \, + \, \tfrac{1}{2} \smallint_{z_4}^{z_7} \ve{\om} \right) \, \tspin \left( \tfrac{1}{2} \smallint_{z_5}^{z_6} \ve{\om} \, + \, \tfrac{1}{2} \smallint_{z_5}^{z_7} \ve{\om} \right) \right\}
\label{7pt2}
\end{align}
To arrive at this nice form of the seven point function, it was also necessary to add a zero of the form
\begin{align}
0 \ \ &= \ \ \si^{[\mu} \, \bar{\si}^{\nu} \, \si^{\la} \, \bar{\si}^{\rho} \, \si^{\tau]} \notag \\
&= \ \ \si^{\mu} \, \bar{\si}^{\nu} \, \si^{\la} \, \bar{\si}^{\rho} \, \si^{\tau} \ + \ \eta^{\mu \nu} \, \si^{\la} \, \bar{\si}^{\rho} \, \si^{\tau} \ - \ \eta^{\mu \la} \, \si^{\nu} \, \bar{\si}^{\rho} \, \si^{\tau} \ + \ \eta^{\mu \rho} \, \si^{\nu} \, \bar{\si}^{\la} \, \si^{\tau} \notag \\
& \ \ \ \ \ \ - \ \eta^{\mu \tau} \, \si^{\nu} \, \bar{\si}^{\la} \, \si^{\rho} \ + \ \eta^{\nu \la} \, \si^{\mu} \, \bar{\si}^{\rho} \, \si^{\tau} \ - \ \eta^{\nu \rho} \, \si^{\mu} \, \bar{\si}^{\la} \, \si^{\tau} \ + \ \eta^{\nu \tau} \, \si^{\mu} \, \bar{\si}^{\la} \, \si^{\rho} \notag \\
& \ \ \ \ \ \ + \ \eta^{\la \rho} \, \si^{\mu} \, \bar{\si}^{\nu} \, \si^{\tau} \ - \ \eta^{\la \tau} \, \si^{\mu} \, \bar{\si}^{\nu} \, \si^{\rho} \ + \ \eta^{\rho \tau} \, \si^{\mu} \, \bar{\si}^{\nu} \, \si^{\la} \notag \\
& \ \ \ \ \ \ + \ \si^{\mu} \, \bigl( \eta^{\nu \la} \, \eta^{\rho \tau} \ - \ \eta^{\nu \rho} \, \eta^{\la \tau} \ + \ \eta^{\nu \tau} \, \eta^{ \la \rho } \bigr)  \ - \ \si^{\nu} \, \bigl( \eta^{\mu \la} \, \eta^{\rho \tau} \ - \ \eta^{\mu \rho} \, \eta^{\la \tau} \ + \ \eta^{\mu \tau} \, \eta^{ \la \rho } \bigr) \notag \\
& \ \ \ \ \ \ + \ \si^{\la} \, \bigl( \eta^{\mu \nu} \, \eta^{\rho \tau} \ - \ \eta^{\mu \rho} \, \eta^{\nu \tau} \ + \ \eta^{\mu \tau} \, \eta^{ \nu \rho } \bigr) \  - \ \si^{\rho} \, \bigl( \eta^{\mu \nu} \, \eta^{\la \tau} \ - \ \eta^{\mu \la} \, \eta^{\nu \tau} \ + \ \eta^{\mu \tau} \, \eta^{ \nu \la } \bigr) \notag \\
& \ \ \ \ \ \ + \ \si^{\tau} \, \bigl( \eta^{\mu \nu} \, \eta^{\la \rho} \ - \ \eta^{\mu \la} \, \eta^{\nu \rho} \ + \ \eta^{\mu \rho} \, \eta^{ \nu \la} \bigr)\,.
\label{zero}
\end{align}
multiplied by a $z$ dependence which can be read off from the last two lines of (\ref{7pt2}).

\subsection{The generalization to higher number of fermions}
\label{sec:gen}

The correlators (\ref{3pt2}), (\ref{4pt2}), (\ref{5pt2}), (\ref{6pt2}) and (\ref{7pt2}) have striking similarities in their structure. Let us denote the arguments of the spin fields as $S_\al(z_{A})$, $S_{\dbe}(z_{B})$ and a generic NS field by $\psi^{\mu_{i}}(z_{i})$ in the following list of observations:
\begin{itemize}
\item Increasing numbers of $\si$ matrices are multiplied by increasing powers of $\frac{E_{AB}}{2 \tspin \left( \tfrac{1}{2} \smallint_{z_B}^{z_A} \ve{\om} \right)}$ (more precisely, each $\si^{\mu_{i}} \bar{\si}^{\mu_{j}}$ within the Lorentz tensor contributes one such factor).
\item The overall prefactor of all the tensor structures contains a $-1/2$ power of each $\psi \leftrightarrow S$ contraction, i.e. the correlators are proportional to $\prod_{i=1}^{N} (E_{iA} E_{iB})^{-1/2}$ where $N$ is the number of $\psi$'s involved.
\item Each Minkowski metric $\eta^{\mu_{i} \mu_{j}}$ with $i<j$ appears in combination with the $z$ dependence $\frac{E_{iA}E_{jB}}{E_{ij}}  \tspin \left( \smallint_{z_j}^{z_i} \ve{\om} \, + \, \tfrac{1}{2} \smallint_{z_B}^{z_A} \ve{\om} \right)$.
\item The sign of each term depends on the ordering of the Lorentz indices, whether they appear as an odd or an even permutation of ${\mu_{1} \mu_{2} ... \mu_{N}}$.
\end{itemize}
These properties lead us to claim analogous expressions for higher point correlators with larger numbers of $\psi^\mu$'s. Firstly, the $2n+1$ point functions with spin fields of opposite chirality reads as follows:

\bigskip
\framebox{\begin{minipage}{6.3in}
\begin{align}
& \ \ \Om_{(n)}^{\mu_{1} ... \mu_{2n-1}}\,_{\al \dbe}(z_{i})  \ \ := \ \
\langle \psi^{\mu_{1}}(z_{1}) \, \psi^{\mu_{2}}(z_{2}) \, ... \, \psi^{\mu_{2n-1}}(z_{2n-1}) \, S_{\al}(z_{A}) \, S_{\dbe}(z_{B}) \rangle \spin \notag \\
& \ \  = \ \ \frac{\left[ \tspin \left( \tfrac{1}{2} \smallint^{z_A} _{z_B} \ve{\om} \right) \right]^{2-n}}{\sqrt{2} \, \tspin ( \ve{0} ) \, \tspin ( \ve{0} )  \, \prod_{i=1}^{2n-1} (E_{iA} \, E_{iB})^{1/2} } \, \sum_{\ell = 0}^{n-1} \, \biggl( \frac{E_{AB}}{2 \, \tspin \left( \tfrac{1}{2} \smallint^{z_A} _{z_B} \ve{\om} \right)} \biggr)^{\ell} \notag \\
& \ \ \ \ \times \sum_{\rho \in S_{2n-1}/{\cal P}_{n,\ell}} \! \! \!  \te{sgn}(\rho) \, \bigl(\si^{\mu_{\rho(1)}} \, \bar{\si}^{\mu_{\rho(2)}} \, ... \, \bar{\si}^{\mu_{\rho(2\ell)}} \, \si^{\mu_{\rho(2\ell+1)}} \bigr)_{\al \dbe} \, \prod_{k=1}^{2\ell+1} \tspin \left( \tfrac{1}{2} \smallint^{z_A} _{z_{\rho(k)}} \ve{\om} \, + \, \tfrac{1}{2} \smallint^{z_B} _{z_{\rho(k)}} \ve{\om} \right) \notag \\
& \ \ \ \ \times \  \prod_{j=1}^{n-\ell-1} \frac{\eta^{\mu_{\rho(2\ell+2j)} \mu_{\rho(2\ell+2j+1)}}}{E_{\rho(2\ell+2j),\rho(2\ell+2j+1)} } \; E_{\rho(2\ell+2j),A} \, E_{\rho(2\ell+2j+1),B}  \, \tspin \left( \smallint^{z_{\rho(2\ell+2j)}}_{z_{\rho(2\ell+2j+1)}} \ve{\om} \, + \, \tfrac{1}{2} \smallint^{z_A} _{z_B} \ve{\om} \right)
\label{Omega}
\end{align}
\end{minipage}}

\bigskip
\noindent
Its relative with even number of NS fermions and two alike spin fields is given by
\bigskip

\framebox{\begin{minipage}{6.3in}
\begin{align}
& \ \ \om_{(n)}^{\mu_{1} ... \mu_{2n-2}}\,_{\al \be}(z_{i})  \ \ := \ \
\langle \psi^{\mu_{1}}(z_{1}) \, \psi^{\mu_{2}}(z_{2}) \, ... \, \psi^{\mu_{2n-2}}(z_{2n-2}) \, S_{\al}(z_{A}) \, S_{\be}(z_{B}) \rangle \spin \notag \\
& \ \  = \ \ \frac{- \, \left[ \tspin \left( \tfrac{1}{2} \smallint^{z_A} _{z_B} \ve{\om} \right) \right]^{3-n}}{ \tspin ( \ve{0} ) \, \tspin ( \ve{0} ) \, E_{AB}^{1/2}  \, \prod_{i=1}^{2n-2} (E_{iA} \, E_{iB})^{1/2} } \, \sum_{\ell = 0}^{n-1} \, \biggl( \frac{E_{AB}}{2 \, \tspin \left( \tfrac{1}{2} \smallint^{z_A} _{z_B} \ve{\om} \right)} \biggr)^{\ell} \notag \\
& \ \ \ \ \times \sum_{\rho \in S_{2n-2}/{\cal Q}_{n,\ell}} \! \! \!  \te{sgn}(\rho) \, \bigl(\si^{\mu_{\rho(1)}} \, \bar{\si}^{\mu_{\rho(2)}} \, ... \, \bar{\si}^{\mu_{\rho(2\ell)} } \, \vep \bigr)_{\al \be} \, \prod_{k=1}^{2\ell} \tspin \left( \tfrac{1}{2} \smallint^{z_A} _{z_{\rho(k)}} \ve{\om} \, + \, \tfrac{1}{2} \smallint^{z_B} _{z_{\rho(k)}} \ve{\om} \right) \notag \\
& \ \ \ \ \times \  \prod_{j=1}^{n-\ell-1} \frac{\eta^{\mu_{\rho(2\ell+2j-1)} \mu_{\rho(2\ell+2j)}}}{E_{\rho(2\ell+2j-1),\rho(2\ell+2j)} } \; E_{\rho(2\ell+2j-1),A} \, E_{\rho(2\ell+2j),B}  \, \tspin \left( \smallint^{z_{\rho(2\ell+2j-1)}}_{z_{\rho(2\ell+2j)}} \ve{\om} \, + \, \tfrac{1}{2} \smallint^{z_A} _{z_B} \ve{\om} \right) \ .
\label{omega}
\end{align}
\end{minipage}}

\bigskip
The proof of these central results is deferred to appendix \ref{sec:appD}.

\medskip
The summation ranges $\rho \in S_{2n-1}/{\cal P}_{n,\ell}$ and $\rho \in S_{2n-2}/{\cal Q}_{n,\ell}$ certainly require some explanation. The conventions are taken from \cite{tree} where a more exhaustive presentation can be found. Formally, we define
\begin{subequations}
\begin{align}
S_{2n-1}/ {\cal P}_{n,\ell} \ \ := \ \ \Bigl\{ &\rho \in S_{2n-1} : \ \rho(1) < \rho(2) < ... < \rho(2\ell + 1) \ , \Bigr. \notag \\
&\rho(2\ell + 2j) < \rho(2\ell + 2j + 1) \ \forall \ j = 1,2,...,n-\ell - 1 \ , \notag \\
\Bigl. &\rho(2\ell + 3) < \rho(2\ell + 5) < ... < \rho(2n -1) \Bigr\} 
\label{npt,4}\ , \\
S_{2n-2}/ {\cal Q}_{n,\ell} \ \ := \ \ \Bigl\{ &\rho \in S_{2n-2} : \ \rho(1) < \rho(2) < ... < \rho(2\ell) \ , \Bigr. \notag \\
&\rho(2\ell + 2j-1) < \rho(2\ell + 2j) \ \forall \ j = 1,2,...,n-\ell - 1 \ , \notag \\
\Bigl. &\rho(2\ell + 2) < \rho(2\ell + 4) < ... < \rho(2n -2) \Bigr\} \ .
\label{npt,5}
\end{align}
\end{subequations}
In other words, the sum runs over these permutations $\rho$ of $(1,2,...,2n-1)$ or $(1,2,...,2n-2)$ which satisfy the following constraints:
\begin{itemize}
\item Only ordered $\si$ products are summed over: The indices $\mu_{\rho(i)}$ attached to a chain of $\si$ matrices are increasingly ordered, e.g. whenever the product $\si^{\mu_{\rho(i)}} \bar{\si}^{\mu_{\rho(j)}} \si^{\mu_{\rho(k)}}$ appears, the subindices satisfy $\rho(i) < \rho(j) < \rho(k)$.
\item On each metric $\eta^{\mu_{\rho(i)} \mu_{\rho(j)}}$ the first index is the ``lower'' one, i.e.\ $\rho(i) < \rho(j)$.
\item Products of several $\eta$'s are not double counted. So once we get $\eta^{\mu_{\rho(i)} \mu_{\rho(j)}}
  \eta^{\mu_{\rho(k)} \mu_{\rho(l)}}$, the term $\eta^{\mu_{\rho(k)} \mu_{\rho(l)}} \eta^{\mu_{\rho(i)} \mu_{\rho(j)}}$
  does not appear.
\end{itemize}
These restrictions to the occurring $S_{2n-1}$ (or $S_{2n-2}$) elements are abbreviated by a quotient ${\cal  P}_{n,\ell}$ and ${\cal Q}_{n,\ell}$. The subgroups removed from $S_{2n-1}$ ($S_{2n-2}$) are $S_{2\ell+1} \times S_{n-\ell - 1} \times (S_{2})^{n-\ell-1}$ and $S_{2\ell} \times S_{n-\ell - 1} \times (S_{2})^{n-\ell-1}$ respectively, therefore the number of terms in (\ref{Omega}) and (\ref{omega}) at fixed $(n,\ell)$ is given by
\begin{subequations}
\begin{align}
\bigl| S_{2n-1} / {\cal P}_{n,\ell} \bigr| \ \ &= \ \ \frac{(2n\, - \, 1)!}{(2\ell \, + \, 1)! \, (n \, - \, \ell \, - \, 1)! \, 2^{n-\ell-1}}\ , \label{npt,7a} \\
\bigl| S_{2n-2} / {\cal Q}_{n,\ell} \bigr| \ \ &= \ \ \frac{(2n\, - \, 2)!}{(2\ell )! \, (n \, - \, \ell \, - \, 1)! \, 2^{n-\ell-1}}\ .
\label{npt,7}
\end{align}
\end{subequations}
Obviously, (\ref{npt,7a}) and (\ref{npt,7}) yield exactly the number of index structures we had in the known $n \leq 3$ correlators at each level of $\si$ chain length. However, this does not mean that we are using minimal sets. The spin field correlators discussed in section \ref{sec:cov} are excellent examples that reducing the $SO(1,3)$ tensors to a minimum might spoil certain symmetries or simplifications in the $z$ dependences.

\medskip
The same is true for the correlators (\ref{Omega}) and (\ref{omega}): Already for the seven point function (\ref{7pt2}), there is one identity (\ref{zero}) relating all the 26 appearing index structures. Eliminating one of the terms in (\ref{7pt2}) by means of (\ref{zero}) would certainly lead to a more complicated $z$ dependence.

\medskip
The overcounting of the basis of Lorentz tensors in (\ref{npt,7a}) and (\ref{npt,7}) grows with $n$. The expression for the eight point function $\om_{(4)}^{\mu_{1}...\mu_{6}} \, _{\al}\,^{\be}(z_{i})$ due to (\ref{omega}) contains 76 terms, but a group theoretic analysis determines the number of scalar representations in the tensor product to be 70. This difference is explained by the six independent identities $0 = \si^{[\mu} \bar{\si}^{ \nu} \si^{\la} \bar{\si}^{\rho} \si^{\tau} \bar{\si}^{\xi ]}$ and $0 = \eta^{\mu [\nu} \vep^{\la \rho \tau \xi]}$
\footnote{Another way to write these equations is: \beq \eta_{\mu \nu} \, \vep^{\la \rho \tau \zeta} \eq \de_{(\mu}^{\la} \, \vep_{\nu)}\, \! ^{\rho \tau \zeta} \ + \ \de_{(\mu}^{\rho} \, \vep^{\la } \, \! _{\nu)}\, \! ^{\tau \zeta} \ + \ \de_{(\mu}^{\tau} \, \vep^{\la \rho} \, \! _{\nu)}\, \! ^{\zeta} \ + \ \de_{(\mu}^{\zeta} \, \vep^{\la \rho \tau} \, \! _{\nu)}\,.  \label{dep} \eeq}. Similarly, for higher point examples $\Om_{(n \geq 4)}$ and $\om_{(n \geq 5)}$, one can find relations of comparable type due to the vanishing of antisymmetrized expressions in $\geq 5$ vector indices.


\subsection{Correlation functions with four spin fields}
\label{sec:4sp}

Correlators with four and more spin fields and three or more NS fermions are quite difficult to handle beyond tree level. With increasing number of fields, one is very soon faced with technical obstructions. The factorization of the fermions is of limited help because of the complicated six point spin field correlator (\ref{(6,0)}) and our incapability to find the right handed completion (\ref{(6,6)}). The method (iii) explained at the beginning of this section (and used before) also becomes awkward when the number of spinor indices and Fierz identities increases: It becomes more and more difficult to choose the indices such that few Lorentz tensors are nonzero.

\medskip
For these reasons, we will only display a few lower order examples here. With the correlators provided in the remainder of this section, it is possible for instance to compute a four fermion amplitude at one loop. Some simple higher order correlation functions such as $\langle \psi^\mu \psi^\nu \psi^\la S_\al S_\be S_\ga S_{\dde} \rangle \spin$ and $\langle \psi^\mu \psi^\nu \psi^\la \psi^\rho S_\al S_\be S_\ga S_\de \rangle \spin$ could be obtained from spin field correlators (\ref{(6,4)}) and the $M=4$ version of (\ref{(2M,4)}) by means of the factorization prescription.

\medskip
The easiest correlation function with four spin fields and at least one NS fermion can be easily derived from (\ref{(4,2)}),
\begin{align}
\langle &\psi^{\mu}(z_{1}) \, S_{\al}(z_{2}) \, S_{\be}(z_{3}) \, S_{\ga}(z_{4}) \, S_{\dde}(z_{5}) \rangle \spin \ \ = \ \ \frac{1}{\sqrt{2} \, (E_{12} \, E_{13} \, E_{14} \, E_{15} \, E_{23} \, E_{24} \, E_{34})^{1/2} \, \tspin ( \ve{0} ) \, \tspin ( \ve{0} )  } \notag \\
& \times \ \left\{ \si^{\mu}_{\ga \dde} \, \vep_{\al \be} \, E_{12} \, E_{34} \, \tspin \left( \tfrac{1}{2} \smallint^{z_1}_{z_3}\ve{\om} \, + \, \tfrac{1}{2} \smallint^{z_1}_{z_4}\ve{\om} \, + \,  \tfrac{1}{2} \smallint^{z_2}_{z_5}\ve{\om} \right) \, \tspin \left( + \, \tfrac{1}{2} \smallint^{z_2}_{z_4}\ve{\om} \, + \, \tfrac{1}{2} \smallint^{z_5}_{z_3}\ve{\om} \right) \right. \notag \\
& \ \ \ \ \ \ \ \left. + \, \si^{\mu}_{\al \dde} \, \vep_{\ga \be} \, E_{14} \, E_{23} \, \tspin \left( \tfrac{1}{2} \smallint^{z_1}_{z_3}\ve{\om} \, + \, \tfrac{1}{2} \smallint^{z_1}_{z_2}\ve{\om} \, + \,  \tfrac{1}{2} \smallint^{z_4}_{z_5}\ve{\om} \right) \, \tspin \left( - \, \tfrac{1}{2} \smallint^{z_2}_{z_4}\ve{\om} \, + \, \tfrac{1}{2} \smallint^{z_5}_{z_3}\ve{\om} \right) 
 \right\} \ .
 \label{5pt4}
\end{align}
The six point function relevant for four fermion amplitudes follows from (\ref{(4,4)}) after factorizing $\psi^\mu$ and $\psi^\nu$:
\begin{align}
&\langle \psi^{\mu}(z_{1}) \, \psi^{\nu} (z_{2}) \, S_{\al}(z_{3}) \, S_{\be}(z_{4})  \, S_{\dga}(z_{5}) \, S_{\dde}(z_{6}) \rangle \spin \notag \\
&= \ \  \frac{1}{2 \, E_{12} \, (E_{13} \, E_{14} \, E_{15} \, E_{16} \, E_{23} \, E_{24} \, E_{25} \, E_{26} \, E_{34} \, E_{56})^{1/2} \, \tspin ( \ve{0} ) \, \tspin ( \ve{0} ) } \notag \\
& \ \ \ \ \times \, \left\{ \si^{\mu}_{\be \dde} \, \si^{\nu}_{\al \dga} \, E_{13} \, E_{15} \, E_{24} \, E_{26} \, \tspin \left( \smallint^{z_1}_{z_2} \ve{\om} \, + \, \tfrac{1}{2}  \smallint^{z_3}_{z_4} \ve{\om} \, + \, \tfrac{1}{2}  \smallint^{z_5}_{z_6} \ve{\om} \right) \, \tspin \left( + \, \tfrac{1}{2}  \smallint^{z_3}_{z_4} \ve{\om} \, - \, \tfrac{1}{2}  \smallint^{z_5}_{z_6} \ve{\om} \right) \right. \notag \\
& \ \ \ \ \ \ \ \! \left. + \  \si^{\mu}_{\al \dga} \, \si^{\nu}_{\be \dde} \, E_{14} \, E_{16} \, E_{23} \, E_{25} \, \tspin \left( \smallint^{z_1}_{z_2} \ve{\om} \, - \, \tfrac{1}{2}  \smallint^{z_3}_{z_4} \ve{\om} \, - \, \tfrac{1}{2}  \smallint^{z_5}_{z_6} \ve{\om} \right) \, \tspin \left(- \, \tfrac{1}{2}  \smallint^{z_3}_{z_4} \ve{\om} \, + \, \tfrac{1}{2}  \smallint^{z_5}_{z_6} \ve{\om} \right) \right. \notag \\
& \ \ \ \ \ \ \ \! \left. - \  \si^{\mu}_{\al \dde} \, \si^{\nu}_{\be \dga} \, E_{14} \, E_{15} \, E_{23} \, E_{26} \, \tspin \left( \smallint^{z_1}_{z_2} \ve{\om} \, - \, \tfrac{1}{2}  \smallint^{z_3}_{z_4} \ve{\om} \, + \, \tfrac{1}{2}  \smallint^{z_5}_{z_6} \ve{\om} \right) \, \tspin \left(- \, \tfrac{1}{2}  \smallint^{z_3}_{z_4} \ve{\om} \, - \, \tfrac{1}{2}  \smallint^{z_5}_{z_6} \ve{\om} \right) \right. \notag \\
& \ \ \ \ \ \ \ \! \left. - \  \si^{\mu}_{\be \dga} \, \si^{\nu}_{\al \dde} \, E_{13} \, E_{16} \, E_{24} \, E_{25} \, \tspin \left( \smallint^{z_1}_{z_2} \ve{\om} \, + \, \tfrac{1}{2}  \smallint^{z_3}_{z_4} \ve{\om} \, - \, \tfrac{1}{2}  \smallint^{z_5}_{z_6} \ve{\om} \right) \, \tspin \left(+ \, \tfrac{1}{2}  \smallint^{z_3}_{z_4} \ve{\om} \, + \, \tfrac{1}{2}  \smallint^{z_5}_{z_6} \ve{\om} \right)
\right\} \label{6pt4a}
\end{align}
Let us finally give the second six point function of $\langle \psi^2 S^4 \rangle$ type. As the underlying pure spin field correlator (\ref{(6,2)}) is used in a non-minimal basis of $\vep$ combinations, there are ambiguities in writing the result in a symmetric fashion with respect to $S_{\al} \leftrightarrow S_\ga$ or $S_{\be} \leftrightarrow S_\de$:
\begin{align}
&\langle \psi^{\mu}(z_{1}) \, \psi^{\nu} (z_{2}) \, S_{\al}(z_{3}) \, S_{\be}(z_{4})  \, S_{\ga}(z_{5}) \, S_{\de}(z_{6}) \rangle \spin \eq \frac{1}{  \tspin ( \ve{0} ) \, \tspin ( \ve{0} ) } \notag \\
&\times \ \frac{1}{(E_{13} \, E_{14} \, E_{15} \, E_{16} \, E_{23} \, E_{24} \, E_{25} \, E_{26} \, E_{34} \, E_{35} \,E_{36} \,E_{45} \, E_{46} \, E_{56} )^{1/2}} \notag \\
&\times \ \biggl\{ \frac{\eta^{\mu \nu}}{E_{12}} \; \vep_{\al \be} \, \vep_{\ga \de} \, E_{36} \, E_{45} \,  \tspin \left( \tfrac{1}{2} \smallint^{z_3}_{z_5} \ve{\om} \, - \, \tfrac{1}{2} \smallint^{z_4}_{z_6} \ve{\om} \right) \biggr. \notag \\
&  \ \ \ \ \ \ \ \ \ \ \ \ \left[ E_{13} \, E_{24} \, E_{25} \, E_{16} \, \tspin \left( \smallint^{z_1}_{z_2} \ve{\om} \, + \, \tfrac{1}{2} \smallint^{z_3}_{z_5} \ve{\om} \, - \, \tfrac{1}{2} \smallint^{z_4}_{z_6} \ve{\om} \right) \right. \notag \\
& \left.   \ \ \ \ \ \ \ \ \ \ \ \   \ \ \ \ \ \ \ \ \ \ \ \  \ + \ E_{23} \, E_{14} \, E_{15} \, E_{26} \, \tspin \left( \smallint^{z_2}_{z_1} \ve{\om} \, + \, \tfrac{1}{2} \smallint^{z_3}_{z_5} \ve{\om} \, - \, \tfrac{1}{2} \smallint^{z_4}_{z_6} \ve{\om} \right) \right] \notag \\
& \ \ \ + \ \frac{\eta^{\mu \nu}}{E_{12}} \; \vep_{\al \de} \, \vep_{\ga \be} \, E_{34} \, E_{56} \,  \tspin \left( \tfrac{1}{2} \smallint^{z_3}_{z_5} \ve{\om} \, + \, \tfrac{1}{2} \smallint^{z_4}_{z_6} \ve{\om} \right) \biggr. \notag \\
&  \ \ \ \ \ \ \ \ \ \ \ \  \left[ E_{13} \, E_{14} \, E_{25} \, E_{26} \, \tspin \left( \smallint^{z_1}_{z_2} \ve{\om} \, + \, \tfrac{1}{2} \smallint^{z_3}_{z_5} \ve{\om} \, + \, \tfrac{1}{2} \smallint^{z_4}_{z_6} \ve{\om} \right)  \right. \notag \\
& \left.  \ \ \ \ \ \ \ \ \ \ \ \  \ \ \ \ \ \ \ \ \ \ \ \  + \ E_{23} \, E_{24} \, E_{15} \, E_{16} \, \tspin \left( \smallint^{z_2}_{z_1} \ve{\om} \, + \, \tfrac{1}{2} \smallint^{z_3}_{z_5} \ve{\om} \, + \, \tfrac{1}{2} \smallint^{z_4}_{z_6} \ve{\om} \right) \right] \notag \\
& \ \ \ + \ \left[ (\si^{\mu \nu} \, \vep)_{\ga \de} \; \frac{E_{56}}{2} \right] \, \vep_{\al \be}  \; \frac{E_{36} \, E_{45} \, E_{13} \, E_{24}}{\tspin \left(   \tfrac{1}{2} \smallint^{z_3}_{z_4} \ve{\om} \, + \, \tfrac{1}{2} \smallint^{z_5}_{z_6} \ve{\om} \right) } \; \tspin \left(- \tfrac{1}{2} \smallint^{z_3}_{z_4} \ve{\om} \, + \, \tfrac{1}{2} \smallint^{z_5}_{z_6} \ve{\om} \right) \notag \\
& \ \ \ \ \ \  \ \ \ \ \ \ \ \ \ \ \, \tspin \left(\tfrac{1}{2} \smallint^{z_1}_{z_5} \ve{\om} \, + \,  \tfrac{1}{2} \smallint^{z_1}_{z_6} \ve{\om} \, + \, \tfrac{1}{2} \smallint^{z_3}_{z_4} \ve{\om} \right) \, \tspin \left(\tfrac{1}{2} \smallint^{z_2}_{z_5} \ve{\om} \, + \,  \tfrac{1}{2} \smallint^{z_2}_{z_6} \ve{\om} \, - \, \tfrac{1}{2} \smallint^{z_3}_{z_4} \ve{\om} \right) \notag \\
& \ \ \ + \ \left[ (\si^{\mu \nu} \, \vep)_{\ga \be} \; \frac{E_{54}}{2} \right] \, \vep_{\al \de} \; \frac{E_{34} \, E_{56} \, E_{13} \, E_{26}}{\tspin \left(  \tfrac{1}{2} \smallint^{z_3}_{z_6} \ve{\om} \, + \, \tfrac{1}{2} \smallint^{z_5}_{z_4} \ve{\om} \right) } \; \tspin \left( - \,  \tfrac{1}{2} \smallint^{z_3}_{z_6} \ve{\om} \, + \, \tfrac{1}{2} \smallint^{z_5}_{z_4} \ve{\om} \right) \notag \\
& \ \ \ \ \ \  \ \ \ \ \ \ \ \ \ \ \, \tspin \left(\tfrac{1}{2} \smallint^{z_1}_{z_4} \ve{\om} \, + \,  \tfrac{1}{2} \smallint^{z_1}_{z_5} \ve{\om} \, + \, \tfrac{1}{2} \smallint^{z_3}_{z_6} \ve{\om} \right) \, \tspin \left(\tfrac{1}{2} \smallint^{z_2}_{z_4} \ve{\om} \, + \,  \tfrac{1}{2} \smallint^{z_2}_{z_5} \ve{\om} \, - \, \tfrac{1}{2} \smallint^{z_3}_{z_6} \ve{\om} \right)  \notag \\
& \ \ \ + \ \left[ (\si^{\mu \nu} \, \vep)_{\al \be} \; \frac{E_{34}}{2} \right] \, \vep_{\ga \de} \; \frac{E_{36} \, E_{45} \, E_{15} \, E_{26}}{\tspin \left(  \tfrac{1}{2} \smallint^{z_3}_{z_4} \ve{\om} \, + \, \tfrac{1}{2} \smallint^{z_5}_{z_6} \ve{\om} \right) } \; \tspin \left(  \tfrac{1}{2} \smallint^{z_3}_{z_4} \ve{\om} \, - \, \tfrac{1}{2} \smallint^{z_5}_{z_6} \ve{\om} \right) \notag \\
& \ \ \ \ \ \  \ \ \ \ \ \ \ \ \ \ \, \tspin \left(\tfrac{1}{2} \smallint^{z_1}_{z_3} \ve{\om} \, + \,  \tfrac{1}{2} \smallint^{z_1}_{z_4} \ve{\om} \, + \, \tfrac{1}{2} \smallint^{z_5}_{z_6} \ve{\om} \right) \, \tspin \left(\tfrac{1}{2} \smallint^{z_2}_{z_3} \ve{\om} \, + \,  \tfrac{1}{2} \smallint^{z_2}_{z_4} \ve{\om} \, - \, \tfrac{1}{2} \smallint^{z_5}_{z_6} \ve{\om} \right) \notag \\
& \ \ \ + \ \left[ (\si^{\mu \nu} \, \vep)_{\al \de} \; \frac{E_{36}}{2} \right] \, \vep_{\ga \be} \; \frac{E_{34} \, E_{56} \, E_{15} \, E_{24}}{\tspin \left(   \tfrac{1}{2} \smallint^{z_3}_{z_6} \ve{\om} \, + \, \tfrac{1}{2} \smallint^{z_5}_{z_4} \ve{\om} \right) } \; \tspin \left(  \tfrac{1}{2} \smallint^{z_3}_{z_6} \ve{\om} \, - \, \tfrac{1}{2} \smallint^{z_5}_{z_4} \ve{\om} \right) \notag \\
& \ \ \ \ \ \ \biggl. \ \ \ \ \ \ \ \ \ \ \, \tspin \left(\tfrac{1}{2} \smallint^{z_1}_{z_3} \ve{\om} \, + \,  \tfrac{1}{2} \smallint^{z_1}_{z_6} \ve{\om} \, + \, \tfrac{1}{2} \smallint^{z_5}_{z_4} \ve{\om} \right) \, \tspin \left(\tfrac{1}{2} \smallint^{z_2}_{z_3} \ve{\om} \, + \,  \tfrac{1}{2} \smallint^{z_2}_{z_6} \ve{\om} \, - \, \tfrac{1}{2} \smallint^{z_5}_{z_4} \ve{\om} \right) \biggr\}
\label{6pt4b}
\end{align}
One could replace $\si^{\mu \nu} = \si^\mu \bar{\si}^\nu + \eta^{\mu \nu}$ according to the philosophy of the previous subsections, but this would also turn the $\eta^{\mu \nu}$ coefficients into a quotient of $\Theta$ functions. The redundancy of one tensor in (\ref{6pt4b}) is reflected in
\beq
(\si^{\mu \nu} \, \vep)_{\al \be} \, \vep_{\ga \de} \ - \ (\si^{\mu \nu} \, \vep)_{\al \de} \, \vep_{\ga \be} \ + \ (\si^{\mu \nu} \, \vep)_{ \ga \de } \, \vep_{\al \be} \ - \ (\si^{\mu \nu} \, \vep)_{\ga \be} \, \vep_{\al \de} \eq 0\ .
\label{sig10} 
\eeq


\vskip1cm
\goodbreak
\centerline{\noindent{\bf Acknowledgments} }\vskip 2mm

I wish to thank Stephan Stieberger for triggering this project and for continuous support. Olaf Lechtenfeld and Stephan Stieberger pointed out some important references and ideas for which I am highly indebted. Special thanks for fruitful discussions as well as proofreading go to Alexander Dobrinevski, Daniel H\"artl, Dieter L\"ust and Stephan Merkle.

\section*{Appendix}
\appendix

\section{Generalized theta function technology}
\label{appB}

In this appendix we give some basic techniques of manipulating generalized $\Theta$ functions, in particular those necessary for deriving and checking the presented results. Firstly, one has to care about periodicity properties at non-zero genus. In contrast to their tree level cousins, $g \geq 1$ correlation functions are not only determined by the singularity structure in their arguments but also by their behaviour when single fields $\psi^\mu$ or $S_{\ga}, S_{\dde}$ are transported around homology cycles. The second part of this appendix discusses addition theorems for $\Theta$ functions which ensure consistency of various correlators with all possible choices of the $SO(1,3)$ indices.

\subsection{Periodicity properties}
\label{sec:appB1}

Any correlation function on a genus $g$ Riemann surface has to respect its $2g$ homology cycles. On a torus with $z \equiv z+1 \equiv z + \tau$, for instance, correlators $\langle \phi_1(z_1) \, ... \, \phi_N(z_N) \rangle \spin$ (with $\phi_i \in \{ \psi^\mu , S_\ga, S_{\dde} \}$) have to satisfy certain periodicity requirements under $z_i \mapsto z_i+1$ and $z_i \mapsto z_i + \tau$ which are specific to the type of field under consideration:

\begin{itemize}
\item [(i)] NS fermions $\phi_i \equiv \psi^\mu$

Worldsheet fermions catch phases $\pm 1$ upon transport about homology cycles $\al_I, b_J$ (where $I,J=1,...,g$). The precise sign configuration under $z \mapsto z + \al_I$ and $z \mapsto z+\be_J$ defines the entries 0 or 1 of the spin structure vectors $\ve{a}, \ve{b}$:
\begin{subequations}
\begin{align}
\langle \psi^\mu(z_1 + \al_J) \, \phi_2(z_2) \, ... \, \phi_N(z_N) \rangle \spin \ \ &= \ \ \exp(- \, i \pi \, a_J ) \, \langle \psi^\mu(z_1) \, \phi_2(z_2) \, ... \, \phi_N(z_N) \rangle \spin \label{per1} \\
\langle \psi^\mu(z_1 + \be_J) \, \phi_2(z_2) \, ... \, \phi_N(z_N) \rangle \spin \ \ &= \ \ \exp(+ \, i \pi \, b_J ) \, \langle \psi^\mu(z_1) \, \phi_2(z_2) \, ... \, \phi_N(z_N) \rangle \spin \label{per2}
\end{align}
\end{subequations}

\item [(ii)] R spin fields $\phi_i \equiv S_{\ga}, S_{\dde}$

Spin fields are responsible for changing the fermions' spin structure by creating branch points. The fermion fields flip sign when going around these branch points, this can be seen from the OPEs (\ref{2,1c}) and (\ref{2,1f}). Translating a spin field once around a cycle extends the brach cut all the way across and thus changes the spin structure:
\begin{subequations}
\begin{align}
\langle S_\ga(z_1 + \al_J) \, \phi_2(z_2) \, ... \, \phi_N(z_N) \rangle \spin \ \ &\sim \ \  \langle S_\ga(z_1) \, \phi_2(z_2) \, ... \, \phi_N(z_N) \rangle _{\ve{b} + \ve{e}_J}^{\ve{a}} \label{per3} \\
\langle S_\ga(z_1 + \be_J) \, \phi_2(z_2) \, ... \, \phi_N(z_N) \rangle \spin \ \ &\sim \ \  \langle S_\ga(z_1) \, \phi_2(z_2) \, ... \, \phi_N(z_N) \rangle ^{\ve{a} + \ve{e}_J}_{\ve{b}} \label{per4}
\end{align}
\end{subequations}
The normalization constants $\bigl[ \tspin(\ve{0}) \bigr]^{-1}$ in (\ref{at1}) and (\ref{at2}) are of course unaffected by $z_i$ transports, so there is a $z$ independent proportionality constant in the relations (\ref{per3}) and (\ref{per4}).
\end{itemize}
To check that the correlators presented in this paper obey (\ref{per1}) to (\ref{per4}) in every variable, one needs the following shift identity for $\Theta$ functions (valid for any $\ve{a},\ve{b}, \ve{t},\ve{s} \in \RR^g$):
\beq
\Theta \spin ( \ve{x} \, + \, \ve{t} \, + \, \Om \, \ve{s} ) \ \ = \ \ \exp \Bigl[ - \, i\pi \, \bigl(\ve{s} \, \Om \, \ve{s} \ + \  \ve{s} \, ( 2 \, \ve{x} \, + \, \ve{b} \, + \, 2 \, \ve{t} ) \bigr) \Bigr] \, \Theta ^{\ve{a} + 2 \ve{s}}_{\ve{b}+ 2\ve{t}} (\ve{x}) \label{per5}
\eeq
Fermion arguments enter via $\Theta \spin \left( \smallint^z_{p} \ve{\om} \right)$, so transporting $z$ around $\al_I$ ($\be_J$) generates a shift by $\ve{t}=\ve{e}_I$ ($\ve{s}= \ve{e}_J$). In this case, the modification in the spin structure can be compensated via
\beq
\Theta^{\ve{a} + 2\ve{e}_I}_{\ve{b} + 2\ve{e}_J}(\ve{x}) \ \ = \ \ \exp \bigl( i\pi \, a_J \bigr) \, \Theta \spin (\ve{x}) \label{per6} \ .
\eeq
Spin field positions on the other hand show up in the form $\Theta \spin \left( \tfrac{1}{2} \smallint^z_{p} \ve{\om} \right)$ with the essential prefactor $\tfrac{1}{2}$ in front of the integral. Then, the shift in the spin structure due to (\ref{per5}) with $\ve{t}=\tfrac{1}{2}\ve{e}_I$ or $\ve{s}= \tfrac{1}{2} \ve{e}_J$ can no longer be removed by means of (\ref{per6}).

\medskip
Transporting worldsheet positions about $\al_I$ is quite harmless because the phase factor in (\ref{per5}) reduces to 1 if $\ve{s} = \ve{0}$. The following applications of (\ref{per5}) are relevant for checking the correlators' periodicity under $\be_J$ translations:
\begin{subequations}
\begin{align}
\Theta \spin \left( \pm \smallint^{z_1 + \be_J}_p \ve{\om} \, + \, \ve{x} \right) \ \ &= \ \ \exp \left[  i \pi \, \left( - \, \Om_{JJ}  \, - \, 2  \smallint^{z_1}_p \om_J \, \mp \,    (2x_J + b_J) \right)  \right] \, \Theta^{\ve{a} }_{\ve{b}} \left( \pm  \smallint^{z_1}_p \ve{\om} \, + \, \ve{x} \right) \label{trans3}  \\
\Theta \spin \left( \pm \, \tfrac{1}{2} \! \! \smallint^{z_1 + \be_J}_p \! \! \ve{\om} \, + \, \ve{x} \right) \ \ &= \ \ \exp \left[ i \pi \, \left( - \, \frac{ \Om_{JJ}}{4} \, - \, \frac{1}{2} \smallint^{z_1}_p \om_J \, \mp \, \frac{  (2x_J + b_J)}{2} \right) \right] \, \Theta^{\ve{a}+\ve{e}_J}_{\ve{b}} \left( \pm \, \tfrac{1}{2} \smallint^{z_1}_p \ve{\om} \, + \, \ve{x} \right)  \label{trans2}
\end{align}
\end{subequations}
The phases due to the prime forms are given by
\begin{subequations}
\begin{align}
E(z + \al_J,w) \ \ &= \ \ E(z,w) \label{per7} \\
E(z + \be_J,w) \ \ &= \ \ \exp \left[ - \, i \pi \, \Om_{JJ} \ - \ 2 \pi i \smallint^{z} _w \om_{J} \right] \, E(z,w) \label{per8} \ .
\end{align}
\end{subequations}

\subsection{Fay trisecant identities}
\label{appA}

This subsection lists several forms of the Fay trisecant identities. From the perspective of this paper, they are generalization of the $z$ crossing identity $z_{ij} z_{kl} = z_{ik} z_{jl} + z_{il} z_{kj}$ relevant at tree level which is an important consistency condition for almost every $g=0$ correlation function. The analogous expressions on higher genus involve the prime form $E(z_i,z_j) := E_{ij}$ and the generalized theta functions $\tspin$.

\medskip 
In its most general form, the Fay trisecant identity reads \cite{fay}
\begin{align}
 \tspin &\left( \sum_{k=1}^N \smallint^{x_k}_{y_k} \ve{\om} \, - \, \ve{e} \right) \, \Bigl[ \tspin (\ve{e}) \Bigr]^{N-1} \; \frac{\prod_{i<j}^N E(x_i,x_j) \, E(y_i,y_j)}{\prod_{i,j=1}^N E(x_i,y_j)} \notag \\
 &= \ \ (-1)^{N(N-1)/2} \, \det_{i,j} \left\{ E(x_{i},y_{j})^{-1} \, \tspin \left( \smallint^{x_i}_{y_j} \ve{\om} \, - \, \ve{e} \right) \right\}
\label{ft1}
\end{align}
where $x_i,y_j$ with $i,j =1,2,...,N$ denote arbitrary positions on a genus $g$ worldsheet and $\ve{e} \in \CC^g$ with $\tspin(\ve{e}) \neq 0$. In the manipulation of spin field correlators, the following version with the particular choice $\ve{e} = \tfrac{1}{2} \sum_{k=1}^N \smallint^{x_k}_{y_k} \ve{\om} - \ve{\De}$ is more helpful:
\begin{align}
\tspin &\left( \tfrac{1}{2} \sum_{k=1}^N \smallint^{x_k}_{y_k} \ve{\om} \, + \, \ve{\De} \right) \, \left[ \tspin \left( \tfrac{1}{2} \sum_{k=1}^N \smallint^{x_k}_{y_k} \ve{\om} \, - \, \ve{\De}  \right) \right]^{N-1} \; \frac{\prod_{i<j}^N E(x_i,x_j) \, E(y_i,y_j)}{\prod_{i,j=1}^N E(x_i,y_j)} \notag \\
&= \ \ (-1)^{N(N-1)/2} \, \det_{i,j} \left\{ E(x_{i},y_{j})^{-1} \, \tspin \left(- \, \tfrac{1}{2} \sum_{k=1}^N \smallint^{x_k}_{y_k} \ve{\om} \, + \, \smallint^{x_i}_{y_j} \ve{\om} \, + \, \ve{\De} \right) \right\}
\label{ft1a}
\end{align}
Let us bring the $N=2$ case into a form which is comparable with the tree level $z$ crossing identity by setting $(z_1,z_2,z_3,z_4) \equiv (x_1,y_1,x_2,y_2)$:
\begin{align}
&E_{13} \, E_{24} \, \tspin \left( \tfrac{1}{2} \smallint^{z_1}_{z_2} \ve{\om} \, + \, \tfrac{1}{2} \smallint^{z_3}_{z_4} \ve{\om} \, + \, \ve{\De} \right) \, \tspin \left( \tfrac{1}{2} \smallint^{z_1}_{z_2} \ve{\om} \, + \, \tfrac{1}{2} \smallint^{z_3}_{z_4} \ve{\om} \, - \, \ve{\De} \right) \notag \\
= \ \ &E_{12} \, E_{34} \, \tspin \left( \tfrac{1}{2} \smallint^{z_1}_{z_3} \ve{\om} \, + \, \tfrac{1}{2} \smallint^{z_2}_{z_4} \ve{\om} \, + \, \ve{\De} \right) \, \tspin \left( \tfrac{1}{2} \smallint^{z_1}_{z_3} \ve{\om} \, + \, \tfrac{1}{2} \smallint^{z_2}_{z_4} \ve{\om} \, - \, \ve{\De} \right) \notag \\
+ \ &E_{14} \, E_{23} \, \tspin \left( \tfrac{1}{2} \smallint^{z_1}_{z_2} \ve{\om} \, + \, \tfrac{1}{2} \smallint^{z_4}_{z_3} \ve{\om} \, + \, \ve{\De} \right) \, \tspin \left( \tfrac{1}{2} \smallint^{z_1}_{z_2} \ve{\om} \, + \, \tfrac{1}{2} \smallint^{z_4}_{z_3} \ve{\om} \, - \, \ve{\De} \right)
\label{ft2}
\end{align}
This is essential for the correlators (\ref{(4,0)}), (\ref{(4,2)}) and (\ref{(4,4)}) discussed in section \ref{sec:2}.

\medskip
The next order version $N=3$ is relevant for spin field correlators with at least six fields with alike chirality:
\begin{align}
&\frac{ E_{13} \, E_{15} \, E_{35} \, E_{24} \, E_{26} \, E_{46} }{ E_{12} \, E_{14} \, E_{16} \, E_{23} \, E_{34} \, E_{36} \, E_{25} \, E_{45} \, E_{56} } \notag \\
& \ \ \ \ \ \ \ \ \times \  \tspin \left( \tfrac{1}{2} \smallint^{z_1}_{z_2} \ve{\om} \, + \, \tfrac{1}{2} \smallint^{z_3}_{z_4} \ve{\om} \, + \, \tfrac{1}{2} \smallint^{z_5}_{z_6} \ve{\om} \, - \, \ve{\De} \right) \, \left[ \tspin \left( \tfrac{1}{2} \smallint^{z_1}_{z_2} \ve{\om} \, + \, \tfrac{1}{2} \smallint^{z_3}_{z_4} \ve{\om} \, + \, \tfrac{1}{2} \smallint^{z_5}_{z_6} \ve{\om} \, + \, \ve{\De} \right) \right]^2 \notag \\
&= \ \ \frac{1}{E_{12} \, E_{34} \, E_{56}} \;   \tspin \left( - \tfrac{1}{2} \smallint^{z_1}_{z_2} \ve{\om} \, + \, \tfrac{1}{2} \smallint^{z_3}_{z_4} \ve{\om} \, + \, \tfrac{1}{2} \smallint^{z_5}_{z_6} \ve{\om} \, + \, \ve{\De} \right) \notag \\
& \ \ \ \ \ \ \times \ \tspin \left(  \tfrac{1}{2} \smallint^{z_1}_{z_2} \ve{\om} \, - \, \tfrac{1}{2} \smallint^{z_3}_{z_4} \ve{\om} \, + \, \tfrac{1}{2} \smallint^{z_5}_{z_6} \ve{\om} \, + \, \ve{\De} \right) \, \tspin \left( \tfrac{1}{2} \smallint^{z_1}_{z_2} \ve{\om} \, + \, \tfrac{1}{2} \smallint^{z_3}_{z_4} \ve{\om} \, - \, \tfrac{1}{2} \smallint^{z_5}_{z_6} \ve{\om} \, + \, \ve{\De} \right) \notag \\
& \ - \ \frac{1}{E_{12} \, E_{36} \, E_{54}} \;   \tspin \left( - \tfrac{1}{2} \smallint^{z_1}_{z_2} \ve{\om} \, + \, \tfrac{1}{2} \smallint^{z_3}_{z_6} \ve{\om} \, + \, \tfrac{1}{2} \smallint^{z_5}_{z_4} \ve{\om} \, + \, \ve{\De} \right) \notag \\
& \ \ \ \ \ \ \times \ \tspin \left(  \tfrac{1}{2} \smallint^{z_1}_{z_2} \ve{\om} \, - \, \tfrac{1}{2} \smallint^{z_3}_{z_6} \ve{\om} \, + \, \tfrac{1}{2} \smallint^{z_5}_{z_4} \ve{\om} \, + \, \ve{\De} \right) \, \tspin \left( \tfrac{1}{2} \smallint^{z_1}_{z_2} \ve{\om} \, + \, \tfrac{1}{2} \smallint^{z_3}_{z_6} \ve{\om} \, - \, \tfrac{1}{2} \smallint^{z_5}_{z_4} \ve{\om} \, + \, \ve{\De} \right) \notag \\
& \ + \ \frac{1}{E_{14} \, E_{36} \, E_{52}} \;   \tspin \left( - \tfrac{1}{2} \smallint^{z_1}_{z_4} \ve{\om} \, + \, \tfrac{1}{2} \smallint^{z_3}_{z_6} \ve{\om} \, + \, \tfrac{1}{2} \smallint^{z_5}_{z_2} \ve{\om} \, + \, \ve{\De} \right) \notag \\
& \ \ \ \ \ \ \times \ \tspin \left( \tfrac{1}{2} \smallint^{z_1}_{z_4} \ve{\om} \, - \, \tfrac{1}{2} \smallint^{z_3}_{z_6} \ve{\om} \, + \, \tfrac{1}{2} \smallint^{z_5}_{z_2} \ve{\om} \, + \, \ve{\De} \right) \, \tspin \left( \tfrac{1}{2} \smallint^{z_1}_{z_4} \ve{\om} \, + \, \tfrac{1}{2} \smallint^{z_3}_{z_6} \ve{\om} \, - \, \tfrac{1}{2} \smallint^{z_5}_{z_2} \ve{\om} \, + \, \ve{\De} \right) \notag \\
& \ - \ \frac{1}{E_{14} \, E_{32} \, E_{56}} \;   \tspin \left( - \tfrac{1}{2} \smallint^{z_1}_{z_4} \ve{\om} \, + \, \tfrac{1}{2} \smallint^{z_3}_{z_2} \ve{\om} \, + \, \tfrac{1}{2} \smallint^{z_5}_{z_6} \ve{\om} \, + \, \ve{\De} \right) \notag \\
& \ \ \ \ \ \ \times \ \tspin \left( \tfrac{1}{2} \smallint^{z_1}_{z_4} \ve{\om} \, - \, \tfrac{1}{2} \smallint^{z_3}_{z_2} \ve{\om} \, + \, \tfrac{1}{2} \smallint^{z_5}_{z_6} \ve{\om} \, + \, \ve{\De} \right) \, \tspin \left( \tfrac{1}{2} \smallint^{z_1}_{z_4} \ve{\om} \, + \, \tfrac{1}{2} \smallint^{z_3}_{z_2} \ve{\om} \, - \, \tfrac{1}{2} \smallint^{z_5}_{z_6} \ve{\om} \, + \, \ve{\De} \right) \notag \\
& \ + \ \frac{1}{E_{16} \, E_{32} \, E_{54}} \;   \tspin \left( - \tfrac{1}{2} \smallint^{z_1}_{z_6} \ve{\om} \, + \, \tfrac{1}{2} \smallint^{z_3}_{z_2} \ve{\om} \, + \, \tfrac{1}{2} \smallint^{z_5}_{z_4} \ve{\om} \, + \, \ve{\De} \right) \notag \\
& \ \ \ \ \ \ \times \ \tspin \left( \tfrac{1}{2} \smallint^{z_1}_{z_6} \ve{\om} \, - \, \tfrac{1}{2} \smallint^{z_3}_{z_2} \ve{\om} \, + \, \tfrac{1}{2} \smallint^{z_5}_{z_4} \ve{\om} \, + \, \ve{\De} \right) \, \tspin \left( \tfrac{1}{2} \smallint^{z_1}_{z_6} \ve{\om} \, + \, \tfrac{1}{2} \smallint^{z_3}_{z_2} \ve{\om} \, - \, \tfrac{1}{2} \smallint^{z_5}_{z_4} \ve{\om} \, + \, \ve{\De} \right) \notag \\
& \ - \ \frac{1}{E_{16} \, E_{34} \, E_{52}} \;   \tspin \left( - \tfrac{1}{2} \smallint^{z_1}_{z_6} \ve{\om} \, + \, \tfrac{1}{2} \smallint^{z_3}_{z_4} \ve{\om} \, + \, \tfrac{1}{2} \smallint^{z_5}_{z_2} \ve{\om} \, + \, \ve{\De} \right) \notag \\
& \ \ \ \  \ \ \times \ \tspin \left( \tfrac{1}{2} \smallint^{z_1}_{z_6} \ve{\om} \, - \, \tfrac{1}{2} \smallint^{z_3}_{z_4} \ve{\om} \, + \, \tfrac{1}{2} \smallint^{z_5}_{z_2} \ve{\om} \, + \, \ve{\De} \right) \, \tspin \left( \tfrac{1}{2} \smallint^{z_1}_{z_6} \ve{\om} \, + \, \tfrac{1}{2} \smallint^{z_3}_{z_4} \ve{\om} \, - \, \tfrac{1}{2} \smallint^{z_5}_{z_2} \ve{\om} \, + \, \ve{\De} \right) 
\label{ft3}
\end{align}
The $N=4$ case is certainly too long to be displayed in full beauty, let us thus abbreviate it as
\begin{align}
& \frac{E_{13} \, E_{15} \, E_{17} \, E_{35} \, E_{37} \, E_{57} \, E_{24} \, E_{26} \, E_{28} \, E_{46} \, E_{48} \, E_{68} }{E_{12} \, E_{14} \, E_{16} \, E_{18} \, E_{23} \, E_{25} \, E_{27} \, E_{34} \, E_{36} \, E_{38} \, E_{45} \, E_{47} \, E_{56} \, E_{58} \, E_{67} \, E_{78}  } 
 \notag \\
 & \ \ \ \ \ \ \times \ \tspin \left(  \frac{1}{2}  \left[  \smallint^{z_1}_{z_2} \ve{\om} + \smallint^{z_3}_{z_4} \ve{\om} +  \smallint^{z_5}_{z_6} \ve{\om} + \smallint^{z_7}_{z_8} \ve{\om}  \right] \, - \, \ve{\De} \right) \, \left\{ \tspin \left(  \frac{1}{2}  \left[  \smallint^{z_1}_{z_2} \ve{\om} + \smallint^{z_3}_{z_4} \ve{\om} +  \smallint^{z_5}_{z_6} \ve{\om} + \smallint^{z_7}_{z_8} \ve{\om} \right]  \, + \, \ve{\De}\right) \right\}^3 \notag \\
&= \ \ \frac{  1 }{ E_{12} \, E_{34} \, E_{56} \, E_{78} } \notag \\
& \ \ \ \ \ \ \ \ \tspin \left(  \frac{1}{2} \, \left[ \smallint^{z_1}_{z_2} \ve{\om} + \smallint^{z_3}_{z_4} \ve{\om} +   \smallint^{z_5}_{z_6} \ve{\om} - \smallint^{z_7}_{z_8} \ve{\om} \right] \, + \, \ve{\De} \right) \,  \tspin \left(  \frac{1}{2} \, \left[ \smallint^{z_1}_{z_2} \ve{\om} +  \smallint^{z_3}_{z_4} \ve{\om} -  \smallint^{z_5}_{z_6} \ve{\om} + \smallint^{z_7}_{z_8} \ve{\om} \right] \, + \, \ve{\De} \right)   \notag \\
& \ \ \ \ \ \ \ \  \tspin \left(  \frac{1}{2} \, \left[ \smallint^{z_1}_{z_2} \ve{\om} -  \smallint^{z_3}_{z_4} \ve{\om} +   \smallint^{z_5}_{z_6} \ve{\om} + \smallint^{z_7}_{z_8} \ve{\om} \right] \, + \, \ve{\De} \right) \, \tspin \left(  \frac{1}{2}  \left[ - \smallint^{z_1}_{z_2} \ve{\om} + \smallint^{z_3}_{z_4} \ve{\om} +  \smallint^{z_5}_{z_6} \ve{\om} + \smallint^{z_7}_{z_8} \ve{\om} \right] \, + \, \ve{\De} \right) \notag \\
& \ \ - \ \frac{ 1 }{ E_{12} \, E_{34} \, E_{58} \, E_{76} } \notag \\
& \ \ \ \ \ \ \ \ \tspin \left(  \frac{1}{2} \, \left[ \smallint^{z_1}_{z_2} \ve{\om} + \smallint^{z_3}_{z_4} \ve{\om} +   \smallint^{z_5}_{z_8} \ve{\om} - \smallint^{z_7}_{z_6} \ve{\om} \right] \, + \, \ve{\De} \right) \,  \tspin \left(  \frac{1}{2} \, \left[ \smallint^{z_1}_{z_2} \ve{\om} +  \smallint^{z_3}_{z_4} \ve{\om} -  \smallint^{z_5}_{z_8} \ve{\om} + \smallint^{z_7}_{z_6} \ve{\om} \right] \, + \, \ve{\De} \right) \biggr. \notag \\
& \ \ \ \ \ \ \ \  \tspin \left(  \frac{1}{2} \, \left[ \smallint^{z_1}_{z_2} \ve{\om} -  \smallint^{z_3}_{z_4} \ve{\om} +   \smallint^{z_5}_{z_8} \ve{\om} + \smallint^{z_7}_{z_6} \ve{\om} \right] \, + \, \ve{\De} \right) \, \tspin \left(  \frac{1}{2} \, \left[ - \smallint^{z_1}_{z_2} \ve{\om} + \smallint^{z_3}_{z_4} \ve{\om} +  \smallint^{z_5}_{z_8} \ve{\om} + \smallint^{z_7}_{z_6} \ve{\om} \right] \, + \, \ve{\De} \right) \notag \\
& \ \ \pm \  \ 22 \ \te{further permutations of} \, (z_2, z_4, z_6, z_8,) \ .
\label{ft4}
\end{align}

\section{Proofs}

The lengthy proofs for the expressions (\ref{(2M,0)}), (\ref{(2M,2)}), (\ref{(2M,4)}) as well as (\ref{Omega}), (\ref{omega}) for correlators of arbitrary size are banished to the following appendix. In each case, we proceed in two steps: First of all demonstrate the behaviour under transporting $z_k \mapsto z_k + \al_J$ or $z_k \mapsto z_k + \be_J$ to agree with the physical expectations, see appendix \ref{sec:appB1}. Secondly, we will show the singularity structure to be determined by the OPEs (\ref{rv,1a}) to (\ref{rv,2b}).

\subsection{Spin fields}
\label{sec:appC}

Clearly, the $2M+4$ point function (\ref{(2M,4)}) is the toughest case among the spin field correlators of variable size, so we will mostly focus on this one. Due to the large extent of symmetry, it suffices to check the properties under $z_1$ and $z_C$ transportation. Recall from (\ref{per5}) that $z_{1} \mapsto z_1 + \al_J$ leaves the prime form $E(z_1,z_k)$ invariant and changes the spin structure of the $\Theta \spin \left( \tfrac{1}{2} \smallint^{z_1}_p \ve{\om} + ... \right)$ functions as $(\ve{a},\ve{b}) \mapsto (\ve{a},\ve{b} + \ve{e}_J)$.

\medskip
The $\be$ cycles are slightly more difficult to handle, see (\ref{trans2}) and (\ref{per8}). Consequently, the right hand side of (\ref{(2M,4)}) catches various phases under $\be_J$ periods (in addition to $\ve{a} \mapsto \ve{a}+\ve{e}_J$). We consider the $z_{ij}$ function associated with $\vep_{\dga \dde} \vep_{\dka \dom}$ (rather than $\vep_{\dga \dom} \vep_{\dka \dde}$) at fixed $\rho \in S_M$ and display the phase contributions from prime forms and from $\Theta \spin$'s separately to demonstrate cancellations of the overall phase:
\begin{itemize}
\item $z_1 \mapsto z_1 + \be_J$
\[
\left. \begin{array}{rc} E_{1k}\te{'s} \ : &
  \exp \left[ \frac{i\pi \, \Om_{JJ}}{2} \, + \,  i \pi \, \left( \smallint^{z_1}_{z_{\rho(2)}} \om_J \, - \, \sum_{k=2}^M \smallint^{z_{2k-1}}_{z_{\rho(2k)}} \om_J \right) \right] \\ \Theta \spin \te{'s} \ : &
  \exp \left[ - \, \frac{i\pi \, \Om_{JJ}}{2} \, + \,  i \pi \, \left( - \, \smallint^{z_1}_{z_{\rho(2)}} \om_J \, + \, \sum_{k=2}^M \smallint^{z_{2k-1}}_{z_{\rho(2k)}} \om_J \right) \right]  \end{array} \right\}  \ \ \rightarrow \ 1 \ \ \te{in total}
\]
\item $z_C \mapsto z_C + \be_J$
\[
\left. \begin{array}{rc} E_{Ck}\te{'s} \ : &
  \exp \left[ \frac{i\pi \, \Om_{JJ}}{2} \, + \,  i \pi \ \smallint^{z_C}_{z_{D}} \om_J \, - \, i \pi \ \smallint^{z_E}_{z_{F}} \om_J  \right] \\ \Theta \spin \te{'s} \ : &
   \exp \left[ - \, \frac{i\pi \, \Om_{JJ}}{2} \, - \,  i \pi \ \smallint^{z_C}_{z_{D}} \om_J \, + \, i \pi \ \smallint^{z_E}_{z_{F}} \om_J  \right] \end{array} \right\}  \ \ \rightarrow \ 1 \ \ \te{in total}
\]
\end{itemize}
Consistency of the singularity structure will now be proven by induction. First of all, the claimed formula can be easily seen to match with the right handed versions of (\ref{(4,0)}), (\ref{(4,2)}) as well as (\ref{(4,4)}), (\ref{(6,4)}) at $M=0,1,2,3$. The inductive step makes use of the fact that the $2M+2$ point correlator should appear from the $2M+4$ point ancestor if we replace two spin fields by the OPE in the corresponding limit $z_i \to z_j$. Symmetries allow to restrict to $z_{2M-1} \rightarrow z_{2M}$ and $z_E \rightarrow z_F$. 
\begin{itemize}
\item limit $z_{2M-1} \rightarrow z_{2M}$:

Assuming (\ref{(2M,4)}) to hold at $M-1$, the OPE (\ref{rv,1d}) requires that
\begin{align}
  \label{inducstep}
  & \langle S_{\al_1}(z_1)\, ... \, S_{\al_{2M-1}}(z_{2M-1})\,S_{\al_{2M}}(z_{2M}) \,  S_{\dga}(z_C) \, S_{\dde}(z_D) \, S_{\dka}(z_E) \, S_{\dom}(z_F) \rangle \spin \, \Bigl.\Bigr|_{z_{2M-1}\ra z_{2M}}   \notag \\
  & \stackrel{!}{=} \ \ - \; \frac{ \vep_{\al_{2M-1} \al_{2M}}}{(z_{2M-1} \, - \, z_{2M})^{1/2}} \; \langle S_{\al_1}(z_1)\, ... \,S_{\al_{2M-2}}(z_{2M-2}) \,  S_{\dga}(z_C) \, S_{\dde}(z_D) \, S_{\dka}(z_E) \, S_{\dom}(z_F) \rangle \spin \notag \\
  & \ \ \ \ \ \ \ \ \ \  \ \ \ \ +\ \mathcal{O}(z_{2M-1,2M}) \notag \\
&= \ \ - \; \frac{\vep_{\al_{2M-1} \al_{2M}}}{(z_{2M-1} - z_{2M})^{1/2}} \;  \frac{ (-1)^{M-1}}{\bigl[\tspin(\ve{0}) \bigr]^2} \, \left( \frac{E_{CD} \, E_{CF} \, E_{DE} \, E_{EF}}{E_{CE} \, E_{DF}}\right)^{1/2} \notag \\
& \ \ \times  \, \left(\prod_{i \leq j}^{M-1} E_{2i-1,2j}\,\prod_{\bi<\bj}^{M-1}  E_{2\bi,2\bj-1}\right)^{1/2} \, \left(\prod_{k<l}^{M-1} E_{2k-1,2l-1}\, E_{2k,2l}\right)^{-1/2} \notag \\
  & \ \ \times \ \Biggl\{ \frac{\vep_{\dga \dde} \, \vep_{\dka \dom}}{E_{CD} \, E_{EF}} \; \left[ \tspin \left( \frac{1}{2} \, \sum_{i=1}^{M-1} \smallint^{z_{2i-1}}_{z_{2i}} \ve{\om} \, \pm \, \frac{1}{2} \smallint^{z_{C}}_{z_{D}} \ve{\om} \, \mp \, \frac{1}{2} \smallint^{z_{E}}_{z_{F}} \ve{\om} \right) \right]^{3-M} \, \sum_{\rho \in S_{M-1}}\te{sgn}(\rho)\Biggr. \notag \\
  & \ \ \ \ \ \ \ \ \ \ \ \ \ \ \ \ \ \Biggl. \prod_{m=1}^{M-1}\frac{\vep_{\al_{2m-1}\al_{\rho(2m)}}}{E_{2m-1,\rho(2m)}} \; \tspin \left( \frac{1}{2} \, \sum_{i=1}^{M-1} \smallint^{z_{2i-1}}_{z_{2i}} \ve{\om} \, - \smallint^{z_{2m-1}}_{z_{\rho(2m)}} \ve{\om} \, \pm \, \frac{1}{2} \smallint^{z_{C}}_{z_{D}} \ve{\om} \, \mp \, \frac{1}{2} \smallint^{z_{E}}_{z_{F}} \ve{\om} \right) \Biggr. \notag \\
  & \ \ \ \ \ \ \ \ \ \ \ \ \Biggl. + \ \bigl[ (z_D,\dde) \, \leftrightarrow \, (z_F, \dom) \bigr] \Biggr \} \ +\ \mathcal{O}(z_{2M-1,2M}) \ .
  \end{align}
This can be reproduced from (\ref{(2M,4)}) at $M$:
\begin{align}
  & \langle S_{\al_1}(z_1)\, ... \, S_{\al_{2M-1}}(z_{2M-1})\,S_{\al_{2M}}(z_{2M}) \,  S_{\dga}(z_C) \, S_{\dde}(z_D) \, S_{\dka}(z_E) \, S_{\dom}(z_F) \rangle \spin \, \Bigl.\Bigr|_{z_{2M-1}\ra z_{2M}}   \notag \\
  &= \ \ \frac{ (-1)^{M}}{\bigl[\tspin(\ve{0}) \bigr]^2} \,  \left( \frac{E_{CD} \, E_{CF} \, E_{DE} \, E_{EF}}{E_{CE} \, E_{DF}}\right)^{1/2} \notag \\
  & \ \ \times \, \left( E_{2M-1,2M} \, \prod_{i \leq j}^{M-1} E_{2i-1,2j} \, \prod_{i =1}^{M-1} E_{2i-1,2M} \, \prod_{\bi<\bj}^{M-1}  E_{2\bi,2\bj-1} \, \prod_{\bi=1}^{M-1}  E_{2\bi,2M-1} \right)^{1/2} \notag \\
  & \ \ \times \,  \left( \prod_{k<l}^{M-1} E_{2k-1,2l-1}\, E_{2k,2l}
 \prod_{i=1}^{M-1} \underbrace{E_{2i-1,2M-1}}_{= \,  E_{2i-1,2M} \, + \, \mathcal{O}(z_{2M-1,2M})}  \, \prod_{\bi=1}^{M-1}  \underbrace{E_{2\bi,2M} }_{= \, E_{2\bi,2M-1} \, + \, \mathcal{O}(z_{2M-1,2M})} \right)^{-1/2} \notag \\
 & \ \ \times \ \Biggl\{ \frac{\vep_{\dga \dde} \, \vep_{\dka \dom}}{E_{CD} \, E_{EF}} \; \left[ \tspin \left( \frac{1}{2} \, \sum_{i=1}^{M-1} \smallint^{z_{2i-1}}_{z_{2i}} \ve{\om} \, \pm \, \frac{1}{2} \smallint^{z_{C}}_{z_{D}} \ve{\om} \, \mp \, \frac{1}{2} \smallint^{z_{E}}_{z_{F}} \ve{\om} \right) \,  + \, \mathcal{O}(z_{2M-1,2M}) \right]^{2-M} \Biggr. \notag \\
 & \ \ \ \ \sum_{\rho \in S_{M}} \! \te{sgn}(\rho) \, \de_{2M,\rho(2M)} \, \prod_{m=1}^{M}\frac{\vep_{\al_{2m-1}\al_{\rho(2m)}}}{E_{2m-1,\rho(2m)}} \; \tspin \left( \frac{1}{2} \, \sum_{i=1}^{M} \smallint^{z_{2i-1}}_{z_{2i}} \ve{\om} \, - \smallint^{z_{2m-1}}_{z_{\rho(2m)}} \ve{\om} \, \pm \, \frac{1}{2} \smallint^{z_{C}}_{z_{D}} \ve{\om} \, \mp \, \frac{1}{2} \smallint^{z_{E}}_{z_{F}} \ve{\om} \right) \Biggr. \notag \\
  & \ \ \ \ \ \ \ \ \ \ \ \ \Biggl. + \ \bigl[ (z_D,\dde) \, \leftrightarrow \, (z_F, \dom) \bigr] \Biggr \}   \ + \ \mathcal{O}(z_{2M-1,2M})  \notag \\
  &= \ \ \frac{ (-1)^{M}}{\bigl[\tspin(\ve{0}) \bigr]^2} \,  \left( \frac{E_{CD} \, E_{CF} \, E_{DE} \, E_{EF}}{E_{CE} \, E_{DF}}\right)^{1/2} \, \left(  \prod_{i \leq j}^{M-1} E_{2i-1,2j} \, \prod_{\bi<\bj}^{M-1}  E_{2\bi,2\bj-1}  \right)^{1/2} \notag \\
  & \ \ \times \,  \left( \prod_{k<l}^{M-1} E_{2k-1,2l-1}\, E_{2k,2l}
 \right)^{-1/2} \, \vep_{\al_{2M-1} \al_{2M}} \, \underbrace{E_{2M-1,2M}^{-1/2}}_{ z_{2M-1,2M}^{-1/2} \, + \,  \mathcal{O}(z_{2M-1,2M}) } \notag \\
 & \ \ \times \ \Biggl\{ \frac{\vep_{\dga \dde} \, \vep_{\dka \dom}}{E_{CD} \, E_{EF}} \; \left[ \tspin \left( \frac{1}{2} \, \sum_{i=1}^{M-1} \smallint^{z_{2i-1}}_{z_{2i}} \ve{\om} \, \pm \, \frac{1}{2} \smallint^{z_{C}}_{z_{D}} \ve{\om} \, \mp \, \frac{1}{2} \smallint^{z_{E}}_{z_{F}} \ve{\om} \right)  \right]^{2-M} \Biggr. \notag \\
 &\ \ \ \ \ \  \sum_{\pi \in S_{M-1}} \! \te{sgn}(\pi) \, \prod_{m=1}^{M-1}\frac{\vep_{\al_{2m-1}\al_{\pi(2m)}}}{E_{2m-1,\pi(2m)}} \; \tspin \left( \frac{1}{2} \, \sum_{i=1}^{M-1} \smallint^{z_{2i-1}}_{z_{2i}} \ve{\om} \, - \smallint^{z_{2m-1}}_{z_{\pi(2m)}} \ve{\om} \, \pm \, \frac{1}{2} \smallint^{z_{C}}_{z_{D}} \ve{\om} \, \mp \, \frac{1}{2} \smallint^{z_{E}}_{z_{F}} \ve{\om} \right) 
 \notag \\
 & \ \ \ \ \ \ \ \ \Biggl. \tspin \left( \frac{1}{2} \, \sum_{i=1}^{M-1} \smallint^{z_{2i-1}}_{z_{2i}} \ve{\om}  \, \pm \, \frac{1}{2} \smallint^{z_{C}}_{z_{D}} \ve{\om} \, \mp \, \frac{1}{2} \smallint^{z_{E}}_{z_{F}} \ve{\om} \right) \  + \ \bigl[ (z_D,\dde) \, \leftrightarrow \, (z_F, \dom) \bigr] \Biggr \}   \ + \ \mathcal{O}(z_{2M-1,2M}) 
  \end{align}
The most singular piece of (\ref{(2M,4)}) in $z_{2M-1,2M}$ is the subset of $S_M$ permutations $\rho$ with $\rho(2M) = 2M$, these are isolated via $\de_{2M,\rho(2M)}$. Since the powers of the permutation independent $\tspin \left( \frac{1}{2} \, \sum_{i=1}^{M-1} \smallint^{z_{2i-1}}_{z_{2i}} \ve{\om} \, \pm \, \frac{1}{2} \smallint^{z_{C}}_{z_{D}} \ve{\om} \, \mp \, \frac{1}{2} \smallint^{z_{E}}_{z_{F}} \ve{\om} \right)$ in the last lines add up to $2-M+1 = 3-M$, we find agreement with (\ref{inducstep}) in the most singular power of $z_{2M-1,2M}$.

\item limit $z_{E} \rightarrow z_{F}$:

Now we reduce the left hand side of (\ref{(2M,4)}) to (\ref{(2M,2)}) by means of the OPE (\ref{rv,1e}).
\begin{align}
&\langle S_{\al_1}(z_1) \, S_{\al_2}(z_2)\, ... \, S_{\al_{2M-1}}(z_{2M-1})\,S_{\al_{2M}}(z_{2M}) \, S_{\dga}(z_C) \, S_{\dde}(z_D) \, S_{\dka}(z_E) \, S_{\dom}(z_F) \rangle\spin \, \Bigl.\Bigr|_{z_{E}\ra z_{F}} \notag \\
& \stackrel{!}{=} \ \  \frac{ \vep_{\dka \dom}}{(z_{EF})^{1/2}} \; \langle S_{\al_1}(z_1) \, S_{\al_2}(z_2)\, ... \, S_{\al_{2M-1}}(z_{2M-1})\,S_{\al_{2M}}(z_{2M}) \, S_{\dga}(z_C) \, S_{\dde}(z_D) \rangle\spin  \ + \ \mathcal{O}(z_{EF})  \notag \\
&= \ \ \frac{ \vep_{\dka \dom}}{(z_{EF})^{1/2}} \; \frac{ (-1)^M}{\bigl[\tspin(\ve{0}) \bigr]^2} \; \left[ \tspin \left( \frac{1}{2} \, \sum_{i=1}^M \smallint^{z_{2i-1}}_{z_{2i}} \ve{\om} \, \pm \, \frac{1}{2} \smallint^{z_{C}}_{z_{D}} \ve{\om} \right) \right]^{2-M} \; \frac{\vep_{\dga \dde}}{E_{CD}^{1/2}} \notag \\
 & \ \ \times \ \bigg(\prod_{i \leq j}^M E_{2i-1,2j}\,\prod_{\bi<\bj}^{M}  E_{2\bi,2\bj-1}\bigg)^{1/2} \, \left(\prod_{k<l}^M E_{2k-1,2l-1}\, E_{2k,2l}\right)^{-1/2} \, \sum_{\rho \in S_M}\te{sgn}(\rho) \notag \\
  & \ \ \times \
  \prod_{m=1}^M\frac{\vep_{\al_{2m-1}\al_{\rho(2m)}}}{E_{2m-1,\rho(2m)}} \; \tspin \left( \frac{1}{2} \, \sum_{i=1}^M \smallint^{z_{2i-1}}_{z_{2i}} \ve{\om} \, - \smallint^{z_{2m-1}}_{z_{\rho(2m)}} \ve{\om} \, \pm \, \frac{1}{2} \smallint^{z_{C}}_{z_{D}} \ve{\om} \right) \ + \ \mathcal{O}(z_{EF})
  \end{align}
The same leading behaviour follows from the right hand side of (\ref{(2M,4)}):
\begin{align}
&\langle S_{\al_1}(z_1) \, S_{\al_2}(z_2)\, ... \, S_{\al_{2M-1}}(z_{2M-1})\,S_{\al_{2M}}(z_{2M}) \, S_{\dga}(z_C) \, S_{\dde}(z_D) \, S_{\dka}(z_E) \, S_{\dom}(z_F) \rangle\spin \, \Bigl.\Bigr|_{z_{E}\ra z_{F}} \notag \\
&= \ \ \frac{ (-1)^M}{\bigl[\tspin(\ve{0}) \bigr]^2} \; (E_{CD} \, E_{EF})^{1/2} \, \underbrace{  \left( \frac{  E_{CF} \, E_{DE}  }{E_{CE} \, E_{DF}}\right)^{1/2} }_{= \, 1 \, + \, \mathcal{O}(z_{EF})} \, \left(\prod_{i \leq j}^M E_{2i-1,2j}\,\prod_{\bi<\bj}^{M}  E_{2\bi,2\bj-1}\right)^{1/2}  \notag \\
  & \ \ \times \ \left(\prod_{k<l}^M E_{2k-1,2l-1}\, E_{2k,2l}\right)^{-1/2} \! \!  \frac{\vep_{\dga \dde} \, \vep_{\dka \dom}}{E_{CD} \, E_{EF}} \; \left[ \tspin \left( \frac{1}{2} \, \sum_{i=1}^M \smallint^{z_{2i-1}}_{z_{2i}} \ve{\om} \, \pm \, \frac{1}{2} \smallint^{z_{C}}_{z_{D}} \ve{\om}  \right) \ + \ \mathcal{O}(z_{EF})\right]^{2-M}  \notag \\
  & \ \ \times \ \sum_{\rho \in S_M}\te{sgn}(\rho) \, \prod_{m=1}^M\frac{\vep_{\al_{2m-1}\al_{\rho(2m)}}}{E_{2m-1,\rho(2m)}} \; \left[ \tspin \left( \frac{1}{2} \, \sum_{i=1}^M \smallint^{z_{2i-1}}_{z_{2i}} \ve{\om} \, - \smallint^{z_{2m-1}}_{z_{\rho(2m)}} \ve{\om} \, \pm \, \frac{1}{2} \smallint^{z_{C}}_{z_{D}} \ve{\om} \right) \ + \ \mathcal{O}(z_{EF}) \right] \notag \\
  &= \ \ \vep_{\dka \dom} \, \underbrace{E_{EF}^{-1/2}}_{z_{EF}^{-1/2} \, + \, \mathcal{O}(z_{EF})} \; \frac{ (-1)^M}{\bigl[\tspin(\ve{0}) \bigr]^2} \; \left[ \tspin \left( \frac{1}{2} \, \sum_{i=1}^M \smallint^{z_{2i-1}}_{z_{2i}} \ve{\om} \, \pm \, \frac{1}{2} \smallint^{z_{C}}_{z_{D}} \ve{\om} \right) \right]^{2-M} \; \frac{\vep_{\dga \dde}}{E_{CD}^{1/2}} \notag \\
 & \ \ \times \ \bigg(\prod_{i \leq j}^M E_{2i-1,2j}\,\prod_{\bi<\bj}^{M}  E_{2\bi,2\bj-1}\bigg)^{1/2} \, \left(\prod_{k<l}^M E_{2k-1,2l-1}\, E_{2k,2l}\right)^{-1/2} \, \sum_{\rho \in S_M}\te{sgn}(\rho) \notag \\
  & \ \ \times \
  \prod_{m=1}^M\frac{\vep_{\al_{2m-1}\al_{\rho(2m)}}}{E_{2m-1,\rho(2m)}} \; \tspin \left( \frac{1}{2} \, \sum_{i=1}^M \smallint^{z_{2i-1}}_{z_{2i}} \ve{\om} \, - \smallint^{z_{2m-1}}_{z_{\rho(2m)}} \ve{\om} \, \pm \, \frac{1}{2} \smallint^{z_{C}}_{z_{D}} \ve{\om} \right) \ + \ \mathcal{O}(z_{EF})
\end{align}
\end{itemize}
Strictly speaking, it is also necessary to check the limits $z_{2M-1} \rightarrow z_{2M}$ of (\ref{(2M,0)}) and (\ref{(2M,2)}) as well as $z_C \rightarrow z_D$ of (\ref{(2M,2)}), but this is an easy exercise compared to the steps shown and will therefore not be displayed explicitly.

\subsection{NS fermions and two spin fields}
\label{sec:appD}

The proof of equations (\ref{Omega}) and (\ref{omega}) for correlators with two spin fields is carried out in this subsection. They naturally hold in the simple cases $n=2,3$ because these lower order examples were the starting point for guessing the general formulae. To check the nontrivial statements at $n \geq 4$, one first of all has to verify the periodicity properties to be correct. Just as in the proof \ref{sec:appC}, the $\al$ cycles do not cause any trouble. As to their $\be$ cousins, the corollary (\ref{trans3}) of (\ref{per5}) will be helpful in addition to (\ref{per8}) and (\ref{trans2}). Let us restrict our attention to the truly inequivalent $z_i$ cases and pick out a generic term from the $\ell$- and $\rho$ sums:
\begin{itemize}
\item $z_{\rho(1)} \mapsto z_{\rho(1)} + \be_J$ in $\Om_{(n)}$
\[
\left. \begin{array}{rc} E_{\rho(1),k}\te{'s} \ : &
  \exp \left[ i\pi \, \Om_{JJ} \, + \,  i \pi \, \left( \smallint^{z_{\rho(1)}}_{z_{A}} \om_J \right) \, + \,  i \pi \, \left( \smallint^{z_{\rho(1)}}_{z_{B}} \om_J \right) \right] \\ \Theta \spin \te{'s} \ : &
  \exp \left[ - \, i\pi \, \Om_{JJ} \, - \,  i \pi \, \left( \smallint^{z_{\rho(1)}}_{z_{A}} \om_J \right) \, - \,  i \pi \, \left( \smallint^{z_{\rho(1)}}_{z_{B}} \om_J \right) \right]  \end{array} \right\}  \ \ \rightarrow \ 1 \ \ \te{in total}
\]
\item $z_{\rho(2n-1)} \mapsto z_{\rho(2n-1)} + \be_J$ in $\Om_{(n)}$
\[
\left. \begin{array}{rc} E_{\rho(2n-1),k}\te{'s} \ : &
  \exp \left[  i\pi \, \Om_{JJ} \, - \,  2 \pi i \, \left( \smallint^{z_{\rho(2n-2)}}_{z_{\rho(2n-1)}} \om_J \right) \, - \,  i \pi \, \left( \smallint^{z_{A}}_{z_{B}} \om_J \right) \right] \\ \Theta \spin \te{'s} \ : &
  \exp \left[ - \, i\pi \, \Om_{JJ} \, + \,  2 \pi i \, \left( \smallint^{z_{\rho(2n-2)}}_{z_{\rho(2n-1)}} \om_J \right) \, + \,  i \pi \, \left( \smallint^{z_{A}}_{z_{B}} \om_J \right) \right]  \end{array} \right\}  \ \ \rightarrow \ 1 \ \ \te{in total}
\]
\item $z_{A} \mapsto z_{A} + \be_J$ in $\Om_{(n)}$
\[
\left. \begin{array}{rc} E_{A,k}\te{'s} \ : &
  \exp \left[  \frac{i\pi \, \Om_{JJ}}{2} \, - \,  \frac{i\pi \, (2\ell-1)}{2} \, \smallint^{z_{A}}_{z_{B}} \om_J \, - \, \frac{i\pi}{2} \sum_{i=1}^{2\ell+1}  \left( \smallint^{z_{\rho(i)}}_{z_{A}} \om_J \, + \, \smallint^{z_{\rho(i)}}_{z_{B}} \om_J \right) \right] \\ & \times \exp \left[  i \pi \, \sum_{j=1}^{n-\ell-1} \smallint^{z_{\rho(2\ell+2j)}}_{z_{\rho(2\ell+2j+1)}} \om_J   \right] \\ \Theta \spin \te{'s} \ : &
  \exp \left[ - \, \frac{i\pi \, \Om_{JJ}}{2} \, + \,  \frac{i\pi \, (2\ell-1)}{2} \, \smallint^{z_{A}}_{z_{B}} \om_J \, + \, \frac{i\pi}{2} \sum_{i=1}^{2\ell+1}  \left( \smallint^{z_{\rho(i)}}_{z_{A}} \om_J \, + \, \smallint^{z_{\rho(i)}}_{z_{B}} \om_J \right) \right] \\ & \times \exp \left[  - \,  i \pi \, \sum_{j=1}^{n-\ell-1} \smallint^{z_{\rho(2\ell+2j)}}_{z_{\rho(2\ell+2j+1)}} \om_J   \right]  \end{array} \right\}  \ \ \rightarrow \ 1\]
\item $z_{\rho(1)} \mapsto z_{\rho(1)} + \be_J$ in $\om_{(n)}$
\[
\left. \begin{array}{rc} E_{\rho(1),k}\te{'s} \ : &
  \exp \left[ i\pi \, \Om_{JJ} \, + \,  i \pi \, \left( \smallint^{z_{\rho(1)}}_{z_{A}} \om_J \right) \, + \,  i \pi \, \left( \smallint^{z_{\rho(1)}}_{z_{B}} \om_J \right) \right] \\ \Theta \spin \te{'s} \ : &
  \exp \left[ - \, i\pi \, \Om_{JJ} \, - \,  i \pi \, \left( \smallint^{z_{\rho(1)}}_{z_{A}} \om_J \right) \, - \,  i \pi \, \left( \smallint^{z_{\rho(1)}}_{z_{B}} \om_J \right) \right]  \end{array} \right\}  \ \ \rightarrow \ 1 \ \ \te{in total}
\]
\item $z_{\rho(2n-2)} \mapsto z_{\rho(2n-2)} + \be_J$ in $\om_{(n)}$
\[
\left. \begin{array}{rc} E_{\rho(2n-2),k}\te{'s} \ : &
  \exp \left[  i\pi \, \Om_{JJ} \, - \,  2 \pi i \, \left( \smallint^{z_{\rho(2n-3)}}_{z_{\rho(2n-2)}} \om_J \right) \, - \,  i \pi \, \left( \smallint^{z_{A}}_{z_{B}} \om_J \right) \right] \\ \Theta \spin \te{'s} \ : &
  \exp \left[ - \, i\pi \, \Om_{JJ} \, + \,  2 \pi i \, \left( \smallint^{z_{\rho(2n-3)}}_{z_{\rho(2n-2)}} \om_J \right) \, + \,  i \pi \, \left( \smallint^{z_{A}}_{z_{B}} \om_J \right) \right]  \end{array} \right\}  \ \ \rightarrow \ 1 \ \ \te{in total}
\]
\item $z_{A} \mapsto z_{A} + \be_J$ in $\om_{(n)}$
\[
\left. \begin{array}{rc} E_{A,k}\te{'s} \ : &
  \exp \left[  \frac{i\pi \, \Om_{JJ}}{2} \, - \, i\pi \, (\ell-1) \smallint^{z_{A}}_{z_{B}} \om_J \, - \, \frac{i\pi}{2} \sum_{i=1}^{2\ell}  \left( \smallint^{z_{\rho(i)}}_{z_{A}} \om_J \, + \, \smallint^{z_{\rho(i)}}_{z_{B}} \om_J \right) \right] \\ & \times \exp \left[  i \pi \, \sum_{j=1}^{n-\ell-1} \smallint^{z_{\rho(2\ell+2j-1)}}_{z_{\rho(2\ell+2j)}} \om_J   \right] \\ \Theta \spin \te{'s} \ : &
  \exp \left[ - \, \frac{i\pi \, \Om_{JJ}}{2} \, + \,  i\pi \, (\ell-1) \smallint^{z_{A}}_{z_{B}} \om_J \, + \, \frac{i\pi}{2} \sum_{i=1}^{2\ell}  \left( \smallint^{z_{\rho(i)}}_{z_{A}} \om_J \, + \, \smallint^{z_{\rho(i)}}_{z_{B}} \om_J \right) \right] \\ & \times \exp \left[  - \,  i \pi \, \sum_{j=1}^{n-\ell-1} \smallint^{z_{\rho(2\ell+2j-1)}}_{z_{\rho(2\ell+2j)}} \om_J   \right]  \end{array} \right\}  \ \ \rightarrow \ 1
\]
\end{itemize}
Let us next turn to the singularities and show them to be consistent with the OPEs (\ref{rv,1a}) to (\ref{rv,2b}), again by induction. The induction step is much more involved than for the spin field correlators: We show that $\Om_{(n)}$ and $\om_{(n)}$ have the correct behaviour for $z_{i} \mto z_{j}$ on basis of the induction hypothesis that the general formulae hold for $\Om_{(n-1)}$ and $\om_{(n-1)}$.

\subsubsection[An auxiliary correlator: $2n$ NS fields]{An auxiliary correlator: $\bm{2n}$ NS fields}
\label{sec:ThePureNSCorrelator}

Due to the OPEs (\ref{rv,1b}), (\ref{rv,1d}), (\ref{rv,1e}) all spin fields vanish from (\ref{Omega}) and (\ref{omega}) in the limit $z_{A} \mto z_{B}$. Both $\Om_{(n)}$ and $\om_{(n+1)}$ leave a $2n$ point function with NS fields only,
\beq
\Psi_{(n)}^{\mu_{1} ... \mu_{2n}} (z_{1},...,z_{2n}) \ \ := \ \ \langle \psi^{\mu_{1}}(z_{1}) \, \psi^{\mu_{2}}(z_{2}) \, ... \, \psi^{\mu_{2n}}(z_{2n}) \rangle \spin \ ,
\label{npt,9}
\eeq
which is important to know for general $n$ in the following. For $n=1$ (\ref{npt,9}) reduces to the standard Szeg\"o kernel:
\beq
\Psi_{(1)}^{\mu_{1} \mu_{2}}(z_{1},z_{2}) \ \ = \ \ \langle \psi^{\mu_{1}}(z_{1}) \, \psi^{\mu_{2}}(z_{2}) \rangle \spin \eq \frac{\eta^{\mu_{1} \mu_{2}} \, \tspin \left( \smallint^{z_1}_{z_2} \ve{\om} \right) }{E_{12} \, \tspin \bigl(\ve{0} \bigr)}\ , \label{npt,10a} 
\eeq
In the absence of spin fields, $\psi^\mu$ is a free field. Hence, we can use Wick's theorem to reduce any (even) higher order correlator $\Psi_{(n)}$ to antisymmetrized products of two point functions (\ref{npt,10a}), e.g.
\begin{align}
\Psi_{(2)}^{\mu_{1} \mu_{2} \mu_3 \mu_4}&(z_{1},z_{2},z_3,z_4) \ \ = \ \ \langle \psi^{\mu_{1}}(z_{1}) \, \psi^{\mu_{2}}(z_{2}) \, \psi^{\mu_{3}}(z_{3}) \, \psi^{\mu_{4}}(z_{4}) \rangle \spin \notag \\
&= \ \ \langle \psi^{\mu_{1}}(z_{1}) \, \psi^{\mu_{2}}(z_{2}) \rangle \spin \, \langle \psi^{\mu_{3}}(z_{3}) \, \psi^{\mu_{4}}(z_{4}) \rangle \spin \ - \ \langle \psi^{\mu_{1}}(z_{1}) \, \psi^{\mu_{3}}(z_{3}) \rangle \spin \, \langle \psi^{\mu_{2}}(z_{2}) \, \psi^{\mu_{4}}(z_{4}) \rangle \spin \notag \\
& \ \ \ \ \ \ \ \ \ \ \ \ \ + \ \langle \psi^{\mu_{1}}(z_{1}) \, \psi^{\mu_{4}}(z_{4}) \rangle \spin \, \langle \psi^{\mu_{2}}(z_{2}) \, \psi^{\mu_{3}}(z_{3}) \rangle \spin \ .
\label{npt,10}
\end{align}
The decomposition of the $(2n)$ point function into pairwise contractions $\frac{\eta^{\mu_{i} \mu_{j}} \, \tspin \left( \smallint^{z_i}_{z_j} \ve{\om} \right) }{E_{ij} \, \tspin \bigl(\ve{0} \bigr)}$ can be written just like the $\ell = 0$ term (i.e. the $\si$ free piece) of the corresponding $\rho \in S_{2n} / {\cal Q}_{n+1,\ell}$ sum of the $\om_{(n+1)}$ correlator. The $\Psi_{(n)}$ can thus be expressed in the notation of subsection \ref{sec:gen} as
\beq
\Psi_{(n)}^{\mu_{1} ... \mu_{2n}} (z_{1},...,z_{2n}) \eq \! \! \! \sum_{\rho \in S_{2n} / {\cal Q}_{n+1,0}} \! \! \te{sgn}(\rho) \prod_{j=1}^{n} \frac{ \eta^{\mu_{\rho(2j-1)} \mu_{\rho(2j)}} \, \tspin \left( \smallint^{z_{\rho(2j-1)} }_{z_{\rho(2j)}} \ve{\om} \right) }{ E_{\rho(2j-1),\rho(2j)} \, \tspin \bigl(\ve{0} \bigr)} \ .
\label{npt,11}
\eeq

\subsubsection{The web of limits}
\label{sec:TheWebOfLimits}

We have just given a closed formula for $\Psi_{(n)}$ in terms of a $S_{2n} / {\cal Q}_{n+1,0}$ sum which is relevant for the limits $z_{A}\mto z_{B}$ of both $\Om_{(n)}$ and $\om_{(n)}$. The asymptotic behaviour of the correlators for $\psi^\mu$ arguments approaching each other does not only relate $\Om_{(n)} \leftrightarrow \Om_{(n-1)}$ and $\om_{(n)} \leftrightarrow \om_{(n-1)}$ but also yields connections between correlators of different type such as $\Om_{(n)}\leftrightarrow \om_{(n)}$ and $\om_{(n)} \leftrightarrow \Om_{(n-1)}$. Due to the OPE \eqref{rv,2a}, the limits $z_{2n-1} \mto z_{A}$ or $z_{1} \mto z_{B}$ intertwine the $\Om_{(n)}$ sequence with its $\om_{(n)}$ cousin. So one cannot prove one of the equations (\ref{Omega}) and (\ref{omega}) separately.

\begin{figure}[H]
    \centering
\includegraphics[width=0.35\textwidth]{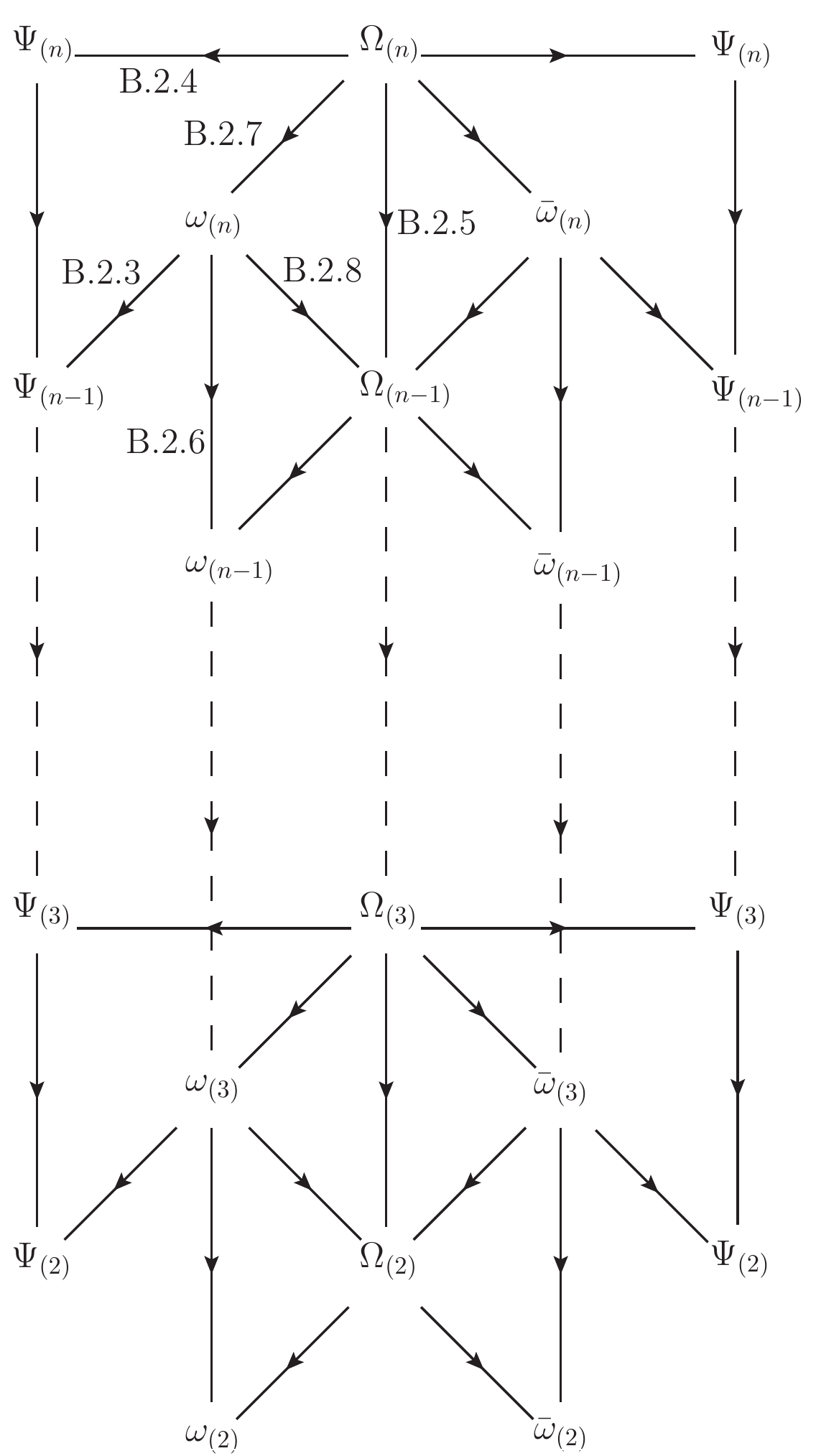}
\caption{Steps necessary for completing the proof by induction}
\label{ind}
\end{figure}

\medskip
There is a web of limiting processes which we have to examine for a complete proof by induction. For completeness, one should also mention the right handed copy of $\om_{(n)}$,
\beq
\bar{\om}_{(n)}^{\mu_1 ... \mu_{2n-2}} \, \! _{\dal \dbe} \ \ := \ \ \langle \psi^{\mu_{1}}(z_{1}) \, \psi^{\mu_{2}}(z_{2}) \, ... \, \psi^{\mu_{2n-2}}(z_{2n-2}) \, S_{\dal}(z_{A}) \, S_{\dbe}(z_{B}) \rangle \spin \ ,
\label{omegabar}
\eeq
which only differs from $\om_{(n)}$ in the overall sign and $(\si^{\mu_{i}} ... \bar{\si}^{\mu_j} \vep)_{\al \be} \mapsto (\vep \bar{\si}^{\mu_{i}} ... \si^{\mu_j} )_{\dal \dbe}$. Figure \ref{ind} summarizes the web of limits including the right handed $\bar{\om}_{(n)}$.
\noindent
The following subsections will verify the consistency of our central results (\ref{Omega}) as well as (\ref{omega}) with every single relation between various $\Om_{(n)}$, $\om_{(n)}$ and $\Psi_{(n)}$ due to the OPEs given in subsection \ref{sec:D=4}. Right handed analogues (e.g. $\Om_{(n)} \rightarrow \bar{\om}_{(n)}$ in addition to $\Om_{(n)} \rightarrow \om_{(n)}$) are suppressed since they can be copied almost literally.

\subsubsection{$\om_{(n)} \ra\Psi_{(n-1)}$ by 
$z_{A} \mto z_{B}$ limit}
\label{sec:i}

Using the OPE \eqref{rv,1d} one finds the following leading $z_{A} \mto z_{B}$ behaviour for the $2n$ point correlators $\om_{(n)}$:
\begin{align}
\om_{(n)}^{\mu_{1} ... \mu_{2n-2}}\, _{\al \be} (z_{i}) \Bigl. \Bigr| \begin{smallmatrix} z_{A} \\ \downarrow \\ z_{B} \end{smallmatrix} \ \ &\stackrel{!}{=} \ \ \frac{- \, \vep_{\al \be}}{(z_{A} \, - \, z_{B})^{1/2}} \; \langle \psi^{\mu_{1}}(z_{1}) \, ... \, \psi^{\mu_{2n-2}} (z_{2n-2}) \rangle \spin \ + \ {\cal O}(z_{AB}) \notag \\
&= \ \  \frac{- \, \vep_{\al \be}}{z_{AB}^{1/2}} \; \Psi^{\mu_{1}...\mu_{2n-2}}_{(n-1)} (z_{i}) \ + \ {\cal O}(z_{AB}) \notag \\
&= \ \ \frac{- \, \vep_{\al \be}}{z_{AB}^{1/2}} \! \sum_{\rho \in S_{2n-2} / {\cal Q}_{n,0}} \! \! \te{sgn}(\rho) \prod_{j=1}^{n-1} \frac{ \eta^{\mu_{\rho(2j-1)} \mu_{\rho(2j)}} \, \tspin \left( \smallint^{z_{\rho(2j-1)} }_{z_{\rho(2j)}} \ve{\om} \right) }{ E_{\rho(2j-1),\rho(2j)} \, \tspin \bigl(\ve{0} \bigr)}  \ + \ {\cal O}(z_{AB}) 
\label{npt,14}
\end{align}
If we start from the expression (\ref{omega}) and isolate the highest $E_{AB}^{-1}$ powers, the sum over $\ell$ breaks down to the $\ell=0$ term where $\bigl(\si^{\mu_{\rho(1)}} ... \bar{\si}^{\mu_{\rho(2\ell)}} \vep \bigr)_{\al \be} \bigl. \bigr|_{\ell =0} = \vep_{\al \be}$,
\begin{align}
&\om_{(n)}^{\mu_{1} ... \mu_{2n-2}}\,_{\al \be} (z_{i}) \Bigl. \Bigr| \begin{smallmatrix} z_{A} \\ \downarrow \\ z_{B} \end{smallmatrix} \ \ \sim \ \ \frac{- \, \left[ \tspin \left( \tfrac{1}{2} \smallint^{z_A} _{z_B} \ve{\om} \right) \right]^{3-n}}{E_{AB}^{1/2} \bigl[ \tspin ( \ve{0} ) \bigr]^2} \; \prod_{i=1}^{2n-2} (E_{iA} \, E_{iB})^{-1/2} \, \sum_{\ell = 0}^{n-1} \, \biggl( \frac{E_{AB}}{2 \, \tspin \left( \tfrac{1}{2} \smallint^{z_A}_{z_B} \ve{\om} \right)} \biggr)^{\ell} \, \de_{\ell,0} \notag \\
& \ \ \ \ \  \ \ \times \, \sum_{\rho \in S_{2n-2}/{\cal Q}_{n,\ell}} \! \! \!  \te{sgn}(\rho) \, \bigl(\si^{\mu_{\rho(1)}} \, \bar{\si}^{\mu_{\rho(2)}} \, ... \, \si^{\mu_{\rho(2\ell-1)}} \, \bar{\si}^{\mu_{\rho(2\ell)}} \, \vep \bigr)_{\al \be} \,
\prod_{k=1}^{2\ell} \tspin \left( \tfrac{1}{2} \smallint^{z_A} _{z_{\rho(k)}} \ve{\om} \, + \, \tfrac{1}{2} \smallint^{z_B} _{z_{\rho(k)}} \ve{\om} \right) \notag \\
& \ \ \ \ \  \ \ \times \,  \prod_{j=1}^{n-\ell-1} \frac{\eta^{\mu_{\rho(2\ell+2j-1)} \mu_{\rho(2\ell+2j)}}}{E_{\rho(2\ell+2j-1),\rho(2\ell+2j)} } \; E_{\rho(2\ell +2j-1),A} \, E_{\rho(2\ell +2j),B} \, \tspin \left( \smallint^{z_{ \rho(2\ell+2j-1) }}_{z_{ \rho(2\ell+2j) }} \ve{\om} \, + \, \tfrac{1}{2} \smallint^{z_A}_{z_B} \ve{\om} \right) \ + \ {\cal O}(z_{AB}) \notag \\
&= \ \ \frac{- \, \vep_{\al \be}}{E_{AB}^{1/2}} \, \left( \left[ \tspin \bigl( \ve{0} \bigr) \right]^{1-n} \ + \ {\cal O}(z_{AB}) \right) \! \sum_{\rho \in S_{2n-2}/{\cal Q}_{n,0}} \! \! \!  \te{sgn}(\rho) \, \prod_{j=1}^{n-1} \frac{\eta^{\mu_{\rho(2j-1)} \mu_{\rho(2j)}}}{E_{\rho(2j-1) ,\rho(2j)}}   \notag \\
& \ \ \ \ \  \ \ \times \, \underbrace{\frac{E_{\rho(2j-1),A} \, E_{\rho(2j),B} }{\bigl( E_{\rho(2j-1),A} \, E_{\rho(2j-1),B} \, E_{\rho(2j),A} \, E_{\rho(2j),B} \bigr)^{1/2} }}_{= \, 1 \ + \ {\cal O}(z_{AB}) } \,  \left[ \tspin \left( \smallint^{z_{ \rho( 2j-1) }}_{z_{ \rho(2j) }} \ve{\om}  \right) \ + \ {\cal O}(z_{AB}) \right]  \ + \ {\cal O}(z_{AB})  \notag \\
&= \ \ - \, \vep_{\al \be} \, \underbrace{  E_{AB}^{-1/2} }_{z_{AB}^{-1/2} \, + \, {\cal O}(z_{AB})} \! \sum_{\rho \in S_{2n-2} / {\cal Q}_{n,0}} \! \! \te{sgn}(\rho) \prod_{j=1}^{n-1} \frac{ \eta^{\mu_{\rho(2j-1)} \mu_{\rho(2j)}} \, \tspin \left( \smallint^{z_{\rho(2j-1)} }_{z_{\rho(2j)}} \ve{\om} \right) }{ E_{\rho(2j-1),\rho(2j)} \, \tspin \bigl(\ve{0} \bigr)}  \ + \ {\cal O}(z_{AB}) \ .
\label{npt,15}
\end{align}
Going to the third line we have rearranged the $\prod_{i=1}^{2n-2}(E_{iA}  E_{iB})^{-1/2}$ product such that the cancellation with the $E_{\rho(2j-1),A}  E_{\rho(2j),B}$ numerators in the $z_{A} \mto z_{B}$ limit becomes obvious. 

\medskip
The argument for $\bar{\om}_{(n)}$ is completely analogous. The agreement (\ref{npt,14}) $\leftrightarrow$ (\ref{npt,15}) in two ways of evaluating the $z_{A} \mto z_{B}$ asymptotics allows to proceed to the next arrow in figure \ref{ind}.

\subsubsection{$\Om_{(n)} \ra \Psi_{(n)}$ by $z_{A} \mto z_{B}$ limit}
\label{sec:ii}

The relevant OPE for this case is \eqref{rv,1b}:
\begin{align}
\Om_{(n)}&^{\mu_{1} ... \mu_{2n-1}}\,  _{\al \dbe} (z_{i}) \Bigl. \Bigr| \begin{smallmatrix} z_{A} \\ \downarrow \\ z_{B} \end{smallmatrix} \ \ \stackrel{!}{=} \ \ \frac{1}{\sqrt{2}} \; (\si_{\mu_{2n}})_{ \al \dbe} \, \langle \psi^{\mu_{1}}(z_{1}) \, ... \, \psi^{\mu_{2n-1}} (z_{2n-1}) \, \psi^{\mu_{2n}}(z_{B}) \rangle \spin \ + \ {\cal O}(z_{AB}) \notag \\
&= \ \ \frac{(\si_{\mu_{2n}})_{ \al \dbe}}{\sqrt{2}} \;  \Psi^{\mu_{1}...\mu_{2n}}_{(n)} (z_{i}) \Bigl. \Bigr|_{z_{2n}=z_{B}} \ + \ {\cal O}(z_{AB}) \notag \\
&= \ \ \frac{(\si_{\mu_{2n}})_{ \al \dbe}}{\sqrt{2}} \;  \! \sum_{\rho \in S_{2n} / {\cal Q}_{n+1,0}} \! \! \te{sgn}(\rho) \prod_{j=1}^{n} \frac{ \eta^{\mu_{\rho(2j-1)} \mu_{\rho(2j)}} \, \tspin \left( \smallint^{z_{\rho(2j-1)} }_{z_{\rho(2j)}} \ve{\om} \right) }{ E_{\rho(2j-1),\rho(2j)} \, \tspin \bigl(\ve{0} \bigr)} \Bigl. \Bigr|_{z_{2n}=z_{B}} \ + \ {\cal O}(z_{AB}) \notag \\
&= \ \ \frac{1}{\sqrt{2}}   \! \sum_{\rho \in S_{2n} / {\cal Q}_{n+1,0}} \! \! \te{sgn}(\rho) \;  \frac{ \overbrace{ (\si_{\mu_{2n}})_{ \al \dbe} \, \eta^{\mu_{\rho(2j_0-1)} \mu_{2n}} }^{\left(\si^{\mu_{\rho(2j_0-1)}} \right)_{ \al \dbe}} \, \tspin \left( \smallint^{z_{\rho(2j_0-1)} }_{z_{B}} \ve{\om} \right) }{ E_{\rho(2j_0-1),B} \, \tspin \bigl(\ve{0} \bigr)} \notag \\
& \ \ \ \ \ \ \ \ \ \ \ \ \times \ \prod_{j=1 \atop {j\neq j_{0} \atop{ \rho(2j_{0}) = 2n}}}^{n}  \frac{ \eta^{\mu_{\rho(2j-1)} \mu_{\rho(2j)}} \, \tspin \left( \smallint^{z_{\rho(2j-1)} }_{z_{\rho(2j)}} \ve{\om} \right) }{ E_{\rho(2j-1),\rho(2j)} \, \tspin \bigl(\ve{0} \bigr)} \ + \ {\cal O}(z_{AB}) \notag \\
&= \ \ \frac{1}{\sqrt{2}} \!  \sum_{\bar{\rho} \in S_{2n-1} / {\cal P}_{n,0}} \! \! \! \! \te{sgn}(\bar{\rho}) \; \frac{\si^{\mu_{\bar{\rho}(1)}}_{\al \dbe} \, \tspin \left( \smallint^{z_{\bar{\rho}(1)} }_{z_{B}} \ve{\om} \right)}{E_{\bar{\rho}(1),B} \, \tspin \bigl(\ve{0} \bigr)} \; \prod_{j=1 }^{n-1}  \frac{\eta^{\mu_{\bar{\rho}(2j)} \mu_{\bar{\rho}(2j+1)}} \, \tspin \left( \smallint^{z_{\bar{\rho}(2j)} }_{z_{\bar{\rho}(2j+1)}} \ve{\om} \right) }{E_{\bar{\rho}(2j),\bar{\rho}(2j+1)} \, \tspin \bigl(\ve{0} \bigr) } \ + \ {\cal O}(z_{AB})
\label{npt,17}
\end{align}
Since the index $\mu_{2n}$ is contracted from the third to the fourth line, one can sum over $S_{2n-1}$ subpermutations $\bar{\rho}$ acting on $(1,2,...,2n-1)$ instead of $\rho \in S_{2n}$. The number of terms is the same in both sums due to (\ref{npt,7a}) and (\ref{npt,7}),
\begin{align}
\bigl| S_{2n} / {\cal Q}_{n+1,0} \bigr| \ \ = \ \ \frac{(2n)!}{n! \, 2^{n}} \ \ &= \ \ \frac{2n \, (2n \, - \, 1)!}{n \, (n \, - \, 1)! \,2 \  2^{n-1}} \notag \\
= \ \ \frac{(2n \, - \, 1)!}{(n \, - \, 1)! \, 2^{n-1}} \ \ &= \ \ \bigl| S_{2n-1} / {\cal P}_{n,0} \bigr| \ ,
\label{npt,18}
\end{align}
and one can check with the help of restrictions (\ref{npt,4}), (\ref{npt,5}) that the last equality of (\ref{npt,17}) holds exactly.

\medskip
The right hand side of (\ref{npt,17}) can be reproduced directly from the expression (\ref{Omega}) for $\Om_{(n)}$: Keeping the most singular $z_{AB}$ dependences only again truncates the $\ell$ sum to the $\ell =0$ term, similar to (\ref{npt,15}):
\begin{align}
&\Om_{(n)}^{\mu_{1} ... \mu_{2n-1}}\,  _{\al \dbe} (z_{i}) \Bigl. \Bigr| \begin{smallmatrix} z_{A} \\ \downarrow \\ z_{B} \end{smallmatrix} \eq \frac{\left[ \tspin \left( \tfrac{1}{2} \smallint^{z_A} _{z_B} \ve{\om} \right) \right]^{2-n}}{\sqrt{2} \, \bigl[ \tspin ( \ve{0} ) \bigr]^2 \, \prod_{i=1}^{2n-1} (E_{iA} \, E_{iB})^{1/2} } \, \sum_{\ell = 0}^{n-1} \, \biggl( \frac{E_{AB}}{2 \, \tspin \left( \tfrac{1}{2} \smallint^{z_A} _{z_B} \ve{\om} \right)} \biggr)^{\ell} \, \de_{\ell,0} \notag \\
& \ \ \ \ \times \sum_{\rho \in S_{2n-1}/{\cal P}_{n,\ell}} \! \! \!  \te{sgn}(\rho) \, \bigl(\si^{\mu_{\rho(1)}} \, \bar{\si}^{\mu_{\rho(2)}} \, ... \, \bar{\si}^{\mu_{\rho(2\ell)}} \, \si^{\mu_{\rho(2\ell+1)}} \bigr)_{\al \dbe} \, \prod_{k=1}^{2\ell+1} \tspin \left( \tfrac{1}{2} \smallint^{z_A} _{z_{\rho(k)}} \ve{\om} \, + \, \tfrac{1}{2} \smallint^{z_B} _{z_{\rho(k)}} \ve{\om} \right) \notag \\
& \ \ \ \ \times \  \prod_{j=1}^{n-\ell-1} \frac{\eta^{\mu_{\rho(2\ell+2j)} \mu_{\rho(2\ell+2j+1)}}}{E_{\rho(2\ell+2j),\rho(2\ell+2j+1)} } \; E_{\rho(2\ell+2j),A} \, E_{\rho(2\ell+2j+1),B}  \, \tspin \left( \smallint^{z_{\rho(2\ell+2j)}}_{z_{\rho(2\ell+2j+1)}} \ve{\om} \, + \, \tfrac{1}{2} \smallint^{z_A} _{z_B} \ve{\om} \right) \ + \ {\cal O}(z_{AB}) \notag \\
&= \ \ \frac{1}{\sqrt{2}} \, \left( \left[ \tspin \bigl( \ve{0} \bigr) \right]^{-n}  + \, {\cal O}(z_{AB}) \right)
 \! \! \! \sum_{\rho \in S_{2n-1}/{\cal P}_{n,0}} \! \!   \;  \frac{\te{sgn}(\rho) \, \si^{\mu_{\rho(1)}}_{\al \dbe}}{ \underbrace{( E_{\rho(1),A} \, E_{\rho(1),B} )^{1/2} }_{E_{\rho(1),B}  \, + \, {\cal O}(z_{AB})} }   \, \underbrace{\tspin \left( \tfrac{1}{2} \smallint^{z_A} _{z_{\rho(1)}} \ve{\om} \, + \, \tfrac{1}{2} \smallint^{z_B} _{z_{\rho(1)}} \ve{\om} \right)}_{= \, \tspin \left(  \smallint^{z_B} _{z_{\rho(1)}} \ve{\om} \right) \, + \, {\cal O}(z_{AB})  } \notag \\
& \ \ \ \ \times \ \prod_{j=1}^{n-1} \frac{\eta^{\mu_{\rho(2j)} \mu_{\rho(2j+1)}}}{E_{\rho(2j) ,\rho(2j+1)}}  \; \underbrace{\frac{E_{\rho(2j),A} \, E_{\rho(2j+1),B} }{\bigl( E_{\rho(2j),A} \, E_{\rho(2j),B} \, E_{\rho(2j+1),A} \, E_{\rho(2j+1),B} \bigr)^{1/2} }}_{ = \, 1 \, + \, {\cal O}(z_{AB})  } \notag \\
& \ \ \ \ \times \ \left[ \tspin \left( \smallint^{z_{ \rho( 2j) }}_{z_{ \rho(2j+1) }} \ve{\om}  \right) \ + \ {\cal O}(z_{AB}) \right]  \ + \ {\cal O}(z_{AB}) \notag \\
&= \ \ \frac{1}{\sqrt{2}} \! \sum_{\rho \in S_{2n-1}/{\cal P}_{n,0}} \! \! \!  \te{sgn}(\rho) \; \frac{\si^{\mu_{\rho(1)}}_{\al \dbe} \, \tspin \left(  \smallint^{z_B} _{z_{\rho(1)}} \ve{\om} \right)}{ E_{\rho(1),B} \, \tspin \bigl( \ve{0} \bigr)} \; \prod_{j=1}^{n-1} \frac{\eta^{\mu_{\rho(2j)} \mu_{\rho(2j+1)}} \, \tspin \left( \smallint^{z_{ \rho( 2j) }}_{z_{ \rho(2j+1) }} \ve{\om}  \right)}{E_{\rho(2j),\rho(2j+1)} \, \tspin \bigl( \ve{0} \bigr)} \ + \ {\cal O}(z_{AB})
\label{npt,19}
\end{align}
Except for $\left( E_{\rho(1),A} E_{\rho(1),B}\right)^{-1/2} = E_{\rho(1),B}^{-1} + {\cal O}(z_{AB})$, the mechanisms are the same as in (\ref{npt,15}).

\subsubsection{$\Om_{(n)} \ra \Om_{(n-1)}$ by $z_{2n-2} \mto z_{2n-1}$ limit}
\label{sec:iii}

This subsection treats the situation when the arguments of two NS field in $\Om_{(n)}$ approach each other. The OPE \eqref{rv,1a} leaves the correlator $\Om_{(n-1)}$: 
\begin{align}
&\Om_{(n)}^{\mu_{1} ... \mu_{2n-1}}\,  _{\al \dbe} (z_{i}) \Bigl. \Bigr| \begin{smallmatrix} z_{2n-2} \\ \downarrow \\ z_{2n-1} \end{smallmatrix} \ \ \stackrel{!}{=} \ \ \frac{\eta^{\mu_{2n-2} \mu_{2n-1}}}{z_{2n-2,2n-1}} \; \langle \psi^{\mu_{1}}(z_{1}) \, ... \, \psi^{\mu_{2n-3}}(z_{2n-3}) \, S_{\al}(z_{A}) \, S_{\dbe}(z_{B}) \rangle \spin \ + \ {\cal O} (z_{2n-2,2n-1}) \notag \\
&= \ \ \frac{\eta^{\mu_{2n-2} \mu_{2n-1}}}{z_{2n-2,2n-1}} \; \Om_{(n-1)}^{\mu_{1} ... \mu_{2n-3}}\,  _{\al \dbe} (z_{i})  \ + \ {\cal O} (z_{2n-2,2n-1}) \notag \\
&= \ \ \frac{\eta^{\mu_{2n-2} \mu_{2n-1}}}{z_{2n-2,2n-1}} \; \frac{\left[ \tspin \left( \tfrac{1}{2} \smallint^{z_A} _{z_B} \ve{\om} \right) \right]^{3-n}}{\sqrt{2} \, \bigl[ \tspin ( \ve{0} ) \bigr]^2 \, \prod_{i=1}^{2n-3} (E_{iA} \, E_{iB})^{1/2} } \, \sum_{\ell = 0}^{n-2} \, \biggl( \frac{E_{AB}}{2 \, \tspin \left( \tfrac{1}{2} \smallint^{z_A} _{z_B} \ve{\om} \right)} \biggr)^{\ell} \sum_{\rho \in S_{2n-3}/{\cal P}_{n-1,\ell}} \! \! \!  \te{sgn}(\rho) \notag \\
& \ \ \ \ \times  \ \bigl(\si^{\mu_{\rho(1)}} \, \bar{\si}^{\mu_{\rho(2)}} \, ... \, \bar{\si}^{\mu_{\rho(2\ell)}} \, \si^{\mu_{\rho(2\ell+1)}} \bigr)_{\al \dbe} \, \prod_{k=1}^{2\ell+1} \tspin \left( \tfrac{1}{2} \smallint^{z_A} _{z_{\rho(k)}} \ve{\om} \, + \, \tfrac{1}{2} \smallint^{z_B} _{z_{\rho(k)}} \ve{\om} \right) \ \prod_{j=1}^{n-\ell-2} \frac{\eta^{\mu_{\rho(2\ell+2j)} \mu_{\rho(2\ell+2j+1)}}}{E_{\rho(2\ell+2j),\rho(2\ell+2j+1)} } \notag \\
& \ \ \ \ \times \  E_{\rho(2\ell+2j),A} \, E_{\rho(2\ell+2j+1),B}  \, \tspin \left( \smallint^{z_{\rho(2\ell+2j)}}_{z_{\rho(2\ell+2j+1)}} \ve{\om} \, + \, \tfrac{1}{2} \smallint^{z_A} _{z_B} \ve{\om} \right)
 \ + \ {\cal O} (z_{2n-2,2n-1})
\label{npt,20}
\end{align}
To bring the expression (\ref{Omega}) in its $z_{2n-2} \mto z_{2n-1}$ asymptotics into the form (\ref{npt,20}) one has to isolate the permutations $S_{2n-1} / {\cal P}_{n,\ell}$ which provide a factor $\frac{\eta^{\mu_{2n-2}    \mu_{2n-1}}}{E_{2n-2,2n-1}}$. A necessary condition for this to occur is $\ell \neq n-1$ because otherwise there would be no $\eta$'s at all. So we skip the $\ell = n-1$ term.

\medskip
Furthermore, the ordering of the $\eta$'s according to their second index (which is rephrased as $\rho(2\ell+3) <
\rho(2\ell+5) < ... < \rho(2n-1)$ in (\ref{npt,4})) makes sure that $\rho(2n-1) = 2n-1$. Consequently, $\eta^{\mu_{2n-2} \mu_{2n-1}}$ appears whenever $\rho(2n-2) = 2n-2$. These arguments lead to
\begin{align}
\Om_{(n)}&^{\mu_{1} ... \mu_{2n-1}}\,  _{\al \dbe} (z_{i}) \Bigl. \Bigr| \begin{smallmatrix} z_{2n-2} \\ \downarrow \\ z_{2n-1} \end{smallmatrix} \ \ = \ \ \frac{\left[ \tspin \left( \tfrac{1}{2} \smallint^{z_A} _{z_B} \ve{\om} \right) \right]^{2-n}}{\sqrt{2} \, \bigl[ \tspin ( \ve{0} ) \bigr]^2} \; \prod_{i=1}^{2n-1} (E_{iA} \, E_{iB})^{-1/2} \, \sum_{\ell = 0}^{n-2} \, \biggl( \frac{E_{AB}}{2 \, \tspin \left( \tfrac{1}{2} \smallint^{z_A} _{z_B} \ve{\om} \right)} \biggr)^{\ell}  \notag \\
& \ \ \ \ \  \ \ \times \ \sum_{\rho \in S_{2n-1}/{\cal P}_{n,\ell}} \! \! \!  \te{sgn}(\rho) \,  \de_{\rho(2n-2),2n-2} \, \bigl(\si^{\mu_{\rho(1)}} \, ... \, \si^{\mu_{\rho(2\ell+1)}} \bigr)_{\al \dbe} \, \prod_{k=1}^{2\ell+1} \tspin \left( \tfrac{1}{2} \smallint^{z_A} _{z_{\rho(k)}} \ve{\om} \, + \, \tfrac{1}{2} \smallint^{z_B} _{z_{\rho(k)}} \ve{\om} \right) \notag \\
& \ \ \ \ \  \ \ \times \ \prod_{j=1}^{n-\ell-1} \frac{\eta^{\mu_{\rho(2\ell+2j)} \mu_{\rho(2\ell+2j+1)}}}{E_{\rho(2\ell+2j),\rho(2\ell+2j+1)} } \; E_{\rho(2\ell+2j),A} \, E_{\rho(2\ell+2j+1),B} \, \tspin \left( \smallint^{z_{\rho(2\ell+2j)}}_{z_{\rho(2\ell+2j+1)}} \ve{\om} \, + \, \tfrac{1}{2} \smallint^{z_A} _{z_B} \ve{\om} \right) \notag \\
& \ \ \ \ \ \ \ \ \ \ \ \ \ \ \ \ \ \ \ \ \ \ \ + \  {\cal O} (z_{2n-2,2n-1}) \notag \\
&= \ \ \frac{\left[ \tspin \left( \tfrac{1}{2} \smallint^{z_A} _{z_B} \ve{\om} \right) \right]^{2-n}}{\sqrt{2} \, \bigl[ \tspin ( \ve{0} ) \bigr]^2} \; \prod_{i=1}^{2n-3} (E_{iA} \, E_{iB})^{-1/2} \, \sum_{\ell = 0}^{n-2} \, \biggl( \frac{E_{AB}}{2 \, \tspin \left( \tfrac{1}{2} \smallint^{z_A} _{z_B} \ve{\om} \right)} \biggr)^{\ell}  \notag \\
& \ \ \ \ \  \ \ \times \  \sum_{\rho \in S_{2n-1}/{\cal P}_{n,\ell}} \! \! \!  \te{sgn}(\rho) \, \de_{\rho(2n-2),2n-2}  \, \bigl(\si^{\mu_{\rho(1)}} \, ... \, \si^{\mu_{\rho(2\ell+1)}} \bigr)_{\al \dbe} \, \prod_{k=1}^{2\ell+1} \tspin \left( \tfrac{1}{2} \smallint^{z_A} _{z_{\rho(k)}} \ve{\om} \, + \, \tfrac{1}{2} \smallint^{z_B} _{z_{\rho(k)}} \ve{\om} \right) \notag \\
& \ \ \ \ \  \ \ \times \ \frac{\eta^{\mu_{2n-2} \mu_{2n-1}}}{E_{2n-2,2n-1}} \; \underbrace{\frac{E_{2n-2,A} \, E_{2n-1,B}}{\bigl(E_{2n-2,A} \, E_{2n-2,B} \, E_{2n-1,A} \, E_{2n-1,B} \bigr)^{1/2} }}_{= \, 1 \, + \,  {\cal O} (z_{2n-2,2n-1})} \, \underbrace{ \tspin \left( \smallint^{z_{2n-2}}_{z_{2n-1}} \ve{\om} \, + \, \tfrac{1}{2} \smallint^{z_A} _{z_B} \ve{\om} \right)}_{\tspin \left( \tfrac{1}{2} \smallint^{z_A} _{z_B} \ve{\om} \right) \, + \,  {\cal O} (z_{2n-2,2n-1})} \notag \\
& \ \ \ \ \  \ \ \times \ \prod_{j=1}^{n-\ell-2} \frac{\eta^{\mu_{\rho(2\ell+2j)} \mu_{\rho(2\ell+2j+1)}}}{E_{\rho(2\ell+2j),\rho(2\ell+2j+1)} } \; E_{\rho(2\ell+2j),A} \, E_{\rho(2\ell+2j+1),B} \, \tspin \left( \smallint^{z_{\rho(2\ell+2j)}}_{z_{\rho(2\ell+2j+1)}} \ve{\om} \, + \, \tfrac{1}{2} \smallint^{z_A} _{z_B} \ve{\om} \right) \notag \\
& \ \ \ \ \ \ \ \ \ \ \ \ \ \ \ \ \ \ \ \ \ \ \ + \  {\cal O} (z_{2n-2,2n-1}) \notag \\
&= \ \ \eta^{\mu_{2n-2} \mu_{2n-1}}  \! \! \! \! \! \! \! \! \underbrace{E_{2n-2,2n-1}^{-1}}_{z_{2n-2,2n-1}^{-1} \, + \, {\cal O} (z_{2n-2,2n-1})} \! \! \! \! \! \! \! \! \frac{\left[ \tspin \left( \tfrac{1}{2} \smallint^{z_A} _{z_B} \ve{\om} \right) \right]^{3-n}}{\sqrt{2} \, \bigl[ \tspin ( \ve{0} ) \bigr]^2} \; \prod_{i=1}^{2n-3} (E_{iA} \, E_{iB})^{-1/2} \, \sum_{\ell = 0}^{n-2} \, \biggl( \frac{E_{AB}}{2 \, \tspin \left( \tfrac{1}{2} \smallint^{z_A} _{z_B} \ve{\om} \right)} \biggr)^{\ell} \notag \\
& \ \ \ \ \ \ \ \times \  \sum_{\bar{\rho} \in S_{2n-3}/{\cal P}_{n-1,\ell}} \! \! \!  \te{sgn}(\bar{\rho}) \, \bigl(\si^{\mu_{\bar{\rho}(1)}} \, ... \, \si^{\mu_{\bar{\rho}(2\ell+1)}} \bigr)_{\al \dbe}  \, \prod_{k=1}^{2\ell+1} \tspin \left( \tfrac{1}{2} \smallint^{z_A} _{z_{\bar{\rho}(k)}} \ve{\om} \, + \, \tfrac{1}{2} \smallint^{z_B} _{z_{\bar{\rho}(k)}} \ve{\om} \right) \notag \\
& \ \ \ \ \  \ \ \times \ \prod_{j=1}^{n-\ell-2} \frac{\eta^{\mu_{\bar{\rho}(2\ell+2j)} \mu_{\bar{\rho}(2\ell+2j+1)}}}{E_{\bar{\rho}(2\ell+2j),\bar{\rho}(2\ell+2j+1)} } \; E_{\bar{\rho}(2\ell+2j),A} \, E_{\bar{\rho}(2\ell+2j+1),B} \, \tspin \left( \smallint^{z_{\bar{\rho}(2\ell+2j)}}_{z_{\bar{\rho}(2\ell+2j+1)}} \ve{\om} \, + \, \tfrac{1}{2} \smallint^{z_A} _{z_B} \ve{\om} \right) \notag \\
& \ \ \ \ \ \ \ \ \ \ \ \ \ \ \ \ \ \ \ \ \ \ \ + \  {\cal O} (z_{2n-2,2n-1}) \ .
\label{npt,21}
\end{align}
In the last step we have used that $S_{2n-1}/{\cal P}_{n,\ell}$ with $\left( \begin{smallmatrix} \rho(2n-1) \\
    \rho(2n-2) \end{smallmatrix} \right) = \left( \begin{smallmatrix} 2n-1 \\ 2n-2 \end{smallmatrix} \right)$ is equivalent to $S_{2n-3}/{\cal P}_{n-1,\ell}$, where the subpermutation $\bar{\rho} \in S_{2n-3}$ is of the same sign as the corresponding $\rho \in S_{2n-1}$.

\subsubsection{$\om_{(n)}  \ra \om_{(n-1)}$ by $z_{2n-3} \mto z_{2n-2}$ limit}
\label{sec:iv}

The singular behaviour of the $\om_{(n)}$ in two NS arguments can be studied in a similar fashion:
\begin{align}
&\om_{(n)}^{\mu_{1} ... \mu_{2n-2}}\,  _{\al  \be} (z_{i}) \Bigl. \Bigr| \begin{smallmatrix} z_{2n-3} \\ \downarrow \\ z_{2n-2} \end{smallmatrix} \ \ \stackrel{!}{=} \ \ \frac{\eta^{\mu_{2n-3} \mu_{2n-2}}}{z_{2n-3,2n-2}} \; \langle \psi^{\mu_{1}}(z_{1}) \, ... \, \psi^{\mu_{2n-4}}(z_{2n-4}) \, S_{\al}(z_{A}) \, S_{\be}(z_{B}) \rangle \spin \ + \  {\cal O} (z_{2n-3,2n-2}) \notag \\
&= \ \ \frac{\eta^{\mu_{2n-3} \mu_{2n-2}}}{z_{2n-3,2n-2}} \; \om_{(n-1)}^{\mu_{1} ... \mu_{2n-4}}\, _{\al  \be} (z_{i}) \ + \  {\cal O} (z_{2n-3,2n-2}) \notag \\
&= \ \   \frac{\eta^{\mu_{2n-3} \mu_{2n-2}}}{z_{2n-3,2n-2}} \; \frac{- \, \left[ \tspin \left( \tfrac{1}{2} \smallint^{z_A} _{z_B} \ve{\om} \right) \right]^{4-n}}{ \tspin ( \ve{0} ) \, \tspin ( \ve{0} ) \, E_{AB}^{1/2}  \, \prod_{i=1}^{2n-4} (E_{iA} \, E_{iB})^{1/2} } \, \sum_{\ell = 0}^{n-2} \, \biggl( \frac{E_{AB}}{2 \, \tspin \left( \tfrac{1}{2} \smallint^{z_A} _{z_B} \ve{\om} \right)} \biggr)^{\ell}  \! \! \! \sum_{\rho \in S_{2n-4}/{\cal Q}_{n-1,\ell}} \! \! \!  \te{sgn}(\rho) \notag \\
& \ \ \ \ \times  \, \bigl(\si^{\mu_{\rho(1)}} \, \bar{\si}^{\mu_{\rho(2)}} \, ... \, \bar{\si}^{\mu_{\rho(2\ell)} } \, \vep \bigr)_{\al \be} \, \prod_{k=1}^{2\ell} \tspin \left( \tfrac{1}{2} \smallint^{z_A} _{z_{\rho(k)}} \ve{\om} \, + \, \tfrac{1}{2} \smallint^{z_B} _{z_{\rho(k)}} \ve{\om} \right) \, \prod_{j=1}^{n-\ell-2} \frac{\eta^{\mu_{\rho(2\ell+2j-1)} \mu_{\rho(2\ell+2j)}}}{E_{\rho(2\ell+2j-1),\rho(2\ell+2j)} } \notag \\
& \ \ \ \ \times \  E_{\rho(2\ell+2j-1),A} \, E_{\rho(2\ell+2j),B}  \, \tspin \left( \smallint^{z_{\rho(2\ell+2j-1)}}_{z_{\rho(2\ell+2j)}} \ve{\om} \, + \, \tfrac{1}{2} \smallint^{z_A} _{z_B} \ve{\om} \right) \ + \  {\cal O} (z_{2n-3,2n-2}) 
\label{npt,22}
\end{align}
When performing the limit $z_{2n-3} \mto z_{2n-2}$ in (\ref{omega}) we have to focus on the permutations leading to a $\frac{\eta^{\mu_{2n-3} \mu_{2n-2}}}{E_{2n-3,2n-2}}$ factor: firstly exclude $\ell = n-1$ (since this would distribute all $^{\mu_{i}}$ indices among the $\si$ matrices) and secondly find the appropriate $S_{2n-2} / {\cal Q}_{n,\ell}$ permutations which really attach $\mu_{2n-3}$ and $\mu_{2n-2}$ to the same Minkowski metric. Equation (\ref{npt,5}) requires $\rho(2\ell +2) < \rho(2\ell+4)<...< \rho(2n-2)$, so the projector to leading $z_{2n-3,2n-2}$ singularities is simply $\de_{\rho(2n-3),2n-3}$.
\begin{align}
&\om_{(n)}^{\mu_{1} ... \mu_{2n-2}}\,  _{\al  \be} (z_{i}) \Bigl. \Bigr| \begin{smallmatrix} z_{2n-3} \\ \downarrow \\ z_{2n-2} \end{smallmatrix} \ \ = \ \ \frac{- \, \left[ \tspin \left( \tfrac{1}{2} \smallint^{z_A} _{z_B} \ve{\om} \right) \right]^{3-n}}{ \bigl[ \tspin ( \ve{0} )  \bigr]^2 \, E_{AB}^{1/2}  \, \prod_{i=1}^{2n-2} (E_{iA} \, E_{iB})^{1/2} } \, \sum_{\ell = 0}^{n-2} \, \biggl( \frac{E_{AB}}{2 \, \tspin \left( \tfrac{1}{2} \smallint^{z_A} _{z_B} \ve{\om} \right)} \biggr)^{\ell} \notag \\
& \ \ \ \ \times \sum_{\rho \in S_{2n-2}/{\cal Q}_{n,\ell}} \! \! \!  \te{sgn}(\rho) \, \de_{\rho(2n-3),2n-3} \, \bigl(\si^{\mu_{\rho(1)}} \, \bar{\si}^{\mu_{\rho(2)}} \, ... \, \bar{\si}^{\mu_{\rho(2\ell)} } \, \vep \bigr)_{\al \be} \, \prod_{k=1}^{2\ell} \tspin \left( \tfrac{1}{2} \smallint^{z_A} _{z_{\rho(k)}} \ve{\om} \, + \, \tfrac{1}{2} \smallint^{z_B} _{z_{\rho(k)}} \ve{\om} \right) \notag \\
& \ \ \ \ \times \  \prod_{j=1}^{n-\ell-1} \frac{\eta^{\mu_{\rho(2\ell+2j-1)} \mu_{\rho(2\ell+2j)}}}{E_{\rho(2\ell+2j-1),\rho(2\ell+2j)} } \; E_{\rho(2\ell+2j-1),A} \, E_{\rho(2\ell+2j),B}  \, \tspin \left( \smallint^{z_{\rho(2\ell+2j-1)}}_{z_{\rho(2\ell+2j)}} \ve{\om} \, + \, \tfrac{1}{2} \smallint^{z_A} _{z_B} \ve{\om} \right) \notag \\
& \ \ \ \ \ \ \ \ \ \ \ \ \ \ \ \ \ \ \ \ \ \ \ + \  {\cal O} (z_{2n-3,2n-2}) \notag \\
&= \ \ \frac{- \, \left[ \tspin \left( \tfrac{1}{2} \smallint^{z_A} _{z_B} \ve{\om} \right) \right]^{3-n}}{ \bigl[ \tspin ( \ve{0} )  \bigr]^2 \, E_{AB}^{1/2}  \, \prod_{i=1}^{2n-4} (E_{iA} \, E_{iB})^{1/2} } \, \sum_{\ell = 0}^{n-2} \, \biggl( \frac{E_{AB}}{2 \, \tspin \left( \tfrac{1}{2} \smallint^{z_A} _{z_B} \ve{\om} \right)} \biggr)^{\ell}    \notag \\
& \ \ \ \ \  \ \ \times \ \sum_{\rho \in S_{2n-2}/{\cal Q}_{n,\ell}} \! \! \!  \te{sgn}(\rho) \, \de_{\rho(2n-3),2n-3} \, \bigl(\si^{\mu_{\rho(1)}} \, ... \, \bar{\si}^{\mu_{\rho(2\ell)}} \, \vep \bigr)_{\al \be} \, \prod_{k=1}^{2\ell} \tspin \left( \tfrac{1}{2} \smallint^{z_A} _{z_{\rho(k)}} \ve{\om} \, + \, \tfrac{1}{2} \smallint^{z_B} _{z_{\rho(k)}} \ve{\om} \right) \notag \\
& \ \ \ \ \  \ \ \times \  \frac{\eta^{\mu_{2n-3} \mu_{2n-2}}}{E_{2n-3,2n-2}} \; \underbrace{\frac{E_{2n-3,A} \, E_{2n-2,B}}{\bigl(E_{2n-3,A} \, E_{2n-3,B} \, E_{2n-2,A} \, E_{2n-2,B} \bigr)^{1/2} }}_{= \, 1  \, + \,  {\cal O} (z_{2n-3,2n-2})} \, \underbrace{ \tspin \left( \smallint^{z_{2n-3}}_{z_{2n-2}} \ve{\om} \, + \, \tfrac{1}{2} \smallint^{z_A} _{z_B} \ve{\om} \right) }_{ \tspin \left(  \tfrac{1}{2} \smallint^{z_A} _{z_B} \ve{\om} \right) \, + \, {\cal O} (z_{2n-3,2n-2}) } \notag \\
& \ \ \ \ \times \  \prod_{j=1}^{n-\ell-2} \frac{\eta^{\mu_{\rho(2\ell+2j-1)} \mu_{\rho(2\ell+2j)}}}{E_{\rho(2\ell+2j-1),\rho(2\ell+2j)} } \; E_{\rho(2\ell+2j-1),A} \, E_{\rho(2\ell+2j),B}  \, \tspin \left( \smallint^{z_{\rho(2\ell+2j-1)}}_{z_{\rho(2\ell+2j)}} \ve{\om} \, + \, \tfrac{1}{2} \smallint^{z_A} _{z_B} \ve{\om} \right) \notag \\
& \ \ \ \ \ \ \ \ \ \ \ \ \ \ \ \ \ \ \ \ \ \ \ + \  {\cal O} (z_{2n-3,2n-2}) \notag \\
&= \ \ - \, \eta^{\mu_{2n-3} \mu_{2n-2}}  \! \! \! \! \! \! \! \! \underbrace{E_{2n-3,2n-2}^{-1}}_{z_{2n-3,2n-2}^{-1} \, + \, {\cal O} (z_{2n-3,2n-2})} \! \! \! \! \! \! \! \! \frac{\left[ \tspin \left( \tfrac{1}{2} \smallint^{z_A} _{z_B} \ve{\om} \right) \right]^{4-n}}{\sqrt{2} \, \bigl[ \tspin ( \ve{0} ) \bigr]^2 \, E_{AB}^{1/2}} \; \prod_{i=1}^{2n-4} (E_{iA} \, E_{iB})^{-1/2} \, \sum_{\ell = 0}^{n-2} \, \biggl( \frac{E_{AB}}{2 \, \tspin \left( \tfrac{1}{2} \smallint^{z_A} _{z_B} \ve{\om} \right)} \biggr)^{\ell} \notag \\
& \ \ \ \ \ \ \ \times \  \sum_{\bar{\rho} \in S_{2n-4}/{\cal Q}_{n-1,\ell}} \! \! \!  \te{sgn}(\bar{\rho}) \, \bigl(\si^{\mu_{\bar{\rho}(1)}} \, ... \, \bar{\si}^{\mu_{\bar{\rho}(2\ell)}} \, \vep \bigr)_{\al \be}  \, \prod_{k=1}^{2\ell} \tspin \left( \tfrac{1}{2} \smallint^{z_A} _{z_{\bar{\rho}(k)}} \ve{\om} \, + \, \tfrac{1}{2} \smallint^{z_B} _{z_{\bar{\rho}(k)}} \ve{\om} \right) \notag \\
& \ \ \ \ \  \ \ \times \ \prod_{j=1}^{n-\ell-2} \frac{\eta^{\mu_{\bar{\rho}(2\ell+2j-1)} \mu_{\bar{\rho}(2\ell+2j)}}}{E_{\bar{\rho}(2\ell+2j-1),\bar{\rho}(2\ell+2j)} } \; E_{\bar{\rho}(2\ell+2j-1),A} \, E_{\bar{\rho}(2\ell+2j),B} \, \tspin \left( \smallint^{z_{\bar{\rho}(2\ell+2j-1)}}_{z_{\bar{\rho}(2\ell+2j)}} \ve{\om} \, + \, \tfrac{1}{2} \smallint^{z_A} _{z_B} \ve{\om} \right) \notag \\
& \ \ \ \ \ \ \ \ \ \ \ \ \ \ \ \ \ \ \ \ \ \ \ + \  {\cal O} (z_{2n-3,2n-2}) 
\label{npt,23}
\end{align}
Having read the arguments for $\Om_{(n)} \ra \Om_{(n-1)}$ in Subsection \ref{sec:iii}, the reader might not be surprised about the $S_{2n-2}/{\cal Q}_{n,\ell}$ subset with $\rho(2n-3)=2n-3$ and $\rho(2n-2)=2n-2$ being equivalent to
$S_{2n-4}/{\cal Q}_{n-1,\ell}$.

\medskip
This analysis is easily extended to $\bar{\om} _{(n)} \ra \bar{\om}_{(n-1)}$, so the analogue of (\ref{npt,22}) and (\ref{npt,23}) will not be displayed explicitly.

\subsubsection{$\Om_{(n)} \ra \om_{(n)}$ by $z_{1} \mto z_{B}$ limit}
\label{sec:v}

Let us now turn to a more sophisticated limiting process where a right handed spin field is converted into a left handed one via OPE \eqref{rv,2b}:
\begin{align}
\Om_{(n)}&^{\mu_{1}... \mu_{2n-1}}\,_{\al \dbe}(z_{i}) \Bigl. \Bigr| \begin{smallmatrix} z_{1} \\ \downarrow \\ z_{B} \end{smallmatrix} \ \ \stackrel{!}{=} \ \ \frac{i \, (\si^{\mu_{1}})_{\ga \dbe} }{\sqrt{2} \, (z_{1 B})^{1/2}} \;  \, \langle \psi^{\mu_{2}}(z_{2}) \, ... \, \psi^{\mu_{2n-1}}(z_{2n-1}) \, S_{\al}(z_{A}) \, S^{\ga}(z_{B}) \rangle \spin \ + \ {\cal O}(z_{1B}) \notag \\
&= \ \ \frac{i}{\sqrt{2} \, (z_{1 B})^{1/2}} \; \om_{(n)}^{\mu_{2}...\mu_{2n-1}} \, _{\al} \, ^{\ga}(z_{i}) \, (\si^{\mu_{1}})_{\ga \dbe}  \ + \ {\cal O}(z_{1B}) \notag \\
&= \ \ \frac{\left[ \tspin \left( \tfrac{1}{2} \smallint^{z_A} _{z_B} \ve{\om} \right) \right]^{3-n}}{\sqrt{2} \, z_{1B}^{1/2} \, \bigl[ \tspin ( \ve{0} ) \bigr]^2}
\! \! \! \! \! \underbrace{\left( \frac{- \, i}{E_{AB}^{1/2}} \right)}_{= \, E_{1A}^{-1/2} \,+ \, {\cal O}(z_{1B})} \! \! \! \! \! \prod_{i=2}^{2n-1} (E_{iA} \, E_{iB})^{-1/2} \, \sum_{\ell = 0}^{n-1} \, \biggl( \frac{E_{AB}}{2 \, \tspin \left( \tfrac{1}{2} \smallint^{z_A} _{z_B} \ve{\om} \right)} \biggr)^{\ell} \! \! \! \! \sum_{\rho \in S_{2n-2}/{\cal Q}_{n,\ell}} \! \! \!  \te{sgn}(\rho)  \notag \\
& \ \ \ \ \ \times \, \bigl(\si^{\mu_{\rho(2)}} \,  ...  \, \bar{\si}^{\mu_{\rho(2\ell+1)}} \bigr)_{\al} \, ^{\ga} \, (\si^{\mu_{1}})_{\ga \dbe} \, \prod_{k=2}^{2\ell+1} \tspin \left( \tfrac{1}{2} \smallint^{z_A} _{z_{\rho(k)}} \ve{\om} \, + \, \tfrac{1}{2} \smallint^{z_B} _{z_{\rho(k)}} \ve{\om} \right) \, \prod_{j=1}^{n-\ell-1} \frac{\eta^{\mu_{\rho(2\ell+2j)} \mu_{\rho(2\ell+2j+1)}}}{E_{\rho(2\ell+2j),\rho(2\ell+2j+1)} } \notag \\
& \ \ \ \ \ \times \, E_{\rho(2\ell+2j),A} \, E_{\rho(2\ell+2j+1),B} \, \tspin \left( \smallint^{z_{\bar{\rho}(2\ell+2j)}}_{z_{\bar{\rho}(2\ell+2j+1)}} \ve{\om} \, + \, \tfrac{1}{2} \smallint^{z_A} _{z_B} \ve{\om} \right)  \ + \ {\cal O}(z_{1B})
\label{npt,24}
\end{align}
The factor $i$ originates from moving the field $\psi^{\mu_{1}}(z_{1})$ across $S_{\al}(z_{A})$ before applying the OPE of $\psi^{\mu_1}(z_1)$ with $S_{\dbe}(z_B)$. A rearrangement of the $\si$ matrices
\begin{align}
\si^{\la_{1}} \, \bar{\si}^{\la_{2}} \, ... \, &\si^{\la_{2n-1}} \, \bar{\si}^{\la_{2n}} \, \si^{\rho} \ \ = \ \ + \ \si^{\rho} \, \bar{\si}^{\la_{1}} \, \si^{\la_{2}} \, ... \, \bar{\si}^{\la_{2n-1}} \, \si^{\la_{2n}} \notag \\
& - \ 2  \, \sum_{r=1}^{n} \, \eta^{\rho \la_{2r}} \, \si^{\la_{1}} \, ... \, \si^{\la_{2r-1}} \, \bar{\si}^{\la_{2r+1}} \, ... \, \si^{\la_{2n}} \notag \\
& + \ 2 \, \sum_{r=1}^{n} \, \eta^{\rho \la_{2r-1}} \, \si^{\la_{1}} \, ... \, \bar{\si}^{\la_{2r-2}} \, \si^{\la_{2r}} \, ... \, \si^{\la_{2n}}
\label{roeven}
\end{align}
turns the result (\ref{npt,24}) into the following:
\begin{align}
&\Om_{(n)}^{\mu_{1}... \mu_{2n-1}}\,_{\al \dbe}(z_{i}) \Bigl. \Bigr| \begin{smallmatrix} z_{1} \\ \downarrow \\ z_{B} \end{smallmatrix} \ \ = \ \ 
\frac{\left[ \tspin \left( \tfrac{1}{2} \smallint^{z_A} _{z_B} \ve{\om} \right) \right]^{3-n}}{\sqrt{2}  \, \bigl[ \tspin ( \ve{0} ) \bigr]^2} \; \! \! \! \! \!  \underbrace{(E_{1A} \, z_{1B})^{-1/2}}_{(E_{1A} \, E_{1B})^{-1/2} \, + \, {\cal O}(z_{1B})}  \! \!    \prod_{i=2}^{2n-1} (E_{iA} \, E_{iB})^{-1/2} \notag \\
&  \ \ \ \ \ \ \ \times \, \sum_{\ell = 0}^{n-1} \, \biggl( \frac{E_{AB}}{2 \, \tspin \left( \tfrac{1}{2} \smallint^{z_A} _{z_B} \ve{\om} \right)} \biggr)^{\ell} \! \! \! \! \sum_{\rho \in S_{2n-2}/{\cal Q}_{n,\ell}} \! \! \!  \te{sgn}(\rho) \, \prod_{k=2}^{2\ell+1} \tspin \left( \tfrac{1}{2} \smallint^{z_A} _{z_{\rho(k)}} \ve{\om} \, + \, \tfrac{1}{2} \smallint^{z_B} _{z_{\rho(k)}} \ve{\om} \right) \notag \\
& \ \ \ \ \ \ \ \times \, \Biggl\{ \bigl(\si^{\mu_{1}} \, \bar{\si}^{\mu_{\rho(2)}} \, ... \, \si^{\mu_{\rho(2\ell+1)}} \bigr)_{\al \dbe} \; \underbrace{\frac{ \tspin \left( \tfrac{1}{2} \smallint^{z_A} _{z_{1}} \ve{\om} \, + \, \tfrac{1}{2} \smallint^{z_B} _{z_{1}} \ve{\om} \right)}{ \tspin \left( \tfrac{1}{2} \smallint^{z_A} _{z_{B}} \ve{\om} \right) }}_{= \, 1 \, + \, {\cal O}(z_{1B})} \Biggr. \notag \\
& \ \ \ \ \ \ \ \ \ \  - \ \underline{2} \, \sum_{r=1}^{\ell} \eta^{\mu_{1} \mu_{\rho(2r+1)}}  \; \bigl( \si^{\mu_{\rho(2)}} \, ... \, \si^{\mu_{\rho(2r)}} \, \bar{\si}^{\mu_{\rho(2r+2)}} \, ... \, \si^{\mu_{\rho(2\ell+1)}} \bigr)_{\al \dbe} \biggr. \notag \\
& \ \ \ \ \ \ \ \ \ \  \ \ \ \ \ \ \ \ \ \  \underbrace{\frac{E_{1A} \, E_{\rho(2r+1),B} \, \tspin \left( \smallint^{z_1} _{z_{\rho(2r+1)}} \ve{\om} \, + \, \tfrac{1}{2} \smallint^{z_A} _{z_{B}} \ve{\om} \right)}{\underline{E_{AB}} \, E_{1,\rho(2r+1)} \, \tspin \left( \tfrac{1}{2} \smallint^{z_A} _{z_{\rho(2r+1)}} \ve{\om} \, + \, \tfrac{1}{2} \smallint^{z_B} _{z_{\rho(2r+1)}} \ve{\om} \right) }}_{= \, 1  \, + \, {\cal O}(z_{1B}) } \notag \\
& \ \ \ \ \ \ \ \ \ \ \  + \ \underline{2} \, \sum_{r=1}^{\ell} \eta^{\mu_{1} \mu_{\rho(2r)}}  \; \bigl( \si^{\mu_{\rho(2)}} \, ... \, \bar{\si}^{\mu_{\rho(2r-1)}} \, \si^{\mu_{\rho(2r+1)}} \, ... \, \si^{\mu_{\rho(2\ell+1)}} \bigr)_{\al \dbe}  \notag \\
& \ \ \ \ \ \ \ \ \ \  \ \ \ \ \ \ \ \ \ \  \Biggl. \underbrace{\frac{E_{1A} \, E_{\rho(2r),B} \, 
\tspin \left( \smallint^{z_1} _{z_{\rho(2r)}} \ve{\om} \, + \, \tfrac{1}{2} \smallint^{z_A} _{z_{B}} \ve{\om} \right) }{\underline{E_{AB}} \, E_{1,\rho(2r)} \, \tspin \left( \tfrac{1}{2} \smallint^{z_A} _{z_{\rho(2r)}} \ve{\om} \, + \, \tfrac{1}{2} \smallint^{z_B} _{z_{\rho(2r)}} \ve{\om} \right)  }}_{= \, 1  \, + \, {\cal O}(z_{1B}) } \Biggr\} \notag \\
& \ \ \ \ \ \ \ \times \, \prod_{j=1}^{n-\ell-1} \frac{\eta^{\mu_{\rho(2\ell+2j)} \mu_{\rho(2\ell+2j+1)}}}{E_{\rho(2\ell+2j),\rho(2\ell+2j+1)} } \;  E_{\rho(2\ell+2j),A} \, E_{\rho(2\ell+2j+1),B}  \, \tspin \left( \smallint^{z_{\bar{\rho}(2\ell+2j)}}_{z_{\bar{\rho}(2\ell+2j+1)}} \ve{\om} \, + \, \tfrac{1}{2} \smallint^{z_A} _{z_B} \ve{\om} \right)  \ + \ {\cal O}(z_{1B}) \notag \\
&= \ \ \frac{\left[ \tspin \left( \tfrac{1}{2} \smallint^{z_A} _{z_B} \ve{\om} \right) \right]^{2-n}}{\sqrt{2}  \, \bigl[ \tspin ( \ve{0} ) \bigr]^2} \; \prod_{i=1}^{2n-1} (E_{iA} \, E_{iB})^{-1/2} \; \sum_{\ell' = 0}^{n-1} \, \biggl( \frac{E_{AB}}{2 \, \tspin \left( \tfrac{1}{2} \smallint^{z_A} _{z_B} \ve{\om} \right)} \biggr)^{\ell'} \notag \\
& \ \ \ \ \times \, \Biggl\{  \sum_{\rho \in S_{2n-2}/{\cal Q}_{n,\ell'}} \! \! \!  \te{sgn}(\rho)  \, \bigl( \si^{\mu_{1}} \, ... \, \si^{\mu_{\rho(2\ell'+1)}} \bigr)_{\al \dbe} \, \tspin \left( \tfrac{1}{2} \smallint^{z_A} _{z_{1}} \ve{\om} \, + \, \tfrac{1}{2} \smallint^{z_B} _{z_{1}} \ve{\om} \right)
 \, \prod_{k=2}^{2\ell'+1} \tspin \left( \tfrac{1}{2} \smallint^{z_A} _{z_{\rho(k)}} \ve{\om} \, + \, \tfrac{1}{2} \smallint^{z_B} _{z_{\rho(k)}} \ve{\om} \right) \notag \\
& \ \ \ \  \times \,  \prod_{j=1}^{n-\ell'-1} \frac{\eta^{\mu_{\rho(2\ell'+2j)} \mu_{\rho(2\ell'+2j+1)}}}{E_{\rho(2\ell'+2j),\rho(2\ell'+2j+1)} } \;  E_{\rho(2\ell'+2j),A} \, E_{\rho(2\ell'+2j+1),B} \, \tspin \left( \smallint^{z_{\rho(2\ell'+2j)}}_{z_{\rho(2\ell'+2j+1)}} \ve{\om} \, + \, \tfrac{1}{2} \smallint^{z_A} _{z_B} \ve{\om} \right) \notag \\
& \ \  + \ \! \! \! \! \! \! \! \! \sum_{\rho \in S_{2n-2}/{\cal Q}_{n,\ell'+1}} \! \! \! \! \! \! \!  \bigl( + \, \te{sgn}(\rho) \bigr) \, \sum_{r=1}^{\ell'+1} \bigl( \si^{\mu_{\rho(2)}} \, ... \, \bar{\si}^{\mu_{\rho(2r-1)}} \, \si^{\mu_{\rho(2r+1)}} \, ... \, \si^{\mu_{\rho(2\ell'+3)}} \bigr)_{\al \dbe} \; \frac{\eta^{\mu_{1} \mu_{\rho(2r)}}}{E_{1,\rho(2r)}} \; E_{1A} \, E_{\rho(2r),B} \notag \\
& \ \ \ \ \times \, \tspin \left( \smallint^{z_{1}}_{z_{\rho(2r)}} \ve{\om} \, + \, \tfrac{1}{2} \smallint^{z_A} _{z_B} \ve{\om} \right) \, \prod_{k=2 \atop{k \neq 2r}}^{2\ell'+3} \tspin \left( \tfrac{1}{2} \smallint^{z_A} _{z_{\rho(k)}} \ve{\om} \, + \, \tfrac{1}{2} \smallint^{z_B} _{z_{\rho(k)}} \ve{\om} \right) \notag \\
& \ \ \ \ \times \, \prod_{j=1}^{n-\ell'-2} \frac{\eta^{\mu_{\rho(2\ell'+2j+2)} \mu_{\rho(2\ell'+2j+3)}}}{E_{\rho(2\ell'+2j+2),\rho(2\ell'+2j+3)} } \;  E_{\rho(2\ell'+2j+2),A} \, E_{\rho(2\ell'+2j+3),B} \, \tspin \left( \smallint^{z_{\rho(2\ell'+2j+2)}}_{z_{\rho(2\ell'+2j+3)}} \ve{\om} \, + \, \tfrac{1}{2} \smallint^{z_A} _{z_B} \ve{\om} \right)\notag \\
& \ \ \  \Biggl. + \ \! \! \! \! \! \! \! \! \sum_{\rho \in S_{2n-2}/{\cal Q}_{n,\ell'+1}} \! \! \! \! \! \! \! \bigl(-\, \te{sgn}(\rho) \bigr) \, \sum_{r=1}^{\ell'+1} \bigl( \si^{\mu_{\rho(2)}} \, ... \, \si^{\mu_{\rho(2r)}} \, \bar{\si}^{\mu_{\rho(2r+2)}} \, ... \, \si^{\mu_{\rho(2\ell'+3)}} \bigr)_{\al \dbe} \; \frac{\eta^{\mu_{1} \mu_{\rho(2r+1)}}}{E_{1,\rho(2r+1)}} \; E_{1A} \, E_{\rho(2r+1),B} \notag \\
& \ \ \ \ \times \, \tspin \left( \smallint^{z_{1}}_{z_{\rho(2r+1)}} \ve{\om} \, + \, \tfrac{1}{2} \smallint^{z_A} _{z_B} \ve{\om} \right) \, \prod_{k=2 \atop{k \neq 2r+1}}^{2\ell'+3} \tspin \left( \tfrac{1}{2} \smallint^{z_A} _{z_{\rho(k)}} \ve{\om} \, + \, \tfrac{1}{2} \smallint^{z_B} _{z_{\rho(k)}} \ve{\om} \right) \notag \\
& \ \ \ \ \ \Biggl. \times \, \prod_{j=1}^{n-\ell'-2} \frac{\eta^{\mu_{\rho(2\ell'+2j+2)} \mu_{\rho(2\ell'+2j+3)}}}{E_{\rho(2\ell'+2j+2),\rho(2\ell'+2j+3)} } \;  E_{\rho(2\ell'+2j+2),A} \, E_{\rho(2\ell'+2j+3),B} \, \tspin \left( \smallint^{z_{\rho(2\ell'+2j+2)}}_{z_{\rho(2\ell'+2j+3)}} \ve{\om} \, + \, \tfrac{1}{2} \smallint^{z_A} _{z_B} \ve{\om} \right) \Biggr\} \notag \\
& \ \ \ \ \ \ \ \ \ \ \ \ \ \ \ \ \ \ \ \ + \ {\cal O}(z_{1B})   \notag \\
&= \ \ \frac{\left[ \tspin \left( \tfrac{1}{2} \smallint^{z_A} _{z_B} \ve{\om} \right) \right]^{2-n}}{\sqrt{2}  \, \bigl[ \tspin ( \ve{0} ) \bigr]^2} \; \prod_{i=1}^{2n-1} (E_{iA} \, E_{iB})^{-1/2} \; \sum_{\ell' = 0}^{n-1} \, \biggl( \frac{E_{AB}}{2 \, \tspin \left( \tfrac{1}{2} \smallint^{z_A} _{z_B} \ve{\om} \right)} \biggr)^{\ell'} \! \! \! \sum_{\bar{\rho} \in S_{2n-1}/{\cal P}_{n,\ell'}} \! \! \!  \te{sgn}(\bar{\rho})  \notag \\
& \ \ \ \ \times \, \bigl(\si^{\mu_{\bar{\rho}(1)}} \,  ...  \, \si^{\mu_{\bar{\rho}(2\ell'+1)}} \bigr)_{\al \dbe} \, \prod_{k=1}^{2\ell'+1} \tspin \left( \tfrac{1}{2} \smallint^{z_A} _{z_{\bar{\rho}(k)}} \ve{\om} \, + \, \tfrac{1}{2} \smallint^{z_B} _{z_{\bar{\rho}(k)}} \ve{\om} \right) \, \prod_{j=0}^{n-\ell'-2} \frac{\eta^{\mu_{\bar{\rho}(2\ell'+2j+2)} \mu_{\bar{\rho}(2\ell'+2j+3)}}}{E_{\bar{\rho}(2\ell'+2j+2),\bar{\rho}(2\ell'+2j+3)} } \notag \\
& \ \ \ \ \times \;  E_{\bar{\rho}(2\ell'+2j+2),A} \, E_{\bar{\rho}(2\ell'+2j+3),B}\, \tspin \left( \smallint^{z_{\bar{\rho}(2\ell'+2j+2)}}_{z_{\bar{\rho}(2\ell'+2j+3)}} \ve{\om} \, + \, \tfrac{1}{2} \smallint^{z_A} _{z_B} \ve{\om} \right)  \ + \ {\cal O}(z_{1B}) 
\label{npt,25}
\end{align}
Several steps might require some further explanation here: Firstly, the underlined factors $\frac{2}{E_{AB}}$ in the third and fourth line together with one $\tspin \left( \tfrac{1}{2} \smallint^{z_A} _{z_B} \ve{\om} \right)$ power of the prefactor shift the summation variable $\ell$ by 1 in the $\eta$ contributions of (\ref{roeven}). Secondly, we have replaced one $S_{2n-2}/{\cal Q}_{n,\ell'}$ sum and $2\ell'+2$ sums over $S_{2n-2}/{\cal Q}_{n,\ell'+1}$ by the $S_{2n-1}/{\cal P}_{n,\ell'}$ sum over larger permutations $\bar{\rho}$ including $\mu_{1}$. The total number of terms is the same since
\begin{align}
\bigl| S_{2n-2}/{\cal Q}_{n,\ell} & \bigr| \ + \ (2\ell \, + \, 2) \, \bigl| S_{2n-2}/{\cal Q}_{n,\ell+1} \bigr| \ \ = \ \ \frac{(2n \, - \, 2)!}{(2\ell)! \, (n \, - \, \ell \, - \, 1)! \, 2^{n-\ell-1}} \notag \\
& \ \ \ \ \ \ + \ \frac{(2\ell \, + \, 2) \, (2n\, - \, 2)!}{(2\ell \, + \, 2)! \, (n \, - \, \ell \, - \, 2)! \, 2^{n-\ell-2}} \notag \\
&= \ \ \frac{(2n \, - \, 2)!}{(2\ell \, + \, 1)! \, (n \, - \, \ell \, - \, 1)! \, 2^{n-\ell-1}} \; \Bigl( 2\ell \, + \, 1 \, + \, 2 \  (n-\ell-1) \Bigr) \notag \\
&= \ \ \frac{(2n \, - \, 1)!}{(2\ell \, + \, 1)! \, (n \, - \, \ell \, - \, 1)! \, 2^{n-\ell-1}} \notag \\
&= \ \ \bigl| S_{2n-1}/{\cal P}_{n,\ell} \bigr| \ .
\label{npt,26}
\end{align}
Moreover, the index $\mu_{1}$ appears in all possible positions in (\ref{npt,25}), i.e.\ attached to $\si$'s as well as to $\eta$'s. The relative sign between $\rho$ and $\bar{\rho}$ is taken into account.

\medskip
Following the general principle of this proof, we should now compare the final lines of (\ref{npt,25}) with the most singular terms in the expression (\ref{Omega}) for $\Om_{(n)}$. Due to our particular arrangement of $\si$ matrices $\rho(1) < ... < \rho(2\ell+1)$, no factors of $E_{1B}$ appear in any denominator. So the right hand side of (\ref{Omega}) keeps all its term of the $\ell$- and $\rho$ sums in the $z_{1} \mto z_{B}$ regime. However, it already matches with (\ref{npt,25}) up to a shift in the $j$ product, so we are done with the $\Om_{(n)} \ra \om_{(n)}$ reduction.

\medskip
Actually, this is the reason we sent $z_{1}$ and not any other $z_{i}$, $i=2,3,...,2n-1$, to $z_{B}$. Only this limit tests every single term in (\ref{Omega}). For the same reasons, the behaviour of $\Om_{(n)}$ under $z_{2n-1} \mto z_{A}$ gives rise to equally rich consistency checks as the right hand side of (\ref{Omega}) does not have any poles in $(z_{2n-1} - z_{A})$ beyond order $1/2$.

\subsubsection{$\om_{(n)} \ra \Om_{(n-1)}$ by $z_{1} \mto z_{B}$ limit}
\label{sec:vii}

The reduction of $\om_{(n)}$ to correlators of $\Om_{(n-1)}$ type is based on the OPE \eqref{rv,2a}:
\begin{align}
&\om_{(n)}^{\mu_{1} ... \mu_{2n-2}}\,  _{\al} \, ^{ \be} (z_{i}) \Bigl. \Bigr| \begin{smallmatrix} z_{1} \\ \downarrow \\ z_{B} \end{smallmatrix} \ \ \stackrel{!}{=} \ \  \frac{i \, (\bar{\si}^{\mu_{1}})^{\dga \be}}{\sqrt{2} \, (z_{1B})^{1/2}} \;  \langle \psi^{\mu_{2}}(z_{2}) \, ... \, \psi^{\mu_{2n-2}}(z_{2n-2}) \, S_{\al}(z_{A}) \, S_{\dga}(z_{B}) \rangle \spin  \ + \ {\cal O}(z_{1B}) \notag \\
&= \ \ \frac{i}{\sqrt{2} \, (z_{1 B})^{1/2}} \; \Om_{(n-1)}^{\mu_{2} ... \mu_{2n-2}}\, _{\al \dga} (z_{i}) \, (\bar{\si}^{\mu_{1}})^{\dga \be}  \ + \ {\cal O}(z_{1B})  \notag \\
&= \ \ \frac{1}{2 \, z_{1B}^{1/2}} \; \underbrace{\frac{i \, E_{AB}}{E_{AB}^{1/2}  \, E_{1A}^{1/2}}}_{= \, 1 \, + \, {\cal O}(z_{1B})} \; \frac{\left[ \tspin \left( \tfrac{1}{2} \smallint^{z_A} _{z_B} \ve{\om} \right) \right]^{2-n}}{\sqrt{2} \, \bigl[ \tspin ( \ve{0} ) \bigr]^2  \, \prod_{i=2}^{2n-2} (E_{iA} \, E_{iB})^{1/2} } \, \sum_{\ell = 0}^{n-2} \, \biggl( \frac{E_{AB}}{2 \, \tspin \left( \tfrac{1}{2} \smallint^{z_A} _{z_B} \ve{\om} \right)} \biggr)^{\ell} \notag \\
& \ \ \ \ \times \sum_{\rho \in S_{2n-3}/{\cal P}_{n-1,\ell}} \! \! \!  \te{sgn}(\rho) \, \bigl(\si^{\mu_{\rho(2)}} \,   ... \, \si^{\mu_{\rho(2\ell+2)}} \bigr)_{\al \dga} \, (\bar{\si}^{\mu_{1}})^{\dga \be} \, \prod_{k=2}^{2\ell+2} \tspin \left( \tfrac{1}{2} \smallint^{z_A} _{z_{\rho(k)}} \ve{\om} \, + \, \tfrac{1}{2} \smallint^{z_B} _{z_{\rho(k)}} \ve{\om} \right) \notag \\
& \ \ \ \ \times \  \prod_{j=1}^{n-\ell-2} \frac{\eta^{\mu_{\rho(2\ell+2j+1)} \mu_{\rho(2\ell+2j+2)}}}{E_{\rho(2\ell+2j+1),\rho(2\ell+2j+2)} } \; E_{\rho(2\ell+2j+1),A} \, E_{\rho(2\ell+2j+2),B}  \, \tspin \left( \smallint^{z_{\rho(2\ell+2j+1)}}_{z_{\rho(2\ell+2j+2)}} \ve{\om} \, + \, \tfrac{1}{2} \smallint^{z_A} _{z_B} \ve{\om} \right) \notag \\
& \ \ \ \ \ \ \ \ \ \ \ \ \ \ \ \ \ \ \ \ + \ {\cal O}(z_{1B}) 
\label{npt,29}
\end{align}
Here, it is necessary to move the $\bar{\si}^{\mu_{1}}$ to the left of $\si^{\mu_{\rho(2)}} ... \si^{\mu_{\rho(2\ell+2)}}$, i.e. across an odd number of $\si$ matrices. With the help of
\begin{align}
\si^{\la_{1}} \, \bar{\si}^{\la_{2}} \, ... \, &\bar{\si}^{\la_{2n-2}} \, \si^{\la_{2n-1}} \, \bar{\si}^{\rho} \ \ = \ \ - \ \si^{\rho} \, \bar{\si}^{\la_{1}} \, \si^{\la_{2}} \, ... \, \si^{\la_{2n-2}} \, \bar{\si}^{\la_{2n-1}} \notag \\
& - \ 2 \, \sum_{r=1}^{n} \, \eta^{\rho \la_{2r-1}} \, \si^{\la_{1}} \, ... \, \bar{\si}^{\la_{2r-2}} \, \si^{\la_{2r}} \, ... \, \bar{\si}^{\la_{2n-1}} \notag \\
& + \ 2 \, \sum_{r=1}^{n-1} \, \eta^{\rho \la_{2r}} \, \si^{\la_{1}} \, ... \, \si^{\la_{2r-1}} \, \bar{\si}^{\la_{2r+1}} \, ... \, \bar{\si}^{\la_{2n-1}} 
\label{roodd}
\end{align}
we obtain:
\begin{align}
&\om_{(n)}^{\mu_{1} ... \mu_{2n-2}}\,  _{\al} \, ^{ \be} (z_{i}) \Bigl. \Bigr| \begin{smallmatrix} z_{1} \\ \downarrow \\ z_{B} \end{smallmatrix} \ \ = \ \ \frac{\left[ \tspin \left( \tfrac{1}{2} \smallint^{z_A} _{z_B} \ve{\om} \right) \right]^{2-n}}{2 \, \bigl[ \tspin ( \ve{0} ) \bigr]^2} \; \frac{\underline{ E_{AB}}}{\underline{2} \, E_{AB}^{1/2}} \; \! \! \! \!   \underbrace{ (E_{1A} \, z_{1B})^{-1/2} }_{(E_{1A} \, E_{1B})^{-1/2} \, + \, {\cal O}(z_{1B})} \! \! \prod_{i=2}^{2n-2} (E_{iA} \, E_{iB})^{-1/2} \notag \\
& \ \ \ \ \times \ \sum_{\ell = 0}^{n-2} \, \biggl( \frac{E_{AB}}{2 \, \tspin \left( \tfrac{1}{2} \smallint^{z_A} _{z_B} \ve{\om} \right)} \biggr)^{\ell}  \ \! \! \! \! \sum_{\rho \in S_{2n-3}/{\cal P}_{n-1,\ell}} \! \! \! \! \! \te{sgn}(\rho) \, \prod_{k=2}^{2\ell+2} \tspin \left( \tfrac{1}{2} \smallint^{z_A} _{z_{\rho(k)}} \ve{\om} \, + \, \tfrac{1}{2} \smallint^{z_B} _{z_{\rho(k)}} \ve{\om} \right) \notag \\
& \ \ \ \ \times \, \Biggl\{ - \ \bigl( \si^{\mu_{1}} \, \bar{\si}^{\mu_{\rho(2)}} \,  ... \,  \bar{\si}^{\mu_{\rho(2\ell+2)}} \bigr)_{\al } \, ^{\be} \, \underbrace{\frac{ \tspin \left( \tfrac{1}{2} \smallint^{z_A} _{z_{1}} \ve{\om} \, + \, \tfrac{1}{2} \smallint^{z_B} _{z_{1}} \ve{\om} \right)}{ \tspin \left( \tfrac{1}{2} \smallint^{z_A} _{z_{B}} \ve{\om} \right) }}_{= \, 1 \, + \, {\cal O}(z_{1B})} \Biggr. \notag \\
&\ \ \ \ \ \ \ \ \ \  - \ \underline{2} \, \sum_{r=1}^{\ell+1} \eta^{\mu_{1} \mu_{\rho(2r)}} \, \bigl( \si^{\mu_{\rho(2)}} \, ... \, \bar{\si}^{\mu_{\rho(2r-1)}} \,  \si^{\mu_{\rho(2r+1)}} \, ... \, \bar{\si}^{\mu_{\rho(2\ell+2)}} \bigr)_{\al } \, ^{\be} \notag \\
& \ \ \ \ \ \ \ \ \ \  \ \ \ \ \ \ \ \ \ \  \underbrace{\frac{E_{1A} \, E_{\rho(2r),B} \, 
\tspin \left( \smallint^{z_1} _{z_{\rho(2r)}} \ve{\om} \, + \, \tfrac{1}{2} \smallint^{z_A} _{z_{B}} \ve{\om} \right) }{\underline{E_{AB}} \, E_{1,\rho(2r)} \, \tspin \left( \tfrac{1}{2} \smallint^{z_A} _{z_{\rho(2r)}} \ve{\om} \, + \, \tfrac{1}{2} \smallint^{z_B} _{z_{\rho(2r)}} \ve{\om} \right)  }}_{= \, 1  \, + \, {\cal O}(z_{1B}) } \notag \\
&\ \ \ \ \ \ \ \ \ \ \  + \ \underline{2} \, \sum_{r=1}^{\ell} \eta^{\mu_{1} \mu_{\rho(2r+1)}} \, \bigl( \si^{\mu_{\rho(2)}} \, ... \, \si^{\mu_{\rho(2r)}} \,  \bar{\si}^{\mu_{\rho(2r+2)}} \, ... \, \bar{\si}^{\mu_{\rho(2\ell+2)}} \bigr)_{\al } \, ^{\be} \notag \\
& \ \ \ \ \ \ \ \ \ \  \ \ \ \ \ \ \ \ \ \ \Biggl. \underbrace{\frac{E_{1A} \, E_{\rho(2r+1),B} \, \tspin \left( \smallint^{z_1} _{z_{\rho(2r+1)}} \ve{\om} \, + \, \tfrac{1}{2} \smallint^{z_A} _{z_{B}} \ve{\om} \right)}{\underline{E_{AB}} \, E_{1,\rho(2r+1)} \, \tspin \left( \tfrac{1}{2} \smallint^{z_A} _{z_{\rho(2r+1)}} \ve{\om} \, + \, \tfrac{1}{2} \smallint^{z_B} _{z_{\rho(2r+1)}} \ve{\om} \right) }}_{= \, 1  \, + \, {\cal O}(z_{1B}) } \Biggr\} \notag \\
& \ \ \ \ \times \ \prod_{j=1}^{n-\ell-2} \frac{\eta^{\mu_{\rho(2\ell+2j+1)} \mu_{\rho(2\ell+2j+2)}}}{E_{\rho(2\ell+2j+1),\rho(2\ell+2j+2)} } \; E_{\rho(2\ell+2j+1),A} \, E_{\rho(2\ell+2j+2),B}  \, \tspin \left( \smallint^{z_{\rho(2\ell+2j+1)}}_{z_{\rho(2\ell+2j+2)}} \ve{\om} \, + \, \tfrac{1}{2} \smallint^{z_A} _{z_B} \ve{\om} \right) \ + \ {\cal O}(z_{1B})
 \notag \\
  &= \ \ \frac{- \, \left[ \tspin \left( \tfrac{1}{2} \smallint^{z_A} _{z_B} \ve{\om} \right) \right]^{3-n}}{  \bigl[ \tspin ( \ve{0} ) \bigr]^2 \, E_{AB}^{1/2}} \; \prod_{i=1}^{2n-2} (E_{iA} \, E_{iB})^{-1/2} \, \sum_{\ell' = 0}^{n-1} \, \biggl( \frac{E_{AB}}{2 \, \tspin \left( \tfrac{1}{2} \smallint^{z_A} _{z_B} \ve{\om} \right)} \biggr)^{\ell'} \notag \\
& \ \ \ \ \times \, \Biggl\{ \sum_{\rho \in S_{2n-3}/{\cal P}_{n-1,\ell'-1}} \! \! \! \! \! \! \! \! \! \! \! \! \te{sgn}(\rho) \, \bigl( \si^{\mu_{1}} \,  ... \,  \bar{\si}^{\mu_{\rho(2\ell')}} \bigr)_{\al } \, ^{\be} \, \tspin \left( \tfrac{1}{2} \smallint^{z_A} _{z_{1}} \ve{\om} \, + \, \tfrac{1}{2} \smallint^{z_B} _{z_{1}} \ve{\om} \right) \, \prod_{k=2}^{2\ell'} \tspin \left( \tfrac{1}{2} \smallint^{z_A} _{z_{\rho(k)}} \ve{\om} \, + \, \tfrac{1}{2} \smallint^{z_B} _{z_{\rho(k)}} \ve{\om} \right) \notag \\
& \ \ \ \ \times \, \prod_{j=1}^{n-\ell'-1} \frac{\eta^{\mu_{\rho(2\ell'+2j-1)} \mu_{\rho(2\ell'+2j)}}}{E_{\rho(2\ell'+2j- 1),\rho(2\ell'+2j)} } \;  E_{\rho(2\ell'+2j-1),A}  \, E_{\rho(2\ell'+2j),B} \, \tspin \left( \smallint^{z_{\rho(2\ell'+2j-1)}}_{z_{\rho(2\ell'+2j)}} \ve{\om} \, + \, \tfrac{1}{2} \smallint^{z_A} _{z_B} \ve{\om} \right) \notag \\
& \ \ + \! \! \! \! \sum_{\rho \in S_{2n-3}/{\cal P}_{n-1,\ell'}} \! \! \! \! \! \bigl( + \,\te{sgn}(\rho) \bigr) \, \sum_{r=1}^{\ell'+1} \bigl( \si^{\mu_{\rho(2)}} \, ... \, \bar{\si}^{\mu_{\rho(2r-1)}} \,  \si^{\mu_{\rho(2r+1)}} \, ... \, \bar{\si}^{\mu_{\rho(2\ell'+2)}} \bigr)_{\al } \, ^{\be} \,  \; \frac{\eta^{\mu_{1} \mu_{\rho(2r)}}}{E_{1,\rho(2r)}} \; E_{1A} \, E_{\rho(2r),B} \notag \\ 
  & \ \ \ \ \times \ \tspin \left( \smallint^{z_1} _{z_{\rho(2r)}} \ve{\om} \, + \, \tfrac{1}{2} \smallint^{z_A} _{z_{B}} \ve{\om} \right) \, \prod_{k=2 \atop{k \neq 2r}}^{2\ell'+2} \tspin \left( \tfrac{1}{2} \smallint^{z_A} _{z_{\rho(k)}} \ve{\om} \, + \, \tfrac{1}{2} \smallint^{z_B} _{z_{\rho(k)}} \ve{\om} \right) \notag \\
& \ \ \ \ \times \, \prod_{j=1}^{n-\ell'-2} \frac{\eta^{\mu_{\rho(2\ell'+2j+1)} \mu_{\rho(2\ell'+2j+2)}}}{E_{\rho(2\ell'+2j+1),\rho(2\ell'+2j+2)} } \;  E_{\rho(2\ell'+2j+1),A}  \, E_{\rho(2\ell'+2j+2),B}  \, \tspin \left( \smallint^{z_{\rho(2\ell'+2j+1)}}_{z_{\rho(2\ell'+2j+2)}} \ve{\om} \, + \, \tfrac{1}{2} \smallint^{z_A} _{z_B} \ve{\om} \right) \notag \\  
  & \ \ + \! \! \! \! \sum_{\rho \in S_{2n-3}/{\cal P}_{n-1,\ell'}} \! \! \! \! \! \bigl( - \, \te{sgn}(\rho) \bigr) \, \sum_{r=1}^{\ell'} \bigl( \si^{\mu_{\rho(2)}} \, ... \, \si^{\mu_{\rho(2r)}} \,  \bar{\si}^{\mu_{\rho(2r+2)}} \, ... \, \bar{\si}^{\mu_{\rho(2\ell'+2)}} \bigr)_{\al } \, ^{\be} \; \frac{\eta^{\mu_{1} \mu_{\rho(2r+1)}}}{E_{1,\rho(2r+1)}} \; E_{1A} \, E_{\rho(2r+1),B} \notag \\
& \ \ \ \ \times \ \tspin \left( \smallint^{z_1} _{z_{\rho(2r+1)}} \ve{\om} \, + \, \tfrac{1}{2} \smallint^{z_A} _{z_{B}} \ve{\om} \right) \, \prod_{k=2 \atop{k \neq 2r+1}}^{2\ell'+2} \tspin \left( \tfrac{1}{2} \smallint^{z_A} _{z_{\rho(k)}} \ve{\om} \, + \, \tfrac{1}{2} \smallint^{z_B} _{z_{\rho(k)}} \ve{\om} \right) \notag \\
& \ \ \ \ \ \Biggl. \times \, \prod_{j=1}^{n-\ell'-2} \frac{\eta^{\mu_{\rho(2\ell'+2j+1)} \mu_{\rho(2\ell'+2j+2)}}}{E_{\rho(2\ell'+2j+1),\rho(2\ell'+2j+2)} } \;  E_{\rho(2\ell'+2j+1),A}  \, E_{\rho(2\ell'+2j+2),B}  \, \tspin \left( \smallint^{z_{\rho(2\ell'+2j+1)}}_{z_{\rho(2\ell'+2j+2)}} \ve{\om} \, + \, \tfrac{1}{2} \smallint^{z_A} _{z_B} \ve{\om} \right) \Biggr\} \notag \\
& \ \ \ \ \ \ \ \ \ \ \ \ \ \ \ \ \ \ \ \ + \ {\cal O}(z_{1B})   \notag \\  
&= \ \ \frac{- \, \left[ \tspin \left( \tfrac{1}{2} \smallint^{z_A} _{z_B} \ve{\om} \right) \right]^{3-n}}{  \bigl[ \tspin ( \ve{0} ) \bigr]^2 \, E_{AB}^{1/2}} \; \prod_{i=1}^{2n-2} (E_{iA} \, E_{iB})^{-1/2} \, \sum_{\ell' = 0}^{n-1} \, \biggl( \frac{E_{AB}}{2 \, \tspin \left( \tfrac{1}{2} \smallint^{z_A} _{z_B} \ve{\om} \right)} \biggr)^{\ell'}  \! \! \! \sum_{\bar{\rho} \in S_{2n-2}/{\cal Q}_{n,\ell'}} \! \! \! \! \! \te{sgn}(\bar{\rho}) \notag \\
& \ \ \ \ \times \, \bigl( \si^{\mu_{\bar{\rho}(1)}} \, ... \, \bar{\si}^{\mu_{\bar{\rho}(2\ell')}} \bigr)_{\al} \,^{\be} \, \prod_{k=1}^{2\ell'} \tspin \left( \tfrac{1}{2} \smallint^{z_A} _{z_{\bar{\rho}(k)}} \ve{\om} \, + \, \tfrac{1}{2} \smallint^{z_B} _{z_{\bar{\rho}(k)}} \ve{\om} \right) \, \prod_{j=0}^{n-\ell'-2} \frac{\eta^{\mu_{\bar{\rho}(2\ell'+2j+1)} \mu_{\bar{\rho}(2\ell'+2j+2)}}}{E_{\bar{\rho}(2\ell'+2j+1),\bar{\rho}(2\ell'+2j+2)} }\notag \\
& \ \ \ \ \times \, E_{\bar{\rho}(2\ell'+2j+1),A}  \, E_{\bar{\rho}(2\ell'+2j+2),B} \, \tspin \left( \smallint^{z_{\bar{\rho}(2\ell'+2j+1)}}_{z_{\bar{\rho}(2\ell'+2j+2)}} \ve{\om} \, + \, \tfrac{1}{2} \smallint^{z_A} _{z_B} \ve{\om} \right)  \ + \ {\cal O}(z_{1B})
\label{npt,30}
\end{align}
The underlined factors of $\frac{2}{E_{AB}}$ in the fourth and sixth line cancel with the$ \frac{E_{AB}}{2}$ from the prefactor, so only the $\si^{\mu_{1}} \bar{\si}^{\mu_{\rho(2)}}  ...  \bar{\si}^{\mu_{\rho(2\ell+2)}}$ term in the third line requires a relabelling $\ell \mto \ell'=\ell+1$. In the last step we have regrouped the $S_{2n-3}$ permutations of total number
\begin{align}
\bigl| S_{2n-3}/{\cal P}_{n-1,\ell-1} & \bigr| \ + \ (2\ell \, + \, 1) \, \bigl| S_{2n-3}/{\cal P}_{n-1,\ell} \bigr| \ \ = \ \ \frac{(2n \, - \, 3)!}{(2\ell \, - \, 1)! \, (n \, - \, \ell \, - \, 1)! \, 2^{n-\ell-1}} \notag \\
& \ \ \ \ \ \ + \ \frac{(2\ell \, + \, 1) \, (2n\, - \, 3)!}{(2\ell \, + \, 1)! \, (n \, - \, \ell \, - \, 2)! \, 2^{n-\ell-2}} \notag \\
&= \ \ \frac{(2n \, - \, 3)!}{(2\ell)! \, (n \, - \, \ell \, - \, 1)! \, 2^{n-\ell-1}} \; \Bigl( 2\ell \,  + \, 2 \  (n-\ell-1) \Bigr) \notag \\
&= \ \ \frac{(2n \, - \, 2)!}{(2\ell )! \, (n \, - \, \ell \, - \, 1)! \, 2^{n-\ell-1}} \notag \\
&= \ \ \bigl| S_{2n-2}/{\cal Q}_{n,\ell} \bigr| \ .
\label{npt,31}
\end{align}
They exhaust all the possible $S_{2n-2} / {\cal Q}_{n,\ell}$ elements $\bar{\rho}$ where $\mu_{1}$ is included. By carefully looking at the $\si$ strings, the reader can check that indeed $\te{sgn}(\rho) = \te{sgn}(\bar{\rho})$ in all cases. The result of (\ref{npt,30}) is exactly what was claimed in (\ref{omega}) for $\om_{(n)}$.

\end{document}